\documentclass[onecolumn,amsmath,amssymb,aps]{revtex4-1}
\usepackage{graphicx}
\usepackage{caption}
\usepackage{float}
\usepackage{dcolumn}
\usepackage{bm}
\usepackage[mathlines]{lineno}
\usepackage{tasks}
\usepackage[margin=1.5in]{geometry}
\usepackage{amsmath}
\usepackage{natbib}
\usepackage[hidelinks]{hyperref}
\usepackage{cleveref}

\begin{document}

%\preprint{APS/123-QED}

\title{Detection techniques and investigation of different neutrino experiments}

\author{Ankur Nath}
\email{ankur04@tezu.ernet.in}
\author{Ng. K. Francis}
 \email{ngkf2010@gmail.com}
\affiliation{Department Of Physics, Tezpur University, Assam-784028, India}
\date{\today}

\begin{abstract}
Neutrino physics is an experimentally driven field. So, we investigate the different detection techniques available in the literature and study the various neutrino oscillation experiments in a chronological manner. Our primary focus is on the construction and detection mechanisms of each experiment. Today, we know a lot about this mysterious ghostly particle by performing different experiments at different times with different neutrino sources viz. solar, atmospheric, reactor, accelerators and high energy astrophysical; and they have contributed in the determination of neutrino parameters. Yet the problems are far from over. We need to determine more precise values of the already known parameters and unravel the completely unknown parameters. Some of the unknowns are absolute masses of neutrino, types of neutrino, mass hierarchy, octant degeneracy and existence of leptonic CP Phase(s). We analyse the neutrino experiments into the past, present and the future (or proposed). We include SNO, Kamiokande, K2K, MINOS, MINOS+, Chooz, NEMO and ICARUS in the past; while Borexino, Double Chooz, Super-K, T2K, IceCube, KamLAND, NO$\nu$A, RENO and Daya Bay in the present; and SNO+, Hyper-K, T2HK, JUNO, RENO-50, INO, DUNE, SuperNEMO, KM3NeT, P2O, LBNO and PINGU in the proposed experiments. We also discuss the necessities of upgrading the present ones to those of the proposed ones thereby summarizing the potentials of the future experiments. We conclude this paper with the current status of the neutrinos.
\begin{description}
\item[PACS numbers]
14.60.Lm, 14.60.Pq, 13.15.+g, 95.55.Vj, 01.65.+g
\item[Keywords] 
neutrino detection techniques, different neutrino experiments, unknown parameters 
\end{description}
\end{abstract}

\pacs{Valid PACS appear here}% PACS, the Physics and Astronomy
%Use showkeys class option if keyword
\maketitle

%\tableofcontents

\section{INTRODUCTION}
\begin{figure}[h]
	\includegraphics[width=5.5in]{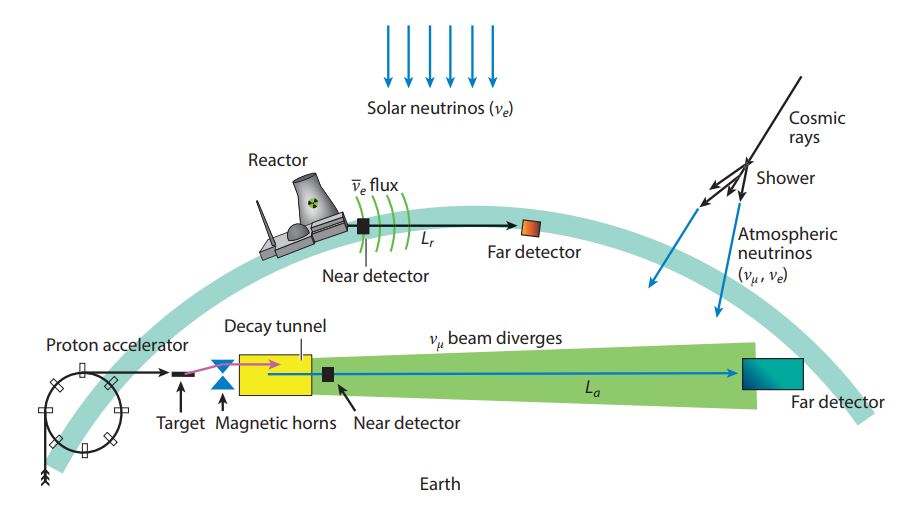}
	\caption{Natural and artificial sources of neutrino in a nutshell\cite{diwan2016long}.}
	\label{Fig.1}
\end{figure}

Wolfgang Pauli, after postulating the existence of the neutrino -- a particle with no net charge and no mass, remarked during a visit to California Institute of Technology\cite{reines1996neutrino}:
\begin{quote}
\textit{I have done a terrible thing: I have postulated a particle that cannot be detected.} 
\end{quote}
Pauli was, however, proved wrong with the result of first-ever direct detection of electron anti-neutrino($\bar{\nu}_e$) by Clyde L. Cowan et al.\cite{cowan1956test} in 1956 from the Savannah river reactor plant in South Carolina. In other words, this discovery validated Pauli's idea\cite{pauli1930letter} of this new particle was assumed in 1930 as a desperate remedy to explain the discrepancy in the beta decay spectrum. Thereafter, the physics community in various corners of the world started conducting underground (\textit{in old mines, excavated beneath the mountains}), under water (\textit{on seabed}) and under ice experiments and a good number of experiments have been upgraded from time to time to understand better about this elusive particle. The other two active neutrinos \textit{viz.} muon neutrino($\nu_{\mu}$) and tau neutrino($\nu_{\tau}$) were directly observed in 1962\cite{danby1962observation} and 2000\cite{kodama2001observation}, respectively. Since then, the development in this field has been exponential. Now, we know that neutrino carries a very tiny but a non-zero mass. All the existing experiments detecting the neutrinos from natural sources (\textit{solar, atmospheric and astrophysical}) and man-made sources (\textit{reactor and accelerator-based})(Figure \ref{Fig.1}) are based on one of the detection techniques \textit{viz.} radiochemical methods, Cherenkov method (\textit{water, ice}), scintillation technique (\textit{solid, liquid}), tracking calorimeter, nuclear emulsions and liquid argon(LAr).\\
The three known neutrino flavour states ($\nu_{e}$, $\nu_{\mu}$, $\nu_{\tau}$) are expressed as quantum superpositions of three massive states $\nu_k$ $(k=1,2,3)$ with different masses $m_k$ with a 3$\times$3 unitary mixing matrix $U_{\alpha k}$($\alpha=e,\mu,\tau$), known as PMNS (Pontecorvo-Maki-Nakagawa-Sakata) matrix $U_{PMNS}$\cite{maki1962remarks,nakamura2010neutrino}, given by
\begin{align*}
U_{PMNS}=U_{\alpha k}&=\begin{bmatrix}
U_{e1}&U_{e2}&U_{e3}\\
U_{\mu 1}&U_{\mu 2}&U_{\mu 3}\\
U_{\tau 1}&U_{\tau 2}&U_{\tau 3}
\end{bmatrix}
\end{align*} 
The $U_{PMNS}$ matrix can be decomposed as:
\begin{equation}
\underbrace{
	\overbrace{\begin{bmatrix}
		\cos\theta_{12}&\sin\theta_{12}&0\\
		-\sin\theta_{12}&\cos\theta_{12}&0\\
		0&0&1
		\end{bmatrix}}^{Solar}
\overbrace{\begin{bmatrix}
		\cos\theta_{13}&0&\sin\theta_{13}e^{-i\delta_{CP}}\\
		0&1&0\\
		-\sin\theta_{13}e^{-i\delta_{CP}}&0&\cos\theta_{13}
		\end{bmatrix}}^{Reactor}
\overbrace{\begin{bmatrix}
		1&0&0\\
		0&\cos\theta_{23}&\sin\theta_{23}\\
		0&-\sin\theta_{23}&\cos\theta_{23}
		\end{bmatrix}}^{Atmospheric}
\cdot M}_{U_{PMNS}}
\end{equation}
and characterized by three non-zero angles $\theta_{kl}\in[0,\frac{\pi}{2}]$ and a charge-parity violating phase $\delta_{CP}\in[0,2\pi]$. The matrix M has a value of $\det M=1$ for the Dirac neutrinos and $M=diag(1,e^{i\alpha_{2}},e^{i\alpha_{3}})$ for Majorana neutrinos\cite{nakamura2010neutrino}. The mixing angles $\theta_{kl}$ are associated with solar, atmospheric and reactor neutrinos given by $\theta_{12}$, $\theta_{23}$ and $\theta_{13}$ respectively, the mass-squared differences i.e. the mass splitting terms being $\Delta m_{21}^2$, $\Delta m_{32}^2$ and $\Delta m_{31}^2$ with $\Delta m_{kl}^{2}=m_{k}^{2}-m_{l}^{2}$. The mass splitting terms can be expressed as:
\begin{align}
\Delta m_{21}^{2}=m_{2}^{2}-m_{1}^{2}, \quad
\Delta m_{3\textit{n}}^{2}=m_{3}^{2}-\frac{(m_{2}^{2}+m_{1}^{2})}{2}
\end{align}
such that, $\Delta m_{21}^2>0$ and $\Delta m_{3n}^2\equiv\Delta m_{31}^2>0$ is positive for Normal Mass Hierarchy (NH) and $\Delta m_{3n}^2\equiv\Delta m_{32}^2<0$ is negative for Inverted Mass Hierarchy (IH) for the neutrino mass spectrum\cite{fogli2002supernova}.\\
The measurements and development in the precision of these parameters can guide the physicists in understanding a few properties of neutrino which include mass ordering of the three neutrino flavours $\nu_{e}$, $\nu_{\mu}$ and $\nu_{\tau}$ associated with the three leptons $e^{-}$, $\mu^{-}$ and $\tau^{-}$ respectively, and CP violating phase $\delta_{CP}$ which can probe the dominance of matter over anti-matter in the universe.\\
The outline of the paper is as follows: \textbf{Section 2} describes the detection mechanisms of neutrino in a brief and concise manner followed by the names of the experiments which exploit these techniques. The paper emphasizes on different neutrino experiments in a chronological pattern : \textit{past, present and future} in \textbf{Section 3}. The section also explains the construction \& detection mechanisms and results \& scopes of the particular experiments included in this paper. In \textbf{Section 4}, we give a picture of the current status on neutrino mentioning about the measured and less known parameters those are dealt by these experiments. We also study the limits on absolute neutrino masses. \textbf{Section 5} contains the summary of the overall descriptions of the paper.
\section{NEUTRINO DETECTION TECHNIQUES}
\subsection{Radiochemical Method}
Inverse Beta Decay (IBD) is the principle used in the radiochemical method of neutrino detection.
\begin{equation}
\nu_{e}+{^A}_{Z}X\longrightarrow e^{-}+{^A}_{Z+1}Y
\end{equation}
In this process, when a neutrino is absorbed in the target of the detection medium, the target is converted into a radioactive element whose decay is further studied and counted. The technique was exploited by the famous Homestake experiment\cite{cleveland1998measurement,davis1994review}, GALLEX\cite{hampel1999gallex}, GNO and SAGE\cite{abdurashitov2009measurement} experiments to detect low energy solar neutrinos. The reactions employed in the above experiments are:
\paragraph{HOMESTAKE-Cl} 
$$\nu_{e}+{^{37}}_{17}Cl (\textit{target})\xrightarrow[\text{Q-value}]{\text{0.814 MeV}} e^{-}+{^{37}}_{18}Ar$$
\paragraph{GALLEX/GNO/SAGE} 
$$\nu_{e}+{^{71}}_{31}Ga (\textit{target}) \xrightarrow[\text{Q-value}]{\text{233 keV}} e^{-}+{^{71}}_{32}Ge$$
\begin{figure}[h]
	\includegraphics[width=4in]{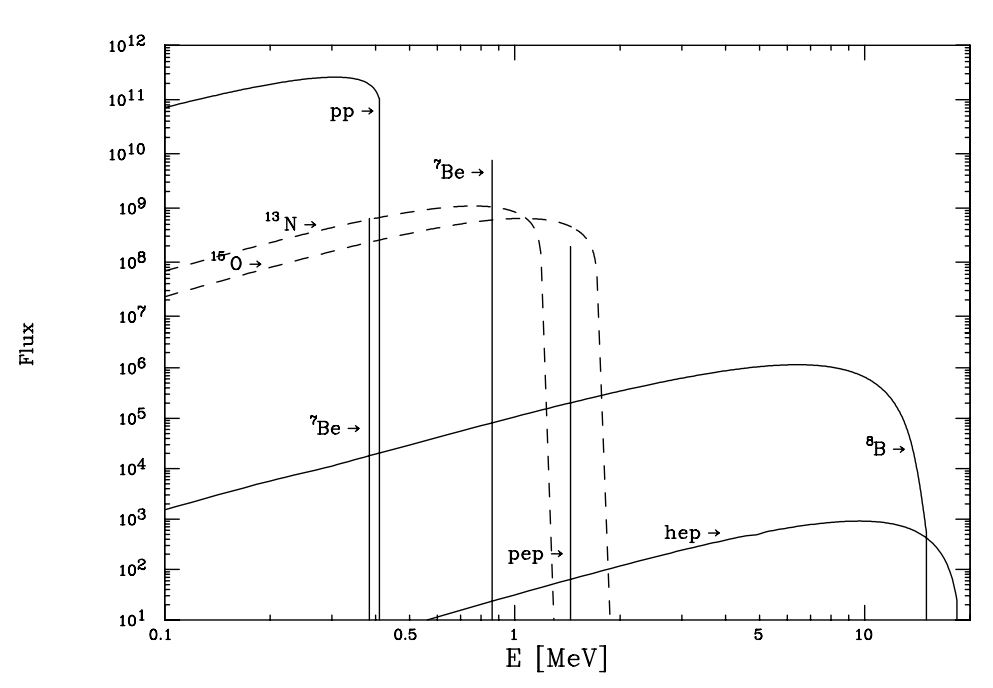}
	\caption{Solar Standard model (SSM). The figure shows the energy spectra of neutrino fluxes from the pp and CNO chains. For continuous sources, the differential flux is in $cm^{-2}s^{-1}MeV^{-1}$. For the lines, the flux is in $cm^{-2}s^{-1}$\cite{castellani1997solar}.}
	\label{Fig.2}
\end{figure}
The advantage of Gallium target of GALLEX/GNO/SAGE over chlorine target of Homestake is that with lower threshold i.e. Q-value, it is possible to detect neutrinos from the initial proton fusion chain(Figure \ref{Fig.2}). 
\subsection{Cherenkov method}
The Cherenkov detection technique\cite{ranucci2016techniques} has been employed in order to investigate neutrinos for a large range of low to high energies.
\begin{figure}[h]
	\includegraphics[width=2in]{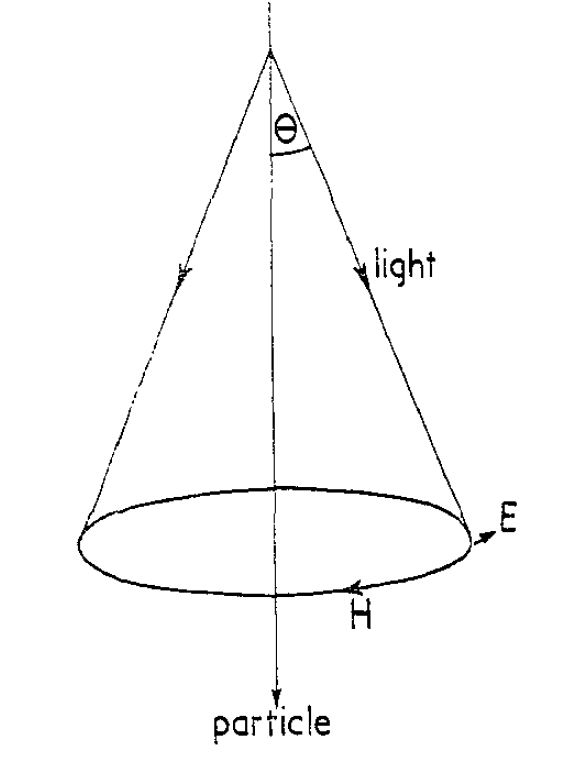}
	\caption{Cherenkov radiation\cite{jelley1955cerenkov}.}
	\label{Fig.3}
\end{figure}
When a charged particle, say an electron, traverses in a medium (\textit{for example,} ordinary water($H_{2}O$) for Super-Kamiokande and heavy water($D_{2}O$) in SNO) of refractive index $n$, polarisation of atoms takes place in the medium resulting in dipole radiation\cite{jelley1955cerenkov}.\\
\paragraph{} If such a particle moves slowly through the medium i.e. $v<\frac{c}{n}$, the radiation from the excited dipoles is emitted symmetrically around the path and sum of all dipoles vanishes\cite{watson2011discovery}.\\
\paragraph{} And if the particle moves with a velocity greater than the local phase velocity of light i.e. $v>\frac{c}{n}$, the dipole distribution is asymmetric. As a result, the sum of all dipoles is non-zero and leads to emission of electromagnetic waves or radiation in the form of a cone known as \textbf{Cherenkov cone} or \textbf{Cherenkov radiation}, named after the Soviet physicist Pavel Alekseyevich Cherenkov. He shared the Nobel Prize in physics in 1958 with I. Frank anf Igor Tamm for the discovery of Cherenkov radiation, made in 1934.\\ 
The Cherenkov radiation angle(Figure \ref{Fig.3}) between the cherenkov photons and the track of particle is given by:
\begin{equation}
\cos\theta=\frac{1}{\beta n}\qquad\text{where, } \beta=\frac{v}{c}
\end{equation} 
$v$ being the velocity of the particle and $c$, the speed of light.\\From equation (4), we learn that there is a threshold value of $\beta$ below which no radiation is emitted coherently w.r.t. the particle position. For a high speed particle i.e. $\beta\equiv1$, there is maximum angle of emission, the Cherenkov angle\cite{watson2011discovery} with
\begin{equation}
\theta_{max}=\cos^{-1}\Big(\frac{1}{n}\Big)
\end{equation}
\begin{figure}{b}
	\includegraphics[width=5in]{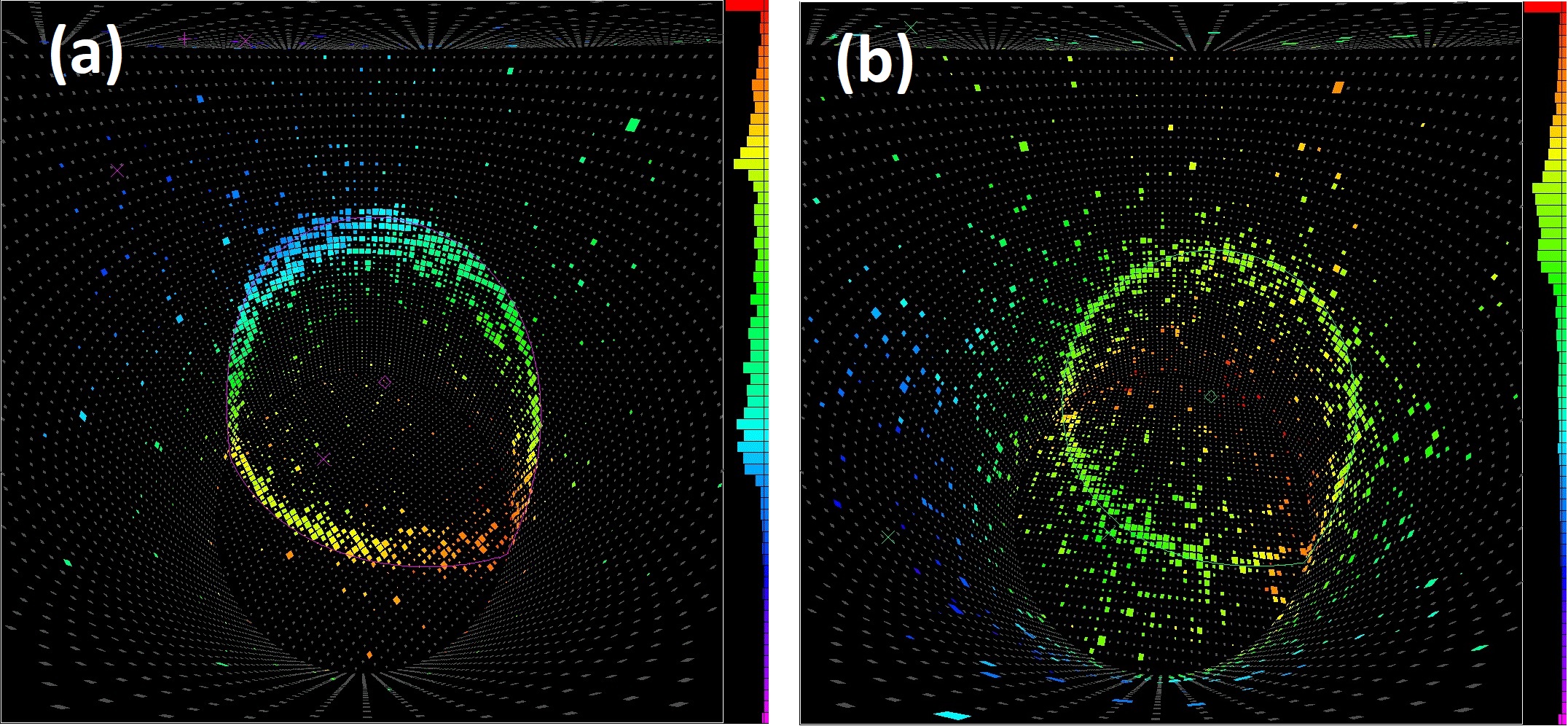}
	\caption{Rings of Cherenkov light detected in 1998 by Super-K, reconstructed as:\\
	(a) muon rings\textit{(sharp)} with momentum of 604 MeV,\\
	(b) electron rings\textit{(fuzzy)} with momentum of 492 MeV.\\			
	 Images are taken from \url{www.ps.uci.edu/~tomba/sk/tscan/compare_mu_e/}}
	\label{Fig.4}
\end{figure}
By detecting the Cherenkov light, in form of a cone, in a large detector with an array of PMTs, the light cone is mapped into a characteristic ring(Figure \ref{Fig.4}). The ring with clean and sharp outer edge is a muon-ring whereas the fuzzy ring produced by scattering of electrons corresponds to electron neutrino. The identification of tau neutrino($\nu_{\tau}$) is more difficult due to the short lifetime of the associated lepton. However, Super-K detected events of tau neutrino($\nu_{\tau}$) using the fact that the $\tau$-lepton decays often produce fast pions in the detector\cite{abe2013evidence}. From the ring, the properties of the incoming particle can be understood precisely. The axis of the cone determines the direction of the particle, and the color of the light gives the particle energy. The events in the Figures \ref{Fig.4}(a) and \ref{Fig.4}(b), recorded in 1998 in the Super-Kamiokande(SK) experiment, were reconstructed as a muon with momentum of 603 MeV and as an electron with momentum of 492 MeV, respectively. A large number of neutrino experiments employ the Cherenkov detection technique and some of them are SNO, Super-Kamiokande, B-DUNT, ICARUS, IceCube, Kamiokande, KM3NeT, MiniBooNE, NEVOD, T2K and UNO. For more details of these experiments, see \cite{adrian2016prototype,aguilarpreprint,avrorin2011gigaton,bellerive2016sudbury,icecube2001icecube,fukuda2003super,haranczyk2017icarus,kindin2015measuring,wilkes2005uno}. 
\subsection{Scintillation Technique}
There are certain organic and inorganic materials which emit light flashes when charged particle, x-rays or gamma rays pass through them. These materials are called scintillators.\\The organic materials are in the form of plastic or liquid and some aromatic polycyclic hydrocarbons, containing one or more benzene ring(s) $C_{6}H_{6}$\cite{brooks1979development}. The second kind is the family of alkali halide crystals\cite{ranucci2016techniques,birks2013theory}.\\The organic materials have the following characteristics:
\begin{tasks}[counter-format={(tsk[r])}, label-align=left, label-offset={1mm}, label-width={6mm}](2)
	\task fast time response,
	\task limited emitted light,
	\task suited for beta spectroscopy,
	\task fast neutron detection.
\end{tasks}
The inorganic materials have the following characteristics:
\begin{tasks}[counter-format={(tsk[r])}, label-align=left, label-offset={1mm}, label-width={6mm}](2)
	\task longer time response,
	\task better light yield,
	\task suited for gamma spectroscopy, %due to high Z and density,
	\task linearity.
\end{tasks}
If a charged particle, a gamma ray or an x-ray is incident on the scintillator, it interacts with matter in three ways:
\begin{tasks}[counter-format={(tsk[r])}, label-align=left, label-offset={1mm}, label-width={6mm}](3)
	\task photoelectric effect,
	\task comptom effect,
	\task pair production.
\end{tasks}
\begin{figure}
	\includegraphics[width=3.2in]{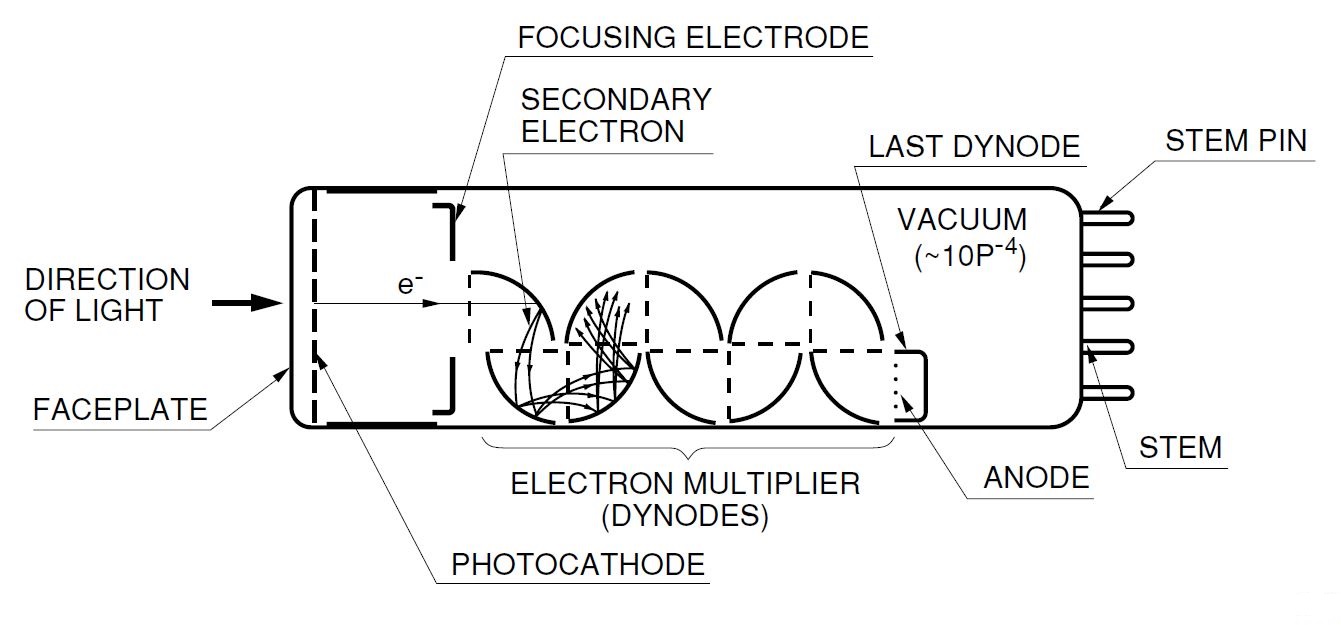}
	\quad
	\includegraphics[width=2in]{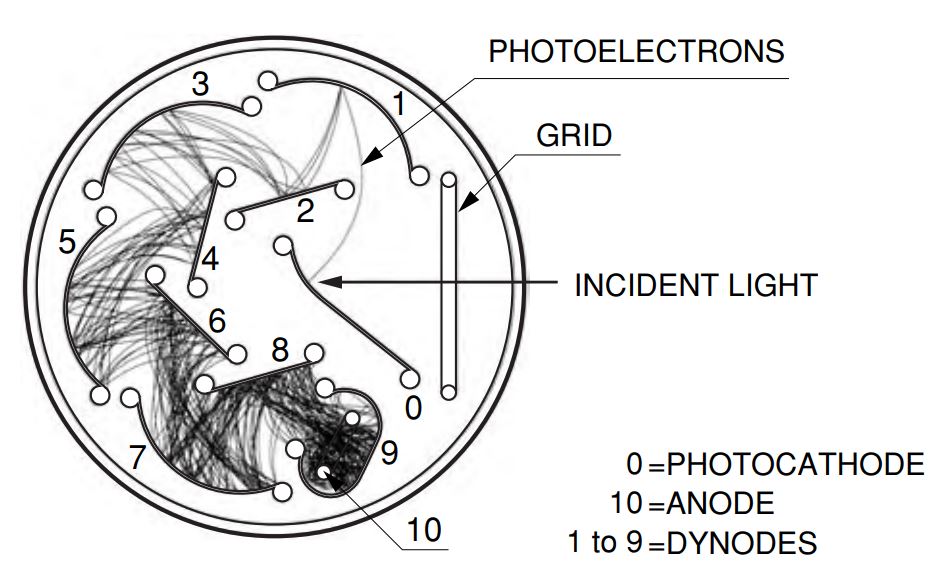}
	\caption{Schematic diagrams of Photo-multiplier Tubes. The right image shows a circular-cage type PMT\cite{hamamatsu2007basics}.}
	\label{Fig.5}
\end{figure}
In each of these effects, electrons of different energy ranges are produced by successive interactions and the kinetic energy of the secondary electrons obtained ionizes the scintillator. The excited states of the scintillator material de-excite to lower states by emission of light flashes. This emitted light is scintillation.\\
The scintillator is optically coupled to a PMT(Figure \ref{Fig.5}). A certain amount of emitted light falls on the photocathode of the PMT producing photo-electrons. The photo-electrons are then accelerated by the electric field to the first dynode producing a bunch of secondary electrons. The electrons are again accelerated to the next dynodes until the electrons reach the anode with a gain of $10^{7}-10^{8}$. Thus, the initial ionizing radiation gives a burst of electrons at the anode where an electrical pulse is taken out for further analysis.\\ The scintillator detectors are most ideal for reactor neutrinos with energy range between 3 MeV and 8 MeV. Some of the experiments like Borexino, CLEAN, Daya Bay, Chooz, Double Chooz, KamLAND, LENS, MINER$\nu$A, MINOS, MINOS+, MOON, NO$\nu$A, RENO SciBooNE, SNO+, SoLiD, STEREO, JUNO, Braidwood, KASKA and Angra have the scintillation technique undertaken for study of neutrinos. The detector specifications of the above experiments will be found in the references \cite{alimonti2009borexino,mckinsey2005neutrino,an2012side,greiner2007double,eguchi2003first,barabanov1999rare,osmanov2011minerva,evans2013minos,ahn2010reno,takei2009scibar,walding2007muon,kamdin2015understanding,labare2017solid,buck2017scintillation,he2015jiangmen,bolton2005braidwood,kuze2005kaska,dornelas2016front}. 
\subsection{Iron Calorimeter (ICAL)}
In this technique, neutrinos from all directions interact with the iron nucleons inside the iron calorimeter by the Neutral-Current (NC) and Charged-Current (CC) interaction channels. In NC interaction, hadrons are generated through exchange of Z-particles. Here, mainly pions are produced, thereby creating the events hit by hadrons only. The reaction involved in the process is:
\begin{equation}
\nu_{\mu}+_{26}^{56}Fe_{30}\rightarrow\nu_{\mu}+X, \quad \text{$X$=hadrons}
\end{equation}
In CC interaction, neutrinos interact weakly through the exchange of a $W^{+}$ or $W^{-}$ boson to form charged particles. In this process, events consist of both muon and hadron hits. The reactions for this process are:
\begin{align}
\nu_{\mu}+Fe_{30}&\rightarrow\mu^{-}+X,\\
\bar{\nu_\mu}+Fe_{30}&\rightarrow\mu^{+}+X, \quad \text{$X$=hadrons}
\end{align}
Apart from NC and CC interactions, low energy neutrinos undergo quasi-elastic scattering and the events are dominated by muon tracks over low energy hadrons. 
INO and the MINOS experiment use the technique of iron calorimeters to detect neutrinos. Some other experiments which take the advantage of this technique include NEMO-3 and SuperNEMO.
\subsection{Liquid Argon Time Projection Chamber (LArTPC)}
D. Nyrgen\cite{nygren1974proposal}, in 1974, introduced an idea of Time Projection Chamber(TPC) where the electron image of an event occuring in a noble gas is drifted towards a collecting multi-electrode array. A three dimensional image $(x,y,z)$ of the event is then reconstructed from the $(x,y)$ information and the drift time $t$.
In 1977, C. Rubbia\cite{rubbia1977liquid} extended the concept to \textit{liquefied noble gas}, considering Argon, naming it as \textit{Liquid Argon Time Projection Chamber (LAr-TPC)}. LAr has the following properties which make it an ideal target for neutrino detection:
\begin{itemize}
	\item It is dense ($1.4 gcc^{-1}$).
	\item It doesn't attach electrons and permits long drift times.
	\item It has high electron mobility.
	\item It is cheap, easy to obtain and to purify.
	\item It is inert and hence, it can be liquefied with liquid nitrogen.
\end{itemize}
The ionization and the drift processes of electrons are the characteristics of LArTPC. The energy required to produce an ion pair in LAr is $23.6\pm0.5 eV$\cite{miyajima1974average}. In high density liquid, a certain number of ionized electrons recombine with the positive  ions. Therefore, an appropriate electric field is applied to separate them quickly\cite{hofmann1976production}. As a result, the electrons are drifted towards a plane of strips. LAr purity is important for detector imaging capability and correct estimation of the energy deposition from the ionization charge signal. The drifting electrons can be easily captured by electronegative impurities, mainly $O_2$, $H_{2}O$ and $CO_2$. The attachment co-efficient for oxygen is a rapidly decreasing function of electron energy. Therefore, attachment of electrons to oxygen or other impurities can be reduced by increasing the electric field\cite{rubbia1977liquid}. \\
Experiments like ICARUS and MicroBooNE used this technique for neutrino detection. The next generation experiments which are proposed to use LAr are DUNE\cite{acciarri2016long} and LBNO\cite{lbne2012long}. 
The reaction engaged in these detectors is:
\begin{equation}
\nu_{x}+e^{-}\longrightarrow \nu_{x}+e^{-}\quad\text{via electron scattering interactions}
\end{equation}
\section{NEUTRINO EXPERIMENTS: A CHRONOLOGICAL ORDER}
\subsection{Past experiments}
\subsubsection{\textbf{Sudbury Neutrino Observatory(SNO)} (Solar experiment)}
\begin{figure}
	\includegraphics[width=1.7in]{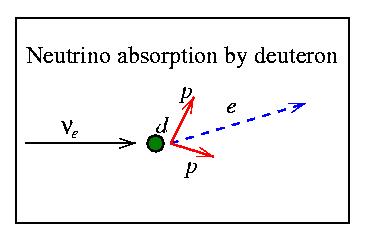}
	\quad
	\includegraphics[width=1.7in]{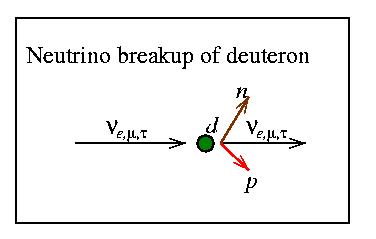}
	\quad
	\includegraphics[width=1.7in]{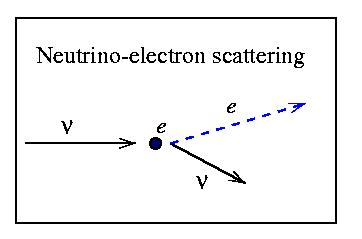}
	\caption{Neutrino interaction with the deuterium nucleus in the SNO experiment in three different ways: \textbf{(i)} Charge Current(CC) interaction(\textit{from left}) sensible to electron neutrinos($\nu_{e}$), \textbf{(ii)} Neutral Current(NC) interaction sensible to all the three flavours($\nu_{e}, \nu_{\mu}, \nu_{\tau}$), and \textbf{(iii)} Electron Scattering (EC), dominated by electron neutrinos($\nu_{e}$). Images are taken from \url{http://www.hep.upenn.edu/}.}
	\label{Fig.17}
\end{figure}

\begin{tasks}[counter-format={(tsk[r])}, label-align=centre, label-offset={1mm}, label-width={6mm}](2)
	\task \textbf{Location:} Creighton Mine, Ontario, Canada
	\task \textbf{Period:} 1999-2006
	\task \textbf{Type:} $\nu_{x}$, x=e, $\mu$, $\tau$
	\task \textbf{Detection Technique:} Cherenkov
\end{tasks}
SNO is a 2100m deep underground detector\cite{ranucci2016techniques} which consists of 1000 tonnes of ultrapure heavy water $D_{2}O$ (99.917\% ${^2}$H by mass) contained in a 12m diameter, 5.6cm thick transparent acrylic vessel. The detection medium is studied by 9438 sensitive photomultiplier tubes (PMTs) on an 18m diameter support structure. The aim of the experiment is to detect  electron neutrinos coming from the core of the Sun through their interactions with the large tank of heavy water. The reactions\cite{jelley2009sudbury} involved in detection of neutrinos(Figure \ref{Fig.17}) are:
\paragraph{Charged-Current (CC):}
$$\nu_{e}+d\longrightarrow p+p+e^{-}-1.44MeV$$
When an incoming neutrino is absorbed by the neutron of each deuterium, the neutron emits an electron and becomes a proton. Then, this emitted electron moves through the heavy water ($D_{2}O$) with a velocity faster than the local speed of light and creates a shock wave in the medium which is detectable as a cone of Cherenkov light. The reaction is sensitive to electron neutrinos only and produces an electron that creates a cone of light observable with an array of PMTs.
%\quad\text{sensitive to electron neutrinos only}$$
\paragraph{Neutral-Current (NC):}
$$\nu_{x}+d\longrightarrow p+n+\nu_{x}-2.2MeV\quad\text{x=e,$\mu$,$\tau$}$$%\quad\text{flavour independent reaction}$$
In the second reaction, each deuteron is split into its constituent parts of proton and neutron when a neutrino is incident on it. In this process, no new charged particle is created and is equally sensitive to all flavours of neutrino. The free neutron, after scattering off of the deuterium nuclei in the heavy water, is eventually captured by another deuteron; creating a tritium nucleus and releasing a high energy gamma-ray. This gamma ray, then, scatters a secondary electron in the heavy water and it is this secondary electron which creates the Cherenkov light detected by the PMTs.  
\paragraph{Electron-Scattering (EC):}
$$\nu_{x}+e^{-}\longrightarrow \nu_{x}+e^{-}\quad\text{x=e,$\mu$,$\tau$}$$%\quad\text{flavour independent reaction yet dominated by $\nu_{e}$}$$
The third reaction is mostly sensitive to electron neutrino by a factor of 6 compared to the other two flavours. It produces an energetic electron that is peaked in the forward direction relative to the incident neutrino and so, can be distinguished from the other two reactions w.r.t. the direction from the sun\cite{mcdonald2016sudbury}.\\
SNO is able to limit the lifetime of nucleon decay of invisible modes (such as $n\rightarrow3\nu$) to $>2\times10^{29}$ years. In May 2002, SNO data for solar neutrinos shows that $\nu_{e}$ from ${^8}B$ decay in the solar core, change their flavour to either muon or tau neutrinos in their way to earth\cite{bellerive2016sudbury}. This process of conversion of flavors of neutrinos from one type to another, called \textbf{neutrino oscillations}, requires the neutrinos to have mass. This result led to the award of 2015 Nobel Prize in physics to Prof. Arthur B. McDonald of SNO along with Prof. Takaki Kajita of Super-Kamiokande collaboration.
\subsubsection{\textbf{Kamioka Nucleon Decay Experiment (Kamiokande)} (atmospheric, astrophysical)}
\begin{tasks}[counter-format={(tsk[r])}, label-align=centre, label-offset={1mm}, label-width={6mm}](2)
	\task \textbf{Location:} Kamioka Japan
	\task \textbf{Period:} 1986-'95
	\task \textbf{Type:} $\nu_{e}$
	\task \textbf{Detection Technique:} Cherenkov
\end{tasks}
The Kamiokande experiment was constructed with the aim to search for the proton decay, atmospheric and astrophysical neutrinos.\\Kamioka was made up of a cylindrical tank of total volume 4.5 ktons, filled with pure ordinary water $H_{2}O$. A total of 980 inward looking PMTs are used to study the inner volume of 2.14 kton to detect the Cherenkov light produced by the particle traversing the water\cite{de2001experimentalist}.\\Kamiokande observed 11 events of thermal neutrinos in February 1987 produced by the supernova 1987A that took place about $1.6\times10^{5}$ light years away in the large Magellanic Cloud\cite{suzuki2009history}. This event gave rise to a new branch in modern physics known as \textbf{\textit{Neutrino Astronomy}}. The first ever detection of astrophyical neutrinos provided Masatoshi Koshiba with the Nobel prize in 2002 along with Raymond Davis Jr. and Riccardo Giacconi.
\subsubsection{\textbf{KEK to Kamioka (K2K)} (Accelerator-based expt.)}
\begin{tasks}[counter-format={(tsk[r])}, label-align=centre, label-offset={1mm}, label-width={6mm}](2)
	\task \textbf{Location:} KEK$\rightarrow$Kamioka, Japan
	\task \textbf{Period:} June, 1999-Nov, 2004
	\task \textbf{Type:} $\nu_{\mu}$
	\task \textbf{Detection Technique:} Cherenkov
\end{tasks}
K2K is the first accelerator-based neutrino experiment where a neutrino beam produced at KEK, the Japanese High Energy accelerator laboratory in Tsukuba, Japan is directed to the Super-Kamiokande detector, the successor of Kamiokande located in the Kamioka mine, 250km apart from the accelerator.\\The main objective of the experiment was to study the neutrino oscillation through interaction with matter. In 2004, the K2K experiment observed the osillation of $\nu_{\mu}$ to $\nu_{\tau}$. K2K experiment confirmed that neutrino oscillation also occur in neutrino beam apart from atmospheric neutrinos.
\subsubsection{\textbf{Main Injector Neutrino Oscillation Search (MINOS)} (Accelerator-based expt.)}
\begin{tasks}[counter-format={(tsk[r])}, label-align=centre, label-offset={1mm}, label-width={6mm}](2)
	\task \textbf{Location:} Fermilab$\rightarrow$Soudan mine, USA
	\task \textbf{Period:} 2005-2012
	\task \textbf{Type:} $\nu_{e}$,$\nu_{\mu}$
	\task \textbf{Detection Technique:} Scintillator, ICAL
\end{tasks}
The aim of MINOS is the precision measurement of oscillation parameters, primarily, the atmospheric mass splitting parameter $\lvert\Delta m^{2}_{32}\rvert$, the corresponding mixing angle $\theta_{23}$ and also the reactor mixing angle $\theta_{13}$. It also aims at precise measurements of the corresponding anti-neutrino parameters.\\
The MINOS experiment\cite{evans2013minos} used protons from the FermiLab Main Injector accelerator to hit a graphite target(NuMI) producing an intense beam of neutrinos of 350kW. The beam is studied in the near detector at FermiLab, Illinois and a far detector at Soudan, Minnesota located 735 km apart. The MINOS detectors are sampling calorimeters with iron absorbers and plastic scintillation strips. A magnetic field of 1.4T is used to energize the iron. The total masses of the far detector and the near detector are 5.4ktons and 0.98kton respectively. The near detector(ND) is 1.04 km from the source. ND measures the energy spectra of the neutrinos before oscillation.\\The muon neutrinos (anti-neutrinos) interact through the charged-current (CC) process--
$$\nu_{\mu}(\bar{\nu}_{\mu})+X\longrightarrow \mu^{-}(\mu^{+})+X^{\prime}$$
The flavour independent reaction is the Neutral-Current (NC) interaction given by--
$$\nu_{x}+X\longrightarrow \nu_{x}+X^{\prime}\quad\text{x=e,$\mu$,$\tau$}$$
And, the electron neutrino undergo Electron-scattering interaction through--
$$\nu_{e}+X\longrightarrow e^{-}+X^{\prime}$$
here, $X$ is the target nucleus where a particular flavor incident and $X^{'}$ is the product nucleus. The electron gives rise to an electromagnetic shower, which produces a much denser and compact shower of energy deposits.\\In 2008\cite{adamson2008measurement}, MINOS released a result of neutrino oscillations in the muon neutrino disappearance mode, using data samples from $3.36\times10^{20}$ protons-on-target(POT) from the NuMI beamline. The atmospheric mass splitting term and the mixing angle are obtained as:
 \begin{align*}
 \lvert\Delta m^{2}_{32}\rvert&=2.43\pm0.13\times10^{-3}eV^{2}\quad&\text{(68\% confidence limit)}\\
 \sin^{2}(2\theta)&>0.90\quad&\text{(90\% confidence limit)}
 \end{align*}
 The above results were updated in 2011 with more than double the data from $7.25\times10^{20}$ POT and improved analysis methodology. The measurement
of the value of the atmospheric mass splitting was found to be $\lvert\Delta m^{2}_{32}\rvert=2.32^{+0.12}_{-0.09}\times10^{-3}eV^{2}$ and the mixing angle as $\sin^{2}(2\theta)>0.90$ at 90\% confidence limit\cite{adamson2011measurement}. This was the then most precise
measurement of this mass splitting by any experiments. 
\subsubsection{\textbf{MINOS+} (Accelerator-based experiment)}
The MINOS experiment\cite{sousa2015first,whitehead2016neutrino} after its operation till 2012 was transitioned to MINOS+ with an extension of 3 years, starting data acquisition from September 2013. The MINOS+ experiment operated with the same detectors and upgraded electronics. The main feature of MINOS+  that differentiates it from MINOS is the medium energy setting of the NuMI beam in the 4-10 GeV range which would deliver about $18\times10^{20}$ protons-on-target during the first three years of operation. Protons with an energy of 120GeV extracted from the Main Injector NuMI were hit on the graphite target. The hadrons \textit{viz.} pions and kaons, produced are focussed by two magnetic horns and directed into the decay pipe. The focussed hadrons decay along its path in the 675m long decay pipe and a neutrino beam is produced. The following processes\cite{backhouse2015results} take place:
$$\pi^{\pm}(K^{\pm})\longrightarrow \mu^{\pm}+\nu_{\mu}(\bar{\nu}_{\mu})\quad \text{(most dominant)}$$  
Secondary decay may take place resulting in $\nu_{e}$, given by
$$\mu^{\pm}\longrightarrow e^{\pm}+\nu_{e}(\bar{\nu}_{e})+\bar{\nu}_{\mu}(\nu_{\mu})$$
$$K^{\pm}\longrightarrow e^{\pm}+\nu_{e}(\bar{\nu}_{e})+\pi^{0}$$

The neutrino interactions in the MINOS+ detector is similar to that of MINOS detector. The collection of high statistics neutrino data (more than 10,000 charged current muon neutrino events and 3,000 neutral current events) would allow the MINOS Far Detector to put a stringent test for the existence of non-standard neutrino interactions(NSI), sterile neutrinos and extra dimensions. In the energy range of 4-10GeV, the experiment would also measure the atmospheric oscillation parameters with precisions higher than that of MINOS(5\% precision), an accuracy of upto 3\% for $\sin^{2}\theta$ and 2\% for $\Delta m^{2}$ combined with the NuMI Off-Axis $\nu_{e}$ Appearance(NO$\nu$A) experiment\cite{tzanankos2011minos+}. MINOS+ would also provide contraints of $\nu_{\mu}$ contamination in $\bar{\nu_\mu}$ running of NO$\nu$A which will substantial for $\delta_{CP}$ search.  \\
In a runtime of eight months from Sep 4, 2013-April 24, 2014, MINOS+ analysed the muon neutrino disappearance and appearance events using data from $1.68\times10^{20}$ protons-on-target(POT) from the upgraded medium energy NuMI beamline setting\cite{sousa2015first}. The best fit values for the atmospheric mass splitting term and the mixing angle are obtained as $\lvert\Delta m^{2}_{32}\rvert=2.37^{+0.11}_{-0.07}\times10^{-3}eV^{2}$ and $\sin^{2}(2\theta_{23})=0.43^{+0.19}_{-0.05}$ for inverted hierarchy\cite{holin2015results}.
\subsubsection{\textbf{Chooz} (Reactor experiment)}
\begin{tasks}[counter-format={(tsk[r])}, label-align=centre, label-offset={1mm}, label-width={6mm}](2)
	\task \textbf{Location:} Ardennes, France
	\task \textbf{Period:} April 1997-July 1998
	\task \textbf{Type:} $\bar{\nu}_{e}$ (disappearance)
	\task \textbf{Detection Technique:} Scintillator
\end{tasks}
The Chooz is a long-baseline reactor neutrino oscillation experiment named after the nuclear power station operated by \'{E}lectricit\'{e} de France (EdF) near the village of Chooz in France.\\The detector is located in an undergorund laboratory at a distance of 1 km from the two pressurised water reactors and with rock overburden of 300 metre water equivalent. The depth reduces the external cosmic ray muon flux by a factor of $\sim$300 to 0.4$m^{-2}s^{-1}$. The neutrino sources i.e. reactors produces a total thermal output of 8.5GW$_{th}$, 4.25GW$_{th}$ from each of the reactors. The detector is in a welded cylindrical steel vessel 5.5m in diameter and 5.5m deep. The inside walls of the vessel are painted with high reflectivity white paint and the vessel is placed in a pit of 7m diameter and 7m depth. The steel vessel is further surrounded by 75 cm of low radioactivity sand and covered by 14cm of cast iron to protect the detector from the natural radioactivity of the rocks. \\The electron anti-neutrino flux emitted by the Chooz reactors depends on the instantaneous fission rate derived from the thermal power of the reactors and also on anti-electron neutrino yield from the four main isotopes-- ${^{235}}U$, ${^{238}}U$, ${^{239}}Pu$ and ${^{241}}Pu$. The electron anti-neutrinos those hit the detector are studied via the Inverse Beta decay (IBD) reaction given by
$$\bar{\nu}_{e}+p\longrightarrow e^{+}+n\quad\text{with}\quad E_{e^{+}}=E_{\bar{\nu}_{e}}-1.804 MeV$$
The target material used in the detector is a hydrogen-rich i.e. free protons paraffinic liquid scintillator loaded with 0.09\% Gadolinium and is contained in an acrylic vessel immersed in a low energy calorimeter made of unloaded liquid scintillator. Gd is selected due to its large neutron capture cross-section and high $\gamma$-ray release after neutron capture of $\sim$8MeV. The detector is composed of three concentric regions\cite{apollonio1998initial,cao2013detection}:
\paragraph{} a central 5-ton target in a transparent plexiglass container with 0.09\% Gd loaded scintillator.
\paragraph{} an intermediate 17-ton region with a thichkness of 70 cm equipped with 192 8" PMTs to contain the $\gamma$-rays from neutron capture.
\paragraph{} an outer 90-ton optically separated active cosmic ray muon veto shield of thickness 80cm equipped with two rings of 24 8"PMTs.\\

For a data taking period of April '97-July '98, a total of $\sim$2500 events gave a neutrino detection rate of $2.5d^{-1}GW^{-1}$, the neutrino detection rate to background ratio is 10:1\cite{mikaelyan2000chooz}. The ratio of the measured to the calculated (for no-oscillation case) neutrino detection rates is found to be $\frac{R_{meas}}{R_{calc}}=1.01\pm2.8\%(stat)+2.7\%(syst)$. The Chooz experiment didn't observe an evidence for neutrino oscillations in the $\bar{\nu}_{e}$ disappearance channel in the mass region given by $\delta m^{2}=7\times10^{-4}eV^{2}$ for maximal mixing, and $\sin^{2}2\theta=0.10$ for large $\delta m^{2}$\cite{apollonio2003search}.
\subsubsection{\textbf{Neutrino Ettore Majorana Observatory (NEMO-3)} (Double-beta decay experiment)}
\begin{tasks}[counter-format={(tsk[r])}, label-align=centre, label-offset={1mm}, label-width={6mm}](2)
	\task \textbf{Location:} Modane Underground Lab, Fr\'{e}jus Road Tunnel, France
	\task \textbf{Period:} Feb, 2003-Jan, 2011
	\task \textbf{Type:} ${\nu}_{e}$
	\task \textbf{Detection Technique:} Calorimeter
\end{tasks}
NEMO-3\cite{pahlka2012nemo} is a tracking calorimeter hosting several double-beta decaying isotopes in thin source foils in a cylindrical shape aimed at studying Majorana nature of neutrino. The experiment is devoted purely to the study of double beta-decay($\beta\beta$) processes. It is located under a rock burden of 4800 metre water equivalent (m.w.e). The decay isotopes used are 6.91 kg of ${^{100}}Mo$, 0.93 kg of ${^{82}}Se$, smaller amounts of ${^{96}}Zr$, ${^{116}}Cd$,${^{130}}Te$ and ${^{150}}Nd$ and 6.99g of ${^{48}}Ca$. The goal of the experiment was to search for neutrinoless double beta decay ($0\nu\beta\beta$) with a half-life sensitivity of $10^{25}$ years\cite{arnold2016measurement}. The detector is capable of identifying  positrons, alphas and gammas.\\Electrons from the decaying isotopes pass through 50cm wide wire chambers on each of the source foils containing in total 6180 geiger cells operating in a gas mixture comprising He with 4\% ethanol quencher, 1\% argon and 0.15\% water vapour. Surrounding the tracker, there is a calorimeter of 1940 plastic scintillators coupled to low radioactivity PMTs. The equations involved in the detection mechanism are:
$${^{100}}Mo\longrightarrow {^{100}}Ru +2e^{-}$$
$${^{82}}Se\longrightarrow {^{82}}Kr +2e^{-}$$
NEMO-3 experiment has made the most precise measurements of the neutrino double beta decay ($2\nu\beta\beta$) half life for all the seven isotopes\cite{saakyan2013two}. However, after 34.7kg-years exposure, no evidence for the $0\nu\beta\beta$ is observed. After considering the statistical and systematic uncertainities, the result\cite{gomez2016latest,waters2017latest} can give a limit to the half-life for the light majorana neutrino mechanism of
$$T^{0\nu}_{\frac{1}{2}}>1.1\times10^{24}\text{ years}\quad\text{at  } 90\% \text{ C.L.}$$
The NEMO-3 experiment also investigated the ${^{48}}Ca$ isotope in the $2\nu\beta\beta$ mode. With an exposure for 5.25 years, the half-life for the standard model $2\nu\beta\beta$ mode is measured as
$$T^{2\nu}_{\frac{1}{2}}=\Big[6.4^{+0.7}_{-0.6}(stat)^{+1.2}_{-0.9}(syst.)\Big]\times10^{19} \text{  years}$$ and is consistent with previous experimental measurement and has significantly smaller uncertainities. No signal has been found and the lower limit on half-life\cite{arnold2016measurement} has been set as 
$$T^{0\nu}_{\frac{1}{2}}>2.0\times10^{22}\text{ years}\quad\text{at  } 90\% \text{ C.L.}$$
\subsubsection{\textbf{Imaging Cosmic And Rare Underground Signals (ICARUS)} (cosmic rays, accelerator-based)}
%Tables should also be cited in the main text in chronological order (\textbf {Table \ref{tab1}}).
\begin{tasks}[counter-format={(tsk[r])}, label-align=centre, label-offset={1mm}, label-width={6mm}](2)
	\task \textbf{Location:} LNGS, Gran Sasso, Italy
	\task \textbf{Period:} summer, 2010-June 2013
	\task \textbf{Type:} muon and electron neutrinos
	\task \textbf{Detection Technique:} LArTPC
\end{tasks}
The ICARUS detector is a single phase Liquid Argon Time Projection Chamber (LAr-TPC) with an active mass of 476 tons and total mass of 760 tons. It is splitted into two identical, adjacent modules, each modules having two TPCs with maximum drift path of 1.5m and sharing a common cathode\cite{rubbia2011underground}. The LAr-TPC technique allows to collect two signals: the ionization electrons at anode wires and scintillation lights by PMTs. A uniform electric field of 500$Vcm^{-1}$ drifts the ionization electrons with a velocity of $\sim1.6mm/\mu s$ towards the three anode wire planes 3mm apart. Using the collected signals, a 3D reconstruction of any ionizing particles crossing the detector can be performed with a spatial resolution of $1mm^{3}$. \\ICARUS was successfully operated in the LNGS(\textit{Laboratori Nazionali del Gran Sasso}) underground laboratory in Italy. The experiment studied  cosmic rays and CNGS(CERN to Gran Sasso) neutrino beam from the Super Proton Synchotron(SPS), CERN located at a distance of 730 km, using the reaction-- $$^{40}Ar+\nu\longrightarrow ^{40}K+e^{-}$$ The CNGS run from 2010 to mid 2013 demonstrated LArTPC as a leading technology for the future short baseline(SBL) and long baseline(LBL) accelerator driven neutrino physics. The liquid argon purity, an indicator for detector performance, has been achieved with free electron lifetime exceeding 12 milliseconds which corresponds to only 25 parts per trillion $O_2$-equivalent contaminations. This opened the way for large future TPC detectors at a scale of tens of kiloton. ICARUS detected six electron neutrino events, providing no evidence of oscillation into sterile neutrinos in the $\frac{L}{E}$ interval. The ICARUS result combined with global fit of all other SBL data limits the parameters $(\Delta m^{2}, \sin^{2}2\theta)$ for a possible LSND anamoly to a very narrow region of $(0.5eV^{2}, \simeq0.005)$\cite{menegolli2016some}. The LSND anomaly and the sterile neutrino hypothesis will be further addressed by the Fermilab laboratory and the ICARUS detector will be used as a far detector in this programme\cite{haranczyk2017icarus}.
\subsection{Present experiments} 
\subsubsection{\textbf{BORon EXperiment(Borexino)} (Solar experiment)}
%Tables should also be cited in the main text in chronological order (\textbf {Table \ref{tab1}}).
\begin{tasks}[counter-format={(tsk[r])}, label-align=centre, label-offset={1mm}, label-width={6mm}](2)
	\task \textbf{Location:} Gran Sasso, Italy
	\task \textbf{Period:} May, 2007-\textit{present}
	\task \textbf{Type:} ${\nu}_{e}$
	\task \textbf{Detection Technique:} Liq. Scintillator
\end{tasks}
Borexino\cite{ranucci2016techniques} is a scintillator detector which employs the active detection medium with a mixture of pseudocumene (PC, 1,2,4-trimethylbenzene) and a fluorescent dye 2,5-diphenyloxazole also known as PPO at a concentration of 1.5g/l. Borexino is based on the principle of \textit{graded shielding} in order to keep the radioactivity at a very low level\cite{ianni2011neutrino}. The experiment was established with the aim to study the entire spectrum of solar neutrinos from low energies(Figure \ref{Fig.2}). The solar fluxes from the \textit{pp}, \textit{pep}, \textit{hep}, \textit{$^{7}Be$} and \textit{$^{8}B$} chains are given in Table \ref{T4}. \\The mass of the scintillator is 278 tons and is contained in a 125$\mu$m thick nylon inner vessel(IV) with a radius of 4.25m. A second nylon outer vessel(OV) with radius 5.50m surrounds the IV and acts as a barrier against $^{222}Rd$(Radon) and other background contaminations ($^{39}Ar$,$^{85}Kr$) from outside. Surrounding them, there is a stainless steel sphere(SSS) covered by 2100 m$^3$ of water tank and supported by 2212 PMTs. The choice of liquid scintillation is due to the high-light yield of the scintillator which differs by a factor of 50 with respect to that of \~{C}erenkov techniques\cite{arpesella2008first}.\\The neutrino signals observed in the Borexino detector are due to the electrons recoiled in the elastic scattering given by--
$$\nu_{x}+e^{-}\longrightarrow \nu_{x}+e^{-}\quad\text{\textit{x} being the three $\nu$-flavors}$$
Hence, the neutrino energy spectrum is the energy spectrum of the electron recoiled from the elastic scattering.
\begin{table}[h]
	\begin{tabular}{c|c|c}
		\centering
		
		\textbf{Reaction}& \textbf{Label}&\textbf{Flux \textit{($cm^{-2}s^{-1}$)}}\\\hline
	$p+p\rightarrow^{2}H+e^{+}+\nu_{e}$&$pp$&$5.95\times10^{10}$\\
	$p+e^{-}+p\rightarrow^{2}H+\nu_{e}$&$pep$&$1.40\times10^{8}$\\
	$^{3}He+p\rightarrow^{4}He+e^{+}+\nu_{e}$&$hep$&$9.3\times10^{3}$\\
	$^{7}Be+e^{-}\rightarrow^{7}Li+\nu_{e}$&$^{7}Be$&$4.77\times10^{9}$\\
	$^{8}B\rightarrow^{8}Be^{*}+e^{+}+\nu_{e}$&$^{8}B$&$5.05\times10^{6}$\\	
	\end{tabular}
	\caption{Production of neutrinos from the fusion reactions in the sun. The total solar flux at the Earth is $6.5\times10^{10}$ neutrinos per $cm^2$ and per second. The \textit{pp} chain corresponds $>91$\%; while the $^{7}Be$, pep, and $^{8}B$ chains correspond to about 7\%, 0.2\%, and 0.008\% of the total flux, respectively\cite{bellerive2004review}.}
	\label{T4}	
\end{table}
The data has been taking place since May 2007 and in its first year of operation, solar neutrino rates from $^{7}Be$\cite{bellini2011precision}, $^{8}B$ with a threshold of 2.8MeV and \textit{pep} flux\cite{bellini2012first} gave the unambiguous signature of the occurrance of solar neutrino detection. Evidences of a null day/night asymmetry in the $^{7}Be$ region\cite{bellini2012absence} and of the solar neutrino flux seasonal variation have been reached.\\Recently, Borexino made the first real time spectroscopic measurement of the fundamental \textit{pp} flux\cite{bellini2014neutrinos}. There has also been evidences about geo-neutrinos obtained at the level of 5.9$\sigma$ C.L.\cite{bellini2016impact}. Borexino is still under operation and aims to study neutrino flux from the CNO cycle and upgrade the precision of the solar neutrino rates already measured\cite{bellerive2004review}.
\subsubsection{\textbf{Double Chooz} (Reactor Experiment)}
The Chooz experiment is upgraded to Double Chooz and is under operation since 2011. The primary objective of Double Chooz is to improve Chooz sensitivity by a factor of  at least 5 to measure $\sin^{2}(2\theta_{13})$, the reactor mixing angle\cite{palomares2011double}.\\The Chooz experiment measured the electron anti-neutrino($\bar{\nu}_{e}$) rate with a 2.8\% statistical error and a 2.7\% systematic error. So, the Double Chooz experiment will employ two almost identical detectors of medium size, each containing 10.3$m^{3}$ of liquid scintillator target of 0.1\% Gd, to reduce the large statistical uncertainity. The detector site of the Chooz experiment which is at about 1 km  from the two pressurized water detector will act as the Far Detector (FD). In order to deduct the systematic uncertainities of the $\bar{\nu}_e$ flux and spectrum originating from the reactors, a second detector; called the Near Detector (ND) is installed close to the nuclear plants at a distance of 409m.\\The anti-neutrino from the reactors collides with a proton in the liquid scintillator and performs via the IBD reaction--
$$\bar{\nu}_{e}+p\longrightarrow e^{+}+n$$
The Double Chooz, during 460.67 live days of data taking and use of 17,351 $\bar{\nu}_{e}$ candidates\cite{carr2016new}; the improved $\theta_{13}$ oscillation parameter after the first indication of non-zero value, now, is found to be $\sin^{2}(2\theta_{13})=0.090^{+0.032}_{-0.023}$\cite{abe2012indication}. The result, though limited by reactor flux uncertainity, is expected to improve with the inclusion of data from the second i.e. the Double Chooz Near detector(ND).
\subsubsection{\textbf{Super Kamiokande(Super-K)} (Solar, atmospheric)}
\begin{tasks}[counter-format={(tsk[r])}, label-align=centre, label-offset={1mm}, label-width={6mm}](2)
	\task \textbf{Location:} Mozumi Mine, Japan
	\task \textbf{Period:} May, 2007-\textit{present}
	\task \textbf{Type:} ${\nu}_{e}$
	\task \textbf{Detection Technique:} Scintillator
\end{tasks}
The Super-Kamiokande experiment is the upgradation of the previous Kamiokande experiment and aims at confirming neutrino oscillation phenomena with source from atmospheric neutrinos and determining more precisely the neutrino oscillation parameters. Super-K is a US-Japan joint collaboration. The detector lies under the peak of Mt. Ikenoyama, with 1000m of rock overburden at geographical co-ordinates $36^{0}25^{\prime}32.6^{\prime\prime} N$ and $137^{0}18^{\prime}37.1^{\prime\prime} E$. The water \~{C}erenkov detector consists of a welded stainless steel tank of 39m diameter and 42m tall with nominal capacity of 50ktons supported by an array of 11,146 50cm-diameter hemispherical inward-facing and 1885 outward facing 20cm diameter hemispherical PMTs\cite{fukuda2003super}.\\Super-K can detect events over a wide range of energy from 4.5MeV to over 1 TeV. The outer detector is useful for identifying entering cosmic ray muons and measuring exiting particles produced by neutrino interactions occuring in the inner detector.\\The experiment has confirmed the apparent deficit in the total flux from the sun. The observed neutrino spectrum reveals the type of distortion expected from neutrino oscillation effect. The solar mixing angle as measured by all the four phases of Super-K\cite{abe2016solar} gives
$$\sin^{2}\theta_{12}=0.334^{+0.027}_{-0.023}$$ with the determined mass splitting term to be $$\Delta m^{2}_{21}=4.8^{+1.5}_{-0.8}\times10^{-5}eV^{2}$$\\During 2055 days of day/night(D/N) spectrum above 3.5 MeV from the Super-K phase-IV, from September, 2008 to April, 2015, the signal of $^{8}B$ solar neutrino flux is obtained as $2.314\pm0.018(stat)\pm0.039(syst)\times10^{6}cm^{-2}s^{-1}$. The measured D/N asymmetry, $A_{DN}$ is $[-3.3\pm1.0(stat)\pm0.5(syst)]\%$ at $3\sigma$ from zero\cite{farzan2017neutrino}. This result combined with the SK I,II, III and IV phases provides an indirect interaction for matter enhanced neutrino oscillations inside the earth. Super-K also searched for proton decay via $p\longrightarrow e^{+}\pi^{0}$ and $p\longrightarrow \mu^{+}\pi^{0}$\cite{abe2017search}. No candidates were seen in the $p\longrightarrow e^{+}\pi^{0}$ but two candidates have been observed for $p\longrightarrow \mu^{+}\pi^{0}$. Lower limits on proton lifetime are set at $\frac{\tau}{B}(p\longrightarrow e^{+}\pi^{0})>1.6\times10^{34}$years and $\frac{\tau}{B}(p\longrightarrow \mu^{+}\pi^{0})>7.7\times10^{33}$years at 90\% confidence levels\cite{abe2017search}.
\begin{figure}[h]
	\includegraphics[width=5in]{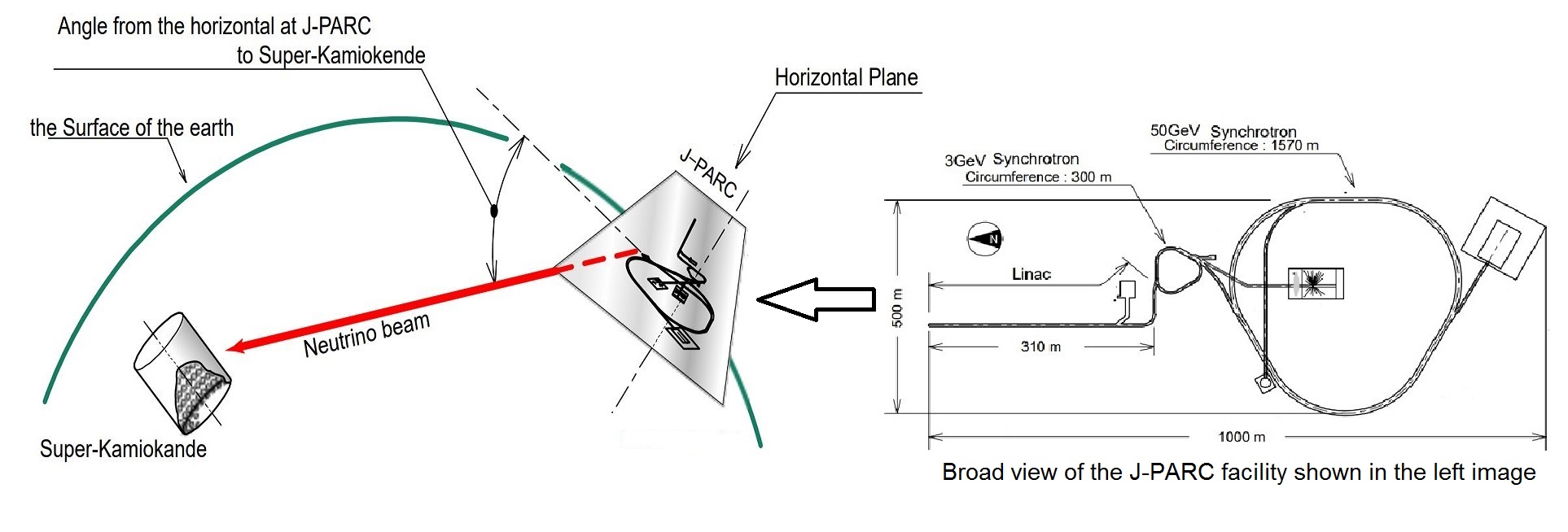}
	\caption{A Schematic diagram of J-PARC, Tokai to Super-K detector, Kamioka. Images taken from \textit{Mishima K, et al. $9^{th}$ International Workshop on Accelerator Alignment, September 26-29, 2006}}
	\label{Fig.7}
\end{figure}
\subsubsection{\textbf{Tokai to Kamioka(T2K)} (Accelerator-based experiment)}
\begin{tasks}[counter-format={(tsk[r])}, label-align=centre, label-offset={1mm}, label-width={6mm}](2)
	\task \textbf{Location:} Tokai$\rightarrow$Kamioka, Japan
	\task \textbf{Period:} 2011-\textit{present}
	\task \textbf{Type:} $\nu_{e},\nu_{\mu},\bar{\nu}_{e},\bar{\nu}_{\mu}$
	\task \textbf{Detection Technique:} Water Cherenkov
\end{tasks}
Another neutrino baseline was set up after K2K with its detector at Super-K and the accelerator at Tokai, Japan. The accelerator facility is Japan Proton Accerator Research Complex(J-PARC). The aim of the experiment is to study neutrino oscillation search among other flavours \textit{i.e.} $\nu_{\mu}\rightarrow\nu_{e}$, after successful study of oscillation of muon-neutrino($\nu_{\mu}$) to tau-neutrino(${\nu_{\tau}}$) by K2K collaboration.\\The J-PARC facility produces an artificial beam of $\nu_{\mu}$ from a proton beam(Figure \ref{Fig.7}). The detector and the accelerator is separated by 295km. The J-PARC accelerator facility consists of three accelerators-- one linear(LINAC) and two circular accelerators called the Main Ring and the Rapid Cycling Synchotron (RCS)\cite{abe2011t2k}. In these accelerators, the protons are accelerated to $99.98\%$ of speed of light and are bent towards Kamioka using super-conducting magnets.\\The T2K collaboration\cite{abe2013measurement}, in July 2013, with its off-axis beam with a peak energy of $0.6GeV$ measures the oscillation parameters from muon neutrino disappearance(\textit{asuming NH}) as:
$$\sin^{2}\theta_{23}=0.514\pm0.082,\quad\lvert\Delta m^{2}_{32}\rvert=2.44^{+0.17}_{-0.15}\times10^{-3}eV^{2}/c^{4}$$In November 2013,  the T2K group made the first observation of electron neutrino appearance in a  muon-neutrino beam \textit{i.e.}  ($\nu_{\mu}\rightarrow\nu_{e}$) with a peak energy of $0.6 GeV$, assuming $\lvert\Delta m^{2}_{32}\rvert=2.4\times10^{-3}eV^{2}$, $\delta_{CP}=0$, $\sin^{2}2\theta_{13}=0.140^{+0.038}_{-0.032}(0.170^{+0.045}_{-0.037})$ is obtained for NH(IH), with a significance of 7.3$\sigma$ over the hypothesis of $\sin^{2}2\theta_{13}=0$\cite{abe2014observation}. Since 2014, the T2K has been dealing with the anti-neutrino beams instead of the original neutrino beams. In July 2015, the experiment completed its first run of anti-neutrino mode collecting 10\% of the anti-neutrino data set and has observed three $\bar{\nu}_e$ candidate events in a $\bar{\nu}_\mu$ beam. On $3^{rd}$ January 2017, T2K released its first results in the search for CP violation using appearance
and disappearance channels for neutrino and anti-neutrino modes. The data collected from Jan 2010 to May 2016 comprise $7.482\times10^{20}$ protons on target (POT) in neutrino mode and $7.471\times10^{20}$ POT in anti-neutrino mode; yielding  32 e-like and 135 $\mu$-like events, and 4 e-like and 66 $\mu$-like events, respectively in the far detector. $\delta_{CP}$ ranges from -3.13 to -0.39 for normal hierarchy with a 1D confidence interval at 90\%. The CP conservation hypothesis for $\delta_{CP}=0,\pi$ is excluded at 90\% confidence level\cite{abe2017combined}.
\subsubsection{\textbf{IceCube} (Atmospheric, astrophysical)}
\begin{tasks}[counter-format={(tsk[r])}, label-align=centre, label-offset={1mm}, label-width={6mm}](2)
	\task \textbf{Location:} South Pole, Antarctica
	\task \textbf{Period:} May, 2011-\textit{present}
	\task \textbf{Type:} $\nu_{e},\nu_{\mu},\nu_{\tau}$
	\task \textbf{Detection Technique:} Cherenkov(Ice)
\end{tasks}
\begin{figure}[h]
	\includegraphics[width=3.5in]{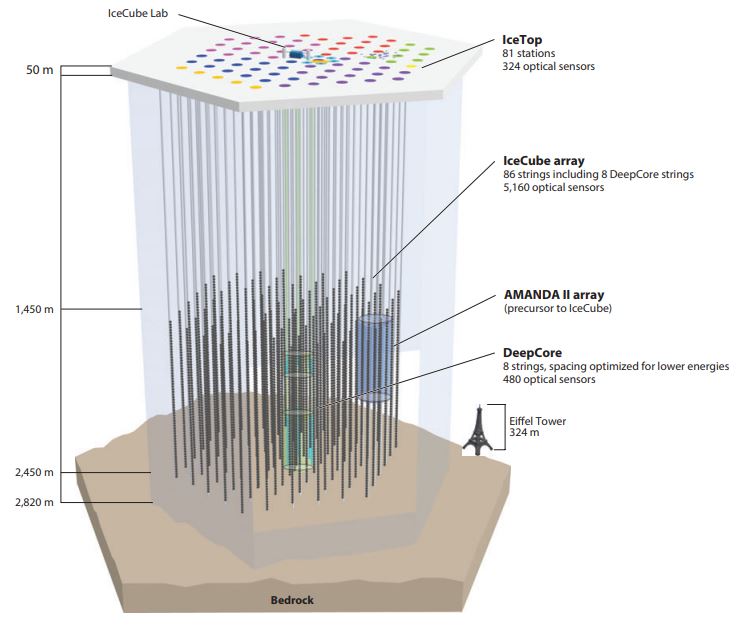}
	\caption{Layout of the IceCube experiment.}
	\label{Fig.18}
\end{figure}
The IceCube experiment\cite{karle2010icecube,icecube2001icecube} aims at studying atmospheric neutrinos and neutrinos of high energy astrophysical sources like supernovae(SN), active galactic nuclei (AGN), gamma ray bursts (GRB), etc.\\IceCube(Figure \ref{Fig.18}) is a $1km^{3}$ neutrino telescope and covers a wide range of neutrino energy from 100 GeV to $10^{9}$ GeV. The detector has 86 strings upto the depth of 2450m below the surface. Between 1450m and 2450m, there are 60 optical sensors mounted on each string, excluding the 8 Deep Core strings on which the sensors more densely fit between 1760m and 2450m. There are 160 Icetop tanks on the surface of the ice directly above the strings in which 324 optical sensors are deployed. Every sensor has a 25cm PMT connected to a waveform recording data acquisition circuit capable of recording pulses with precision upto nanoseconds.\\
The updated atmospheric data which uses neutrinos with reconstructed energies from $5.6-56GeV$\cite{aartsen2018measurement} provides a best fit of paramaters as:
\begin{align*}
\Delta m^{2}_{32}&=2.31^{+0.11}_{-0.13}\times10^{-3}eV^{2} &\text{(Normal Hierarchy)}\\
&=-2.32\times10^{-3}eV^{2} &\text{(Inverted Hierarchy)}
\end{align*}
and the atmospheric mixing angle to be
\begin{align*}
\sin^{2}\theta_{23}&=0.51^{+0.07}_{-0.09}&\text{(Normal Hierarchy)}\\
&=0.51 &\text{(Inverted Hierarchy)}
\end{align*}
\subsubsection{\textbf{Kamioka Liquid-scintillation Anti-Neutrino Detector (KamLAND)} (Reactor)}
\begin{tasks}[counter-format={(tsk[r])}, label-align=centre, label-offset={1mm}, label-width={6mm}](2)
	\task \textbf{Location:} Kamioka, Japan
	\task \textbf{Period:} 2002-\textit{present}
	\task \textbf{Type:} $\bar{\nu}_{e}$ (disappearance)
	\task \textbf{Detection Technique:} Liquid Scintillator
\end{tasks}
KamLAND searches for oscillation of anti-neutrinos emitted from distant power reactors. It also aims at studying geoneutrinos i.e. $\bar{\nu}_e$ from the Earth's interior. It's estimation of the radiogenic heat produced inside the Earth agrees with the current Earth models. KamLAND replaces the site of the earlier Kamiokande experiment under 2700 metre water equivalent of rock\cite{eguchi2003first}. The neutrino detector is 1 kton of ultrapure liquid scintillation contained in a 13m diameter spherical balloon made of 135$\mu$m thick transparent nylon/EVOD (\textit{ethylene vinyl alcohol copolymer}) composite film supported by a network of Kelvar. A 1879 PMTs array-- 1325 newly developed fast $17^{\prime\prime}$ diameter PMTs and 554 older Kamiokande 20$^{\prime\prime}$, mounted on the vessel completes the inner detector. This vessel is surrounded by a 3.2 kton water-\~{C}erenkov detector with 225 20$^{\prime\prime}$ PMTs. The outer detector absorbs gamma rays and neutrons from surrounding rocks.\\KamLAND demonstrated reactor $\bar{\nu}_e$ disappearance at long baselines at high confidence level(99.95\%) for the first time. It also sets limit on the neutrinoless double beta decay($0\nu\beta\beta$) half life\cite{eguchi2003first} of $$T^{0\nu}_{\frac{1}{2}}>1.9\times10^{25} \text{ years at 90\% confidence level.}$$
\subsubsection{\textbf{NuMI Off-Axis $\nu_{e}$ Appearance(NO$\nu$A)} (Accelerator-based)}
\begin{tasks}[counter-format={(tsk[r])}, label-align=centre, label-offset={1mm}, label-width={6mm}](2)
	\task \textbf{Location:} FermiLab $\rightarrow$ Ash River, Minesota, US 
	\task \textbf{Period:} 2011-\textit{present}
	\task \textbf{Type:} $\nu_{\mu}\rightarrow\nu_{e}$, $\bar{\nu}_{\mu}\rightarrow\bar{\nu}_{e}$
	\task \textbf{Detection Technique:} Scintillator, Calorimeter
\end{tasks}
The NO$\nu$A is the upgradation of the MINOS experiment where the power of the FermiLab accelerator is increased to 700kW. The experiment aims at measuring  the transition probability $P(\nu_{\mu}\rightarrow \nu_{e})$(or $\nu_{e}$-appearance) and survival probability $P(\nu_{\mu}\rightarrow \nu_{\mu})$(or $\nu_{\mu}$-disappearance), thereby determining the atmospheric oscillation parameters $\Delta m^{2}_{32}$ and $\theta_{23}$, CP violating phase, $\delta_{CP}$ and the neutrino mass hierarchy.\\The NO$\nu$A\cite{jediny2017nova} project managed by Fermi National Accelerator Laboratory(FNAL) or FermiLab is an off-axis long baseline neutrino experiment aimed downward at an angle of 14 milliradian to the incident beam direction to achieve peak energy of 2GeV. It has two identical detectors-- a near detector at FermiLab of dimension $4.2\times4.2\times15m^{3}$ and a far detector in Minnesota of dimension $15.6\times15.6\times60m^{3}$ to study the oscillation of the muon neutrino flavor to electrino neutrino type. A neutrino beam produced at FermiLab is sent to the 14kton far detector in Ash river, Minnesota located 810 km away. The near detector is 1 km from the source with a mass of about 300 tons.\\The neutrino interaction mechanism is similar to that of MINOS. The NO$\nu$A detectors are huge tracking calorimeter consisting of many PVC cells of $6cm\times4cm$ filled with mineral oil liquid scintillator with 5\% mixture of pseudocumene.\\The results obtained obtained between July 2013 and March 2015 are divided into $\nu_{\mu}$-disappearance and $\nu_{e}$-appearance results. The $\nu_{\mu}$-disappearance mode is sensitive to $\theta_{23}$ and $\Delta m^{2}_{23}$ measurement. A total of 78 events have been observed in the disappearance channel with the prediction of 82 events at the best-fit value. 33 events have been detected in the appearance channel. The obtained results are $$\Delta m^{2}_{32}=2.52^{+0.20}_{-0.18}\times10^{-3}eV^{2}$$ for $0.38<\sin^{2}\theta_{23}<0.65$ at 68\% C.L. for Normal Hierarchy(NH), and $$\Delta m^{2}_{32}=-2.56\pm0.19\times10^{-3}eV^{2}$$ for $0.37<\sin^{2}\theta_{23}<0.64$ at 68\% C.L. for Inverted Hierarchy(IH). \\The analysis od NO$\nu$A data slightly disfavours inverted mass heirarchy with $\Delta\chi^{2}=-0.47$ and maximal atmospheric mixing is disfavoured at around 2.6$\sigma$. The most recent update of the oscillation results are announced at the January 2018 JETP talk \textquotedblleft Latest Oscillation Results from NO$\nu$A" by A. Radovic\cite{radovic2018latest} where at $8.85\times10^{20}$ POT(\textit{protons on target}), NO$\nu$A obtains:
\paragraph{}For $\nu_{\mu}$ disappearance, $\Delta m^{2}_{32}=+2.444^{+0.079}_{-0.077}\times10^{-3}eV^{2}$ with Normal Hierarchy preferred at 0.2$\sigma$, and the atmospheric mixing angle to be
\begin{align*}
\sin^{2}\theta_{23}&=0.558^{+0.041}_{-0.033}&(NH)\\
&=0.475^{+0.036}_{-0.044}&(IH)
\end{align*} 
\paragraph{}For $\nu_{e}$ appearance, inverted Hierarchy at $\delta_{CP}=\frac{\pi}{2}$ is disfavored at greater than 3$\sigma$. Further study includes approaching towards $2\sigma$ rejection.
The reactor mixing angle is also obtained as $\sin^{2}\theta_{13}=0.082$. The experiment is performing with excellent detectors and beam performance. 
\subsubsection{\textbf{Reactor Experiment for Neutrino Oscillation (RENO)} (Reactor)}
\begin{tasks}[counter-format={(tsk[r])}, label-align=centre, label-offset={1mm}, label-width={6mm}](2)
	\task \textbf{Location:} Yonggwang, South Korea
	\task \textbf{Period:} Aug, 2011-\textit{present}
	\task \textbf{Type:} $\bar{\nu}_{e}$ (disappearance)
	\task \textbf{Detection Technique:} Scintillator
\end{tasks}
The RENO experiment was set-up to measure the reactor neutrino mixing angle $\theta_{13}$. The experiment detects electron anti-neutrinos from the six reactors at Hanbit nuclear power plant located in the west coast of South Korea at Yonggwang, about 400km from Seoul, producing a total thermal output of 16.4GWhr\cite{kim2016measurement}.\\ Two identical detectors-- a near and a far detectors, each having 16.5 tons of Gadolinium (Gd) loaded liquid scintillator (LS), are located at 294m and 1383m respectively from the centre of the reactor array in the opposite direction to each other. The objective of placing two identical detectors is to minimize the systematic uncertainities arising due to the uncertainities in the number of $\bar{\nu}_e$ from the sources\cite{ahn2012observation}. Each RENO detector has a cylindrical shape of 8.8m in height and  8.4m in diameter; consists of a main Inner Detector(ID) and Outer veto Detector(OD). The ID is in a cylindrical stainless vessel of 5.4m diameter and 5.8m in height which houses two nested cylindrical acrylic vessels\cite{joo2012status}. The principle used in the detection process is the inverse beta decay (IBD) given by:
$$\bar{\nu_e}+p\rightarrow e^{+}+n$$In 1500 days of data taken between $11^{th}$ August 2011 and September 2015\cite{seo2017talk}, RENO observes an excess of $\sim 5 MeV$ in reactor neutrino spectrum. The absolute reactor neutrino flux is obtained to be $R=0.946\pm0.021$\cite{seo2017talk}. Moreover, using these 1500 days of data and observation of energy dependent disappearance of reactor neutrino, the parameters\cite{seo2017talk} obtained are- 
\begin{align*}
\sin^{2}2\theta_{13}&=0.086\pm0.006\text{\textit{(stat)}}\pm0.005\text{\textit{(sys)}}\quad&\text{with a precision of 9\%.}  \\
\lvert\Delta m^{2}_{ee}\rvert&=2.61^{+0.15}_{-0.16}(stat)\pm0.09(sys)\quad&\text{with a precision of 7\%}
\end{align*}
RENO is expected to take data for $\sim3500$ days in order to improve the precision of $\theta_{13}$ to 6-7\% and that of $\lvert\Delta m^{2}_{ee}\rvert$ to 4-5\%\cite{seo2017talk}.
\subsubsection{\textbf{Daya Bay Reactor Neutrino Experiment} (Reactor)}
\begin{figure}[h]
	\includegraphics[width=3in]{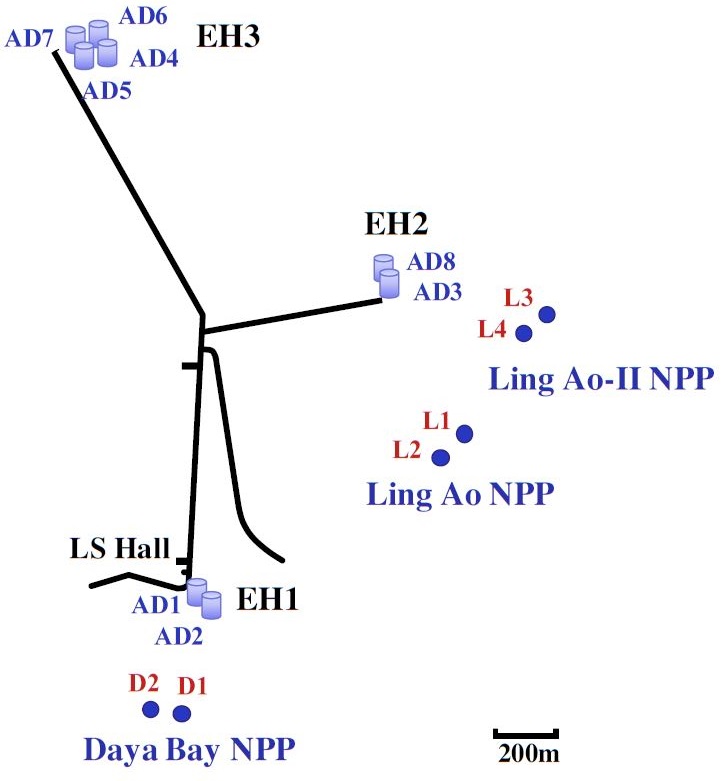}
	\caption{Layout of the Daya Bay experiment with eight anti-neutrino detectors(ADs) divided in three underground experimental halls(EHs) \& D1, D2, L1, L2, L3 and L4 are the nuclear reactors represented by dots in the image\cite{an2017improved}.}
	\label{Fig.16}
\end{figure}
\begin{tasks}[counter-format={(tsk[r])}, label-align=centre, label-offset={1mm}, label-width={6mm}](2)
	\task \textbf{Location:} Daya bay, China
	\task \textbf{Period:} 2011-\textit{present}
	\task \textbf{Type:} $\bar{\nu}_{e}$ (disappearance)
	\task \textbf{Detection Technique:} Scintillation
\end{tasks}
The Daya Bay\cite{cao2016overview,kim2013reactor} nuclear power complex is located on the southern coast of China, approximately 55km to the north-east of Hongkong. The experiment primarily aims at determining the reactor mixing angle parameter $\theta_{13}$. The layout of the Daya Bay experiment is depicted in Figure \ref{Fig.16}.\\The anti-neutrinos are generated by six nuclear reactors, producing a total of 2.9GW of thermal power, deployed in two near underground experimental halls (EHs) with baselines of 560m and 600m and one far underground EH with baseline of 1640m\cite{an2017improved}. There are three kinds of detectors in every experimental hall-- the Anti-neutrino Detectors(ADs), the water \~{C}erenkov detectors and the RPC detectors. In total, there are eight 110tons ADs, 3 water pools filled with a total of 4400tons of purified water, and three arrays of RPCs covering a total of $800m^{2}$\cite{an2012side}. Each AD has three nested cylindrical volumes separated by concentric acrylic vessels. The ouermost vessel is constructed of stainless steel known as SSV.\\In March 2012, Daya Bay collaboration announced the observation of non-zero value of $\sin^{2}\theta_{13}$ with a significance of 5.2$\sigma$\cite{an2012observation}. Recently in April 2017, the Daya Bay experiment measured the rate and energy spectrum of electron antineutrinos emitted by the six reactors. Combining 217 days of data collected using six antineutrino detectors with 1013 days of data using eight detectors,
a total of $2.5\times10^6$ electron anti-neutrino inverse beta-decay interactions were observed\cite{an2017measurement}. The oscillation parameters\cite{an2017measurement} are found to be, 
\begin{align*}
\sin^{2}2\theta_{13}&=0.0841\pm0.0027(stat)\pm0.0019(syst)\\
\lvert\Delta m^{2}_{ee}\rvert&=(2.50\pm0.06(stat)\pm0.06(syst))\times10^{-3}eV^{2}\\
\Delta m^{2}_{32}&=(2.45\pm0.06(stat)\pm0.06(syst))\times10^{-3}eV^{2},\quad&\text{assuming the normal ordering}\\
&=(-2.56\pm0.06(stat)\pm0.06(syst))\times10^{-3}eV^{2},\quad&\text{assuming the inverted ordering}
\end{align*}
\subsection{Future experiments}
\subsubsection{\textbf{SNO+} (Solar, geo-neutrinos, reactor, astrophysical)}
\begin{tasks}[counter-format={(tsk[r])}, label-align=centre, label-offset={1mm}, label-width={5mm}](2)
	\task \textbf{Location:} Creighton Mine, Vale, Canada
	\task \textbf{Period:} under construction (\textit{Feb, 2017})
	\task \textbf{Type:} $\nu_{x}$, x=e,$\mu$,$\tau$
	\task \textbf{Detection Technique:} Liq. Scintillator
\end{tasks}
SNO+, the successor of the 2015 Nobel prize winning SNO Collaboration, is a multi-purpose large liquid scintillator located at a depth of about 6000 metre water equivalent in SNOLAB.\\SNO+\cite{andringa2016current,maneira2016status,maneira2013sno+} will reuse the SNO detector holding the Tellurium-loaded liquid scintillator in a 12m diameter acrylic vessel. Near the target, the scintillator consists of linear alkyl benzene (LAB) and 2g/L of the flour 2-5 diphenyloxazole(PPO). LAB has good optical transparency, low scattering and fast decay time. The light emitted by charged particles passing through the scintilator is detected by $\sim$9,500 PMTs that view the acrylic vessel. The energy and the position of each element are reconstructed from the PMT signal. Also, new scintillator purification systems have been designed to reach the required $U$ and $Th$ contamination levels.\\The physics goals of SNO+ are the searches for the $0\nu\beta\beta$ of $^{130}Te$. The discovery of such a rare process will confirm the Majorana nature of this elusive particle and play a key ingredient in the theory of leptogenesis. The depth of SNOLAB also exposes to measure low energy solar neutrinos in the \textit{pep} and CNO regions. The precise measurement of the \textit{pep} flux can probe the MSW (Mikheyev, Smirnov and Wolfenstein) effect of neutrino mixing and the solar metallicity. The large volume and the high radiopurity of the SNO+ experiment can probe the understanding of heat production mechanism in the Earth by observation of geo-neutrinos. The study of reactor neutrino from the Bruce, Pickering and Darlington nuclear generating stations will improve the neutrino oscillation parameters. The experiment also aims at studying neutrinos from supernovae explosions and the low background allows search for invisible nucleon decay and axion-like particle search (\textit{exotic physics})\cite{chen2008sno+}.\\The data taking of SNO+ will be divided into three phases\cite{kamdin2015understanding}:
\paragraph{\underline{Water phase:}} The acrylic vessel is fitted with 905 tonnes of pure water and is aimed at search for exotic physics and supernovae explosions.
\paragraph{\underline{Scintillator phase without $^{130}Te$:}} The detector will be filled with about 780t of LAB-PPO liquid scintillator and the goals cover measurement of neutrinos from \textit{pep} and CNO regions in SSM, study of geo and reactor anti-neutrino via IBD process.
\paragraph{\underline{Scintillator with $^{130}Te$ loaded:}} This will be the longest phase and expected to last for 5 years for search of $0\nu\beta\beta$-decay of $^{130}Te$ specifically.
\subsubsection{\textbf{Hyper Kamiokande(Hyper-K)} (Solar, atmospheric, astrophysical)}
\begin{figure}[h]
	\includegraphics[width=5.3in]{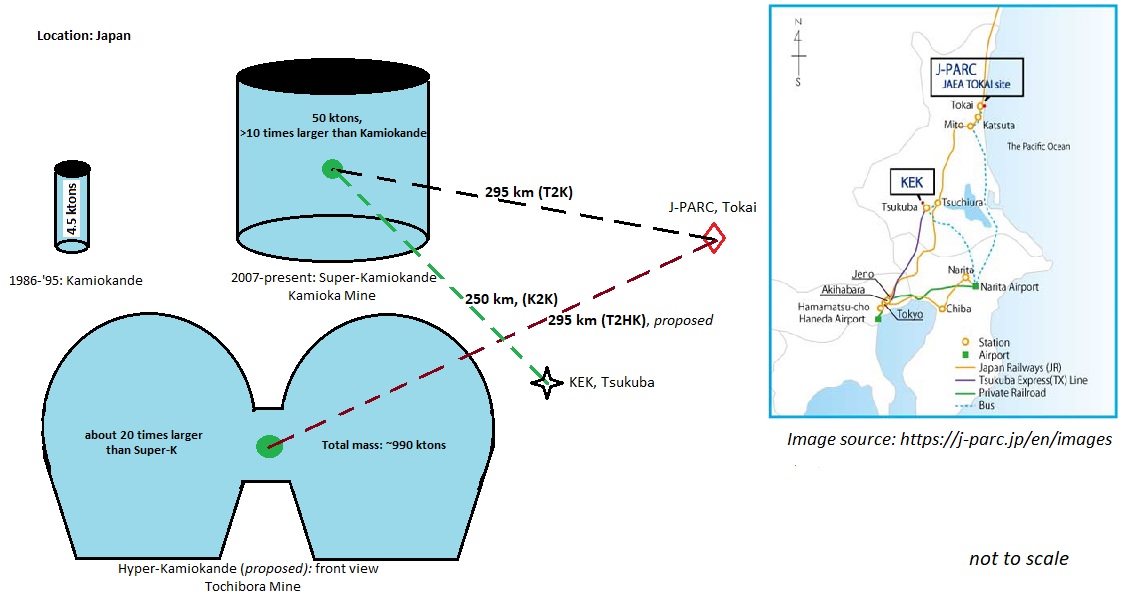}
	\caption{A rough sketch of the experiments in Japan}
	\label{Fig.8}
\end{figure}

\begin{tasks}[counter-format={(tsk[r])}, label-align=centre, label-offset={1mm}, label-width={6mm}](2)
	\task \textbf{Location:} Tochibora Mine, Gifu, Japan
	\task \textbf{Period:} $\sim2025$ (\textit{expected})
	\task \textbf{Type:} $\nu_{x}$, x=e,$\mu$,$\tau$
	\task \textbf{Detection Technique:} Water Cherenkov
\end{tasks}
Hyper Kamiokande\cite{abe2011letter} is a proposed next generation water Cherenkov detector that is going to success the present Super-K experiment(Figure \ref{Fig.8}). It will serve as the detector of the long baseline neutrino oscillation experiment planned for the upgraded J-PARC.\\Hyper-K consists of two cylindrical tanks lying side-by-side, the outer dimensions of each tank being $48m(W)\times54m(H)\times250m(L)$. The total fiducial mass of the detector is 0.99 million metric tons which is about 20-25 times larger than that of Super-K. Hyper-K is planned to be constructed at about 8km south of Super-K and 295km away from J-PARC at an underground depth of 1750 m.w.e.(648m). The inner detector design will be viewed by 99,000 $20^{\prime\prime}$ PMTs\cite{cremonesi2015sensitivity}.\\The scope of study in Hyper-K covers high precision measurements of solar neutrinos having sensitivity to \textit{hep} solar neutrinos, observation of supernova burst neutrinos, dark matter searches and possible flare neutrinos, neutrino mass hierarchy i.e. to select $\Delta m^{2}_{32}>0\text{ or } \Delta m^{2}_{32}<0$ with more than 3$\sigma$ significance provided $\sin^{2}\theta_{23}<0.4$. The day-night asymmetry of the solar neutrino flux-- concrete evidence of the matter effects on oscillation could be discovered\cite{di2017hyper}. The experiment has potential for precision measurements of neutrino oscillation parameters and reach for CPV in the leptonic sector. The prospects of geoneutrinos are also mentioned.
\subsubsection{\textbf{Tokai to Hyper-Kamiokande (T2HK)} (Accelerator-based experiment)}
\begin{tasks}[counter-format={(tsk[r])}, label-align=centre, label-offset={1mm}, label-width={6mm}](2)
	\task \textbf{Location:} J-PARC, Tokai$\rightarrow$Hyper-K
	\task \textbf{Period:} $\sim2025$ (\textit{operation})
	\task \textbf{Type:} $\nu_{\mu}$, $\bar{\nu}_{\mu}$
	\task \textbf{Detection Technique:} Water Cherenkov
\end{tasks}
The T2K collaboration with its far detector at Super-K was successful in observing neutrino oscillation via $\nu_{\mu}$ to $\nu_{e}$ appearance with significance over $7\sigma$. Another experiment T2HK has been proposed as the successor to T2K with the objectives\cite{group2014long} of discovering CP asymmetry, determination of Mass hierarchy(NH or IH), octant of $\theta_{23}$ from atmospheric neutrino data with $>3\sigma$ C.L. at $\sin^{2}\theta_{23}>0.4$ \& $\sin^{2}\theta_{23}<0.46$ or $\sin^{2}\theta_{23}>0.56$, and searches for nucleon decay via $p\longrightarrow e^{+}\pi^{0}$ and $p\longrightarrow\bar{\nu}K^{+}$.\\ 
T2HK is a proposed next-generation long baseline(LBL) accelerator neutrino oscillation experiment\cite{ballett2017sensitivities}, with baseline of 295 km, where the protons beams of energy 30 GeV will be produced at J-PARC, Tokai and will be sent to the water Cherenkov detector Hyper-K in the Tochibora Mine(Figure \ref{Fig.8}). It has the same off-axis angle 2.5$^{0}$ and the baseline length as of Super-K. The beam power of the J-PARC is expected to be upgraded to 1.3 MW and the time period of operation of the experiment will be 10 years with 2.5 years in neutrino mode and 7.5 years in anti-neutrino mode. The differences in the oscillation behaviour between neutrinos and anti-neutrinos can give an estimate of CP violation.\\There is also a plan for T2HKK\cite{raut2017matter} which includes one detector at Hyper-K at a distance of 295 km from J-PARC as discussed earlier and the second detector  is proposed in the southern part of the Korean peninsula at a distance of $1100-1300km$ from the accelerator with off-axis angles varying from 1$^{0}$ to 2.5$^{0}$.
%Due to shorter baseline length in comparison to other proposed future experiments with more than 1000km baseline like DUNE, the CPV effect will be more prominent than the matter effect. 
\subsubsection{\textbf{Jiangmen Underground Neutrino Observatory(JUNO)} (Reactor experiment)}
\begin{tasks}[counter-format={(tsk[r])}, label-align=centre, label-offset={1mm}, label-width={6mm}](2)
	\task \textbf{Location:} Kaiping, China
	\task \textbf{Period:} 2019 (\textit{expected to be completed})
	\task \textbf{Type:} $\bar{\nu}_{e}$ (disappearance)
	\task \textbf{Detection Technique:} Liquid Scintillator
\end{tasks}
JUNO\cite{he2015jiangmen,brugiere2017jiangmen} is a multi-purpose neutrino oscillation experiment with the objectives to determine mass hierarchy by precisely measuring the energy spectrum of reactor $\bar{\nu}_e$ from the reactors, observe supernova neutrinos, study the atmospheric, solar and geo- neutrinos. It also aims at improving the precision of $\Delta m^{2}_{21}$ and $\Delta m^{2}_{32}$ to better than 1\% as compared to that of $\sim$4\% by its predecessor Daya Bay.\\The principle of neutrino detection to be used is the Inverse Beta Decay process given by--
$$\bar{\nu}_{e}\text{ (\textit{from reactors})}+\text{p}\longrightarrow e^{+}+n$$
The JUNO detector at Jiangmen, Kaiping is about 53 km from Yangjiang and Taishan nuclear power plants, producing a total thermal power of 36GW. The configuration of 20ktons of liquid scintillator with $\sim$15,000 20$^{\prime\prime}$ high detection efficiency PMTs(at 1MeV) makes it the largest liquid scintillation detector ever built. It is planned to locate at a depth of 700m through a tunnel to suppress muon induced background. Also, a 270 m high granite mountain will shield the cosmic muons. JUNO includes an underground experiment hall, a water pool, a central detector and muon tracking detector. The water pool will protect the central detector from natural activity and also serves as a water \~{C}erenkov detector with $\sim$1500 20$^{\prime\prime}$ PMTs to track cosmic muons\cite{an2016neutrino}. The experiment is anticipated to take data from 2020. 
\subsubsection{\textbf{RENO-50} (Reactor experiment)}
RENO-50\cite{kim2015review,djurcic2013review}, the successor of RENO, is a proposed underground detector which will consist of 18 ktons of ultra-low radioactivity liquid scintillator and 15,000 high quantum efficiency 20$^{\prime\prime}$ PMTs located beneath Mt. Guemseong at a depth of 450m, 47km away from Hanbit/Yonggkang nuclear power plant in South Korea. The detector is expected to detect neutrinos from the nuclear reactors, the Sun and any possible stellar events. The detector also aims in analyzing $\sim$5600 events of a neutrino bursts from a supernova in the galaxy, $\sim$1000 geoneutrino events for 6 years and $\sim$200 events of $\nu_{\mu}$ from J-PARC neutrino beam every year. The expected year of operation of RENO-50 is 2021.\\The proposed site has the maximum neutrino oscillation due to solar mixing angle $\theta_{12}$. The goals include determination of neutrino mass ordering and measurement of $\theta_{12}$ and $\Delta m^{2}_{21}$ with unprecedented accuracy at $<0.5\%$.
\subsubsection{\textbf{India-based Neutrino Observatory (INO)} (Atmospheric experiment)}
\begin{tasks}[counter-format={(tsk[r])}, label-align=centre, label-offset={1mm}, label-width={6mm}](2)
	\task \textbf{Location:} Theni, TN, India
	\task \textbf{Period:} 2018 (\textit{construction})
	\task \textbf{Type:} $\nu_{\mu}$, $\bar{\nu_\mu}$ (disappearance)
	\task \textbf{Detection Technique:} Magnetised Iron Calorimeter
\end{tasks}

India-based Neutrino Observatory\cite{datar2017india,kumar2017invited} is a proposed multi-institutional mega science project to be set up under the Bodi West Hills near Pottipuram village in the Theni District of Tamil Nadu. INO is surrounded by a rock cover of at least one kilometer which restricts the background cosmic muons.\\The underground laboratory consists of Iron Calorimeters (ICAL) which will have Resistive Plate Chambers\cite{bhattacharya2014error} as the main detector elements. The proposed ICAL@INO will have 5.6 cm thick iron plates and RPCs of area 1.84m$\times$1.84m weighing a total mass of 50ktons. The 29,000 RPCs will be used as active medium for detection\cite{kumar2017invited}. ICAL detectors generally detect muon neutrinos ($\nu_{\mu}$) only, the INO-ICAL being sensitive to atmospheric $\nu_{\mu}$ in the 1-15 GeV range. It will provide the timing and position information of the particle generated by interaction of atmospheric neutrinos (anti-neutrinos) with the iron plates. The experiment will study the neutrinos and anti-neutrinos separately using the 1.5 Tesla strength of magnetic field. This will enable to determine the charges of muon($\mu^-$, $\mu^+$) particles produced as a result of the charge-current interactions of $\nu_{\mu}$ and $\bar{\nu_\mu}$ inside the calorimeter\cite{khatun2018can}. The reaction involved in the process is 
$$\nu_{\mu} \text{ (\textit{atmospheric})}+_{26}^{56}Fe_{30}\longrightarrow \mu^{-}+X(\textit{hadrons})$$ 
The aims\cite{bhattacharya2006india} of INO-ICAL include determination of $\nu$-mass hierarchy and deviation of $\theta_{23}$ from the maximal value. The experiment also has long-term plan of exploring the CP violation in the leptonic sector using long baseline accelerator neutrinos in its second phase.
\subsubsection{\textbf{Deep Underground Neutrino Experiment (DUNE)} (Accelerator-based)}
\begin{tasks}[counter-format={(tsk[r])}, label-align=centre, label-offset={1mm}, label-width={6mm}](2)
	\task \textbf{Location:} FermiLab$\rightarrow$Soudan Lab, US
	\task \textbf{Period:} 2017 (\textit{construction})
	\task \textbf{Type:} $\nu_{\mu}$, $\nu_{e}$, $\bar{\nu}_\mu$, $\bar{\nu}_e$
	\task \textbf{Detection Technique:} LAr-TPC
\end{tasks}
The DUNE\cite{diwan2016long,adams2013long} is a long baseline neutrino oscillation experiment that consists of a horn-produced proton beams of 60-120 GeV, a beam power of $\sim$1.2MW from FermiLab, Illinois will be sent to a 40ktons fiducial volume liquid argon time-projection far detector(FD) at a depth of $\sim$1450m at Sanford Underground Research Facilty (SURF) in South Dakota passing through a high resolution near detector(ND) at FermiLab. The baseline is $\sim$1300km and is ideal to measure the matter effects and the CP violations simultaneously.\\DUNE\cite{acciarri2016long} aims at probing the following fundamental problems:
\begin{enumerate}
	\item precision measurements of the parameters that govern the neutrino oscillation $\nu_{\mu}\rightarrow\nu_{e}$($\nu_{e}$ appearance) and $\bar{\nu}_{\mu}\rightarrow\bar{\nu}_{e}$($\bar{\nu}_{e}$ appearance), thereby--
	\begin{itemize}
		\item contributing to the CP violating phase. $\delta_{CP}\neq0,\pi$ will correspond to the discovery of CPV in the leptonic sector. Further, presence of matter-antimatter asymmetry can be explained.
		\item determining the neutrino mass hierarchy (\textit{whether normal or inverted?}) by confirming the sign of $\Delta m^{2}_{31}=m^{2}_{3}-m^{2}_{1}$.
		\item determining the octant of atmospheric mixing angle $\theta_{23}$ i.e. whether $\theta_{23}<45^{0}$ or $\theta_{23}>45^{0}$? 
	\end{itemize}
	\item search for proton decay, providing a path to Grand Unified Theory (GUT).
	\item detection and measurement of the $\nu_{e}$-flux from a core-collapse supernovae within our galaxy, if any occurs in the operation period of the DUNE project. 
\end{enumerate}
\begin{figure}[h]
	\includegraphics[width=5in]{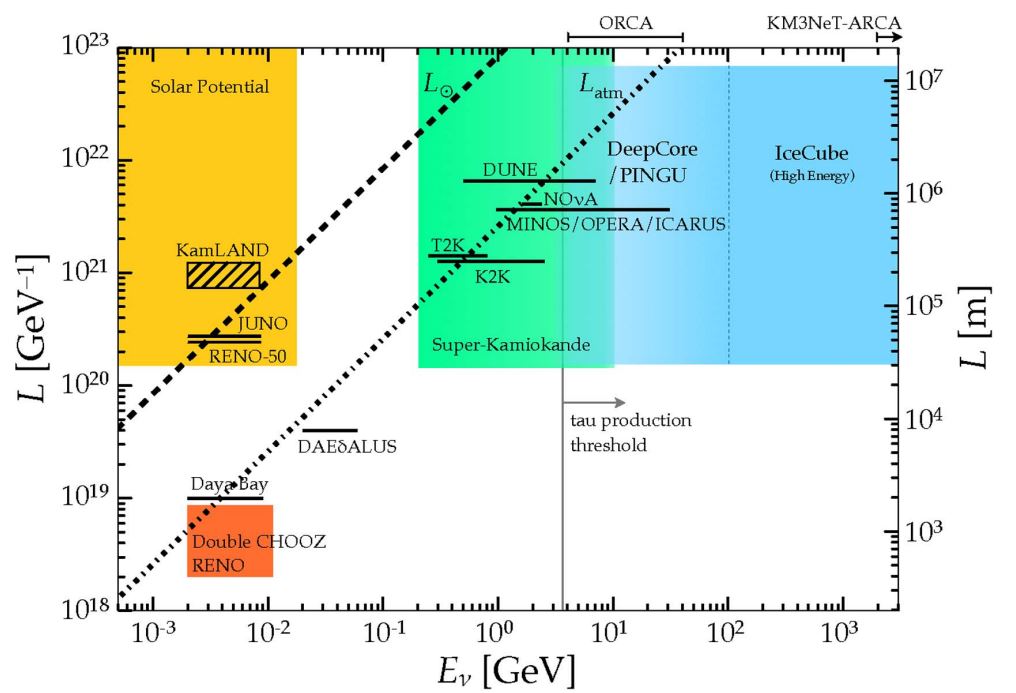}
	\caption{The axes represents the energy ranges and baselines of the existing and proposed neutrino oscillation experiments. The \textit{dashed diagonal} line indicates the oscillation scale $L_{\odot}$ set by solar mass-squared difference $\Delta m^{2}_{21}$ and the \textit{dotted-dashed} are those set by atmospheric mass-squared difference $\Delta m^{2}_{32}$. . The vertical line at 3.5 GeV represents the threshold for tau-lepton production in $\nu_{\tau}$-CC events. The bars above the plot-box indicate the energy ranges covered by the KM3NeT-ORCA \& ARCA detectors.\cite[p.9]{aartsen2017pingu} }
	\label{Fig.14}
\end{figure}
\subsubsection{\textbf{SuperNEMO} (Neutrinoless double beta decay experiment)}
\begin{tasks}[counter-format={(tsk[r])}, label-align=centre, label-offset={1mm}, label-width={6mm}](2)
	\task \textbf{Location:} LSM, Fr\'{e}jus Road Tunnel, France
	\task \textbf{Period:} 2017 (\textit{final stage})
	\task \textbf{Type:} $\nu_{e}$
	\task \textbf{Detection Technique:} Calorimeter 
\end{tasks}
SuperNEMO\cite{vilela2015supernemo} is the successor to the NEMO-3 experiment and the detector replaces the NEMO-3 in the Modane Underground Laboratory(LSM). It develops the strategy proven by the NEMO-3 experiment of a tracker, calorimeter and seven double-beta decaying isotopes.\\NEMO-3 produced the best  measurements of $2\nu\beta\beta$ for all the 7 isotopes but failed to observe $0\nu\beta\beta$ events in any of them. SuperNEMO will employ 100kg of $^{82}Se$ isotopes to probe $0\nu\beta\beta$ which is a direct indication of leptogenesis. The experiment also has a unique capability to search for alternative $0\nu\beta\beta$ decay mechanisms \textit{e.g.} right-handed currents and Majoron emission. The increase in the source mass and stringent background requirements fulfillment gives SuperNEMO a sensitiveness of two orders better than that of NEMO-3\cite{waters2017latest}.\\The SuperNEMO detector will consist of 20 modules, each one using 5-10 kg of $^{82}Se$ in a $4\times3.7m^{2}$ foil sheet. A magnetized tracker volume will occupy the space on both sides of the foil. The tracker will be made of 113 rows of 9 tracker cells per side. A calorimeter will surround the tracker, formed of 520 scintillator blocks coupled to 8$^{\prime\prime}$ Hamamatsu PMTs\cite{guzowski2015construction}.\\The aim of the calorimeter is to measure the energy and time of the $\beta$-decay electrons and other background particles. The tracker will discriminate among electrons, positrons, gamma and alphas and also measure event kinematics. The reactions involved in the $\beta\beta$ decay process of the sources are--
\begin{align*}
^{82}Se&\longrightarrow^{82}Kr+2e^{-}\\
^{150}Nd&\longrightarrow^{150}Sm+2e^{-}
\end{align*}
\subsubsection{\textbf{Cubic Kilometer Neutrino Telescope (KM3NeT)} (Atmospheric,astrophysical)}
\begin{tasks}[counter-format={(tsk[r])}, label-align=centre, label-offset={1mm}, label-width={6mm}](2)
	\task \textbf{Location:} Mediterranean Sea
	\task \textbf{Period:} 2021 (\textit{operation})
	\task \textbf{Type:} $\nu_{e}$, $\nu_{\mu}$, $\nu_{\tau}$
	\task \textbf{Detection Technique:} Cherenkov (Sea)  
\end{tasks}
KM3NeT\cite{zaborov2018km3net} is one of the largest multi-research infrastructure dedicated to neutrinos. It comprises of two underwater neutrino detectors\cite{collaboration2016km3net} in the Mediterranean sea:
\begin{enumerate}
	\item Oscillation Research with Cosmics in the Abyss (ORCA)
	\item Astroparticle Research with Cosmics in the Abyss (ARCA)
\end{enumerate}
Both ORCA and ARCA will detect neutrinos through the detection of the Cherenkov light induced by secondary particles produced from interactions with sea water. They will consists of several thousands Digital Optical Modules (DOMs)\cite{bruijn2016km3net}, each DOM consisting 31 PMTs and associated electronics packed in a pressure resistant glass sphere. The DOMS are arranged in vertical strings called Detection Units (DUs), with 18 DOMS on each DU. The DUs are again anchored to the seabed and then connected to the shore station using underwater cable network. Table \ref{T3} shows the spacing between two DOMs and two DUs and describes an idea of total number of DOMs to be used in the KM3NeT experiment\cite{zaborov2018km3net}.
    \paragraph{ORCA}\cite{katz2014orca} is optimized for the study of atmospheric neutrino oscillation in the energy range between 3 GeV and 20 GeV, with the primary goal to determine the neutrino mass hierarchy\cite{kouchner2016km3net} considering  Mikheyev-Smirnov-Wolfenstein (MSW) effect\cite{wolfenstein1978neutrino,mikheyev1985sp} in the core, mantle and and crust of the Earth. It has a mass of 8 Mton of water and is located 40 km off-shore Toulon, France at a depth of 2450m below the sea level. It's large volume and low energy threshold($\sim$3GeV) will enable to detect about $5\times10^{3}$ atmospheric neutrinos per year with muon and electron neutrinos in majority.
    \begin{table}[b]    	
    	\begin{tabular}{l|cc}
    		\centering
    		%\hline\hline
    		\textbf{Specifications}& \textbf{ORCA}&\textbf{ARCA}\\\hline\\
    		a. The vertical spacing between DOMs:&9m&36m\\\\
    		b. The average horizontal spacing between DUs:&23m&90m\\\\
    		c. Total DUs:&115&230				
    	\end{tabular}
    	\caption{Specifications of the ORCA and ARCA of KM3NeT.}
    	\label{T3}	
    \end{table}
    \paragraph{ARCA}\cite{migliozzi2016high} is optimized for the purpose of neutrino astronomy in the $TeV-PeV$ energy range. It is located 100km off-shore Capo Passero, Sicily in Italy at a depth of 3500m below the sea level. The main objective of ARCA is the detection of high energy neutrinos of cosmic origins, particularly from sources within the galaxy. The experiment will also probe how the neutrinos are accelerated to PeV energies. It is complementary to the IceCube experiment that observes mostly the northern sky. KM3NeT-ARCA will observe mostly the southern sky by identifying the upward-moving muons. Moreover, the expected angular resolution is better than $0.3^{\circ}$ at $E>10TeV$ which is better than that of the IceCube. ARCA also includes study of the flavor composition of cosmic neutrino flux, indirect dark matter searches, searches for exotic particles and Lorentz invariance violation\cite{zaborov2018km3net}.
\subsubsection{\textbf{Protvino to ORCA (P20)} (Accelerator-based)}
\begin{tasks}[counter-format={(tsk[r])}, label-align=centre, label-offset={1mm}, label-width={6mm}](2)
	\task \textbf{Location:} Protvino, Russia$\rightarrow$ORCA, Mediterranean Sea 
	\task \textbf{Period:} proposed
	\task \textbf{Type:} $\nu_{\mu}\rightarrow\nu_{e}$,
$\bar{\nu_\mu}\rightarrow\bar{\nu_e}$	
	\task \textbf{Detection Technique:} Cherenkov  
\end{tasks}
P2O\cite{zaborov2018km3net,brunner2013measurement}, a proposed accelerator-based long baseline neutrino experiment, is an extended project of KM3NeT where a neutrino beam from Russia to the Mediterranean sea of baseline length 2590 km will be used for a very sensitive probe of leptonic CP violation\cite{sakharov1967violation}. The U-70 synchotron\cite{гаркушаисследование} in the Protvino accelerator facility, located at a distance of 100km south of Moscow will accelerate the protons upto 70GeV producing an intense beam of neutrinos and anti-neutrinos with energies upto 7GeV. The first oscillation maxima will be at 5 GeV which is close to the matter resonance energy of 4GeV in the Earth's crust. However, the upgradation of beam power of the synchotron is necessary for the operation of the experiment. Currently, the U-70 operates at a time-averaged beam power of 15kW which can be increased\cite{U-70} to:
\begin{itemize}
	\item $\approx90kW$ by new ion injection scheme and optimization of the accelerator cycle from 9s to $\approx4.5$s.
	\item 450kW by a chain of new booster accelerators.
\end{itemize}
Other necessary requirements are the construction of a neutrino beam line for a focussed beam in the direction of ORCA and a near detector (ND) to accurately monitor the neutrino beam intensity, energy spectrum and flavor composition before oscillations take place.
\begin{table}[b]
	\begin{tabular}{l|cc}
		\centering
		%\hline\hline
		\textbf{Parameters}& \textbf{Sensitivity}&\textbf{Running Time (\textit{in years})}\\\hline\\
		Mass Hierarchy&5-10$\sigma$&1 year(450kW), 5 years(90kW)\\\\
		CP violation&2-3$\sigma$&3 years(450kW), 5 years(90kW)					
	\end{tabular}
	\caption{Sentivity of the P2O experiment to mass hierarchy and CP violation.}
	\label{T2}	
\end{table}
The sensitivity of the experiment to the unknown mass hierarchy and leptonic CPV is summarised in Table \ref{T2}. A beam power of 90kW will produce 3000 neutrino events every year in ORCA in contrast to only 1000 such events with 1MW beam, 40kton LArTPC detector and baseline of 1300km of DUNE. Also, DUNE is expected to reach $3\sigma$ sensitivity to CP violation using 15 years of operation with the beam\cite{acciarri2016long}. Thus, P2O will be competitive to DUNE even when using the low intensity beam \textit{i.e.} 90 kW.
\subsubsection{\textbf{Long Baseline Neutrino Observatory (LBNO)} (Accelerator-based)}
\begin{tasks}[counter-format={(tsk[r])}, label-align=centre, label-offset={1mm}, label-width={6mm}](2)
	\task \textbf{Location:} CERN$\rightarrow$Pyh\"{a}salmi, Finland
	\task \textbf{Period:} prototypes under construction
	\task \textbf{Type:} $\nu_{\mu}\rightarrow\nu_{e}$, $\bar{\nu}_\mu\rightarrow\bar{\nu}_e$, $\nu_{\mu}\rightarrow\nu_{\tau}$, $\bar{\nu}_\mu\rightarrow\bar{\nu}_\tau$, $\nu_{\mu}$ and $\bar{\nu}_\mu$ disappearances.
	\task \textbf{Detection Technique:} LAr-TPC, ICAL 
\end{tasks}
LBNO\cite{stahl2012expression} is a proposed next-generation long baseline neutrino and anti-neutrino oscillation experiment. A wide-band neutrino beam will be observed at a distance of 2300 km from the Super Proton Synchotron (SPS), CERN of 750kW power with a high-pressure argon gas time projection chamber as the near detector\cite{rubbia2013laguna}. The far site of LBNO would consist of a deep underground laboratory hosting a 35 kilotons magnetised iron
calorimeter \textbf{(MIND)} and the Giant Liquid Argon Charge Imaging ExpeRiment \textbf{(GLACIER)}\cite{murphy2015glacier}, with a mass of 20ktons for the first phase and extendable to 100ktons for the second phase of operation. \\ With 5 years of operation, the first phase of LBNO has the potential to determine unambiguously the mass hierarchy to more than 5$\sigma$ confidence limit over the whole phase space. The experiment can also give evidence for CP violation at a significance of $3\sigma$ by exploiting the $\frac{L}{E}$ dependence of $\nu_{\mu}\rightarrow\nu_{e}$ and $\bar{\nu_\mu}\rightarrow\bar{\nu_e}$ appearance probabilities over their first and second oscillation maxima, distinguishing effects arising from $\delta_{CP}$ and matter\cite{agarwalla2014mass}. The neutrino beam can be changed into anti-neutrino beam by changing the polarity of the magnetised horns.
\begin{itemize}
	\item The $\nu_{\mu}\rightarrow\nu_{\mu}$ and  $\bar{\nu_\mu}\rightarrow\bar{\nu_\mu}$ disappearance channels would determine the atmospheric parameters.
	\item The $\nu_{\mu}\rightarrow\nu_{\tau}$ and $\bar{\nu_\mu}\rightarrow\bar{\nu_\tau}$ appearance channels will also be studied in an unprecedented precesion.
\end{itemize}  
LBNO would also provide complementary studies of the three active flavor transitions charge current(CC) events over a wide range of neutrino energies from sub-GeV to multi-GeV and can also probe the active-sterile transitions ny measuring neutral current(NC) events. The experiment will also search for proton decay. It is expected that after 10 years of exposure, the sensitivity of the proton lifetime will reach $\tau_{p}\geq2\times10^{34}years$ at 90\% C.L. in the $p\rightarrow K\bar{\nu}$ channel. The $p\rightarrow e^{+}\pi^{0}$ and $p\rightarrow\mu^{+}\pi^{0}$ channels may also be investigated. LBNO also can measure 5600 atmospheric neutrino events every year and detect unknown source of astrophysical neutrinos like that from the annihilation processes of WIMP particles. The neutrino burst from a galactic supernovae(SN) would also be observed during operation with high statistics in the electron neutrino channel, providing valuable information about the inner mechanism of the supernovae\cite{avanzini2015laguna}.
\begin{figure}[h]
	\includegraphics[width=4.5in]{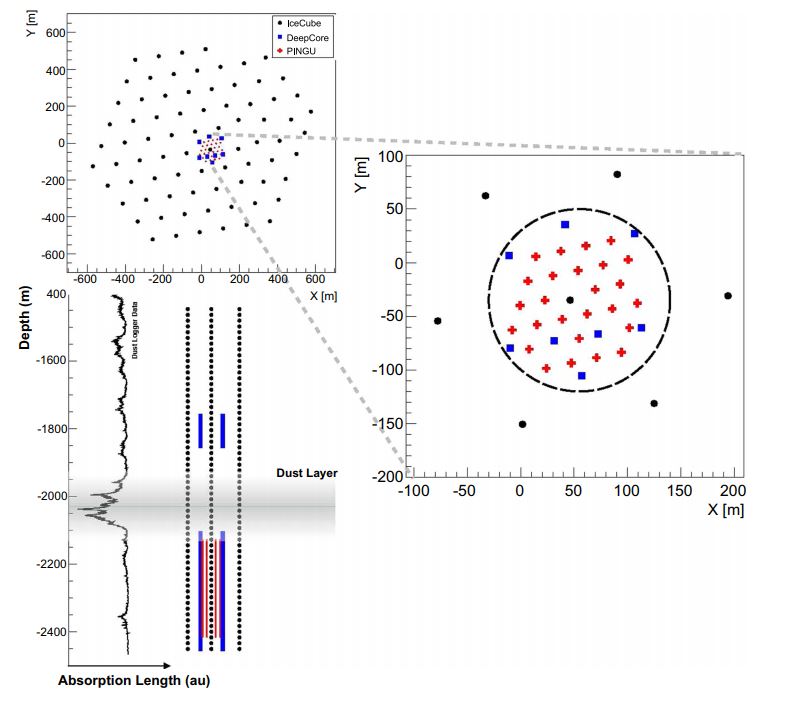}
	\caption{The image shows the top view of PINGU detectors within the IceCube DeepCore detector. The inset at right has:\\ (a) \textbf{Black circles} indicating the \textit{IceCube strings}, on a 125 m hexagonal grid,\\ (b) \textbf{Blue squares} indicate the existing \textit{DeepCore strings}, and \\(c) \textbf{Red crosses} show \textit{proposed PINGU string} locations.\\ PINGU modules would be deployed at the bottom of the detector with vertical spacing several times denser than DeepCore(\textit{see the depth-absorption length plot})\cite[p.7]{aartsen2017pingu}.}
	\label{Fig.15}
\end{figure}
\subsubsection{\textbf{Precision IceCube Next Generation Upgrade (PINGU)} (Atmospheric)}
\begin{tasks}[counter-format={(tsk[r])}, label-align=centre, label-offset={1mm}, label-width={6mm}](2)
	\task \textbf{Location:} Amundson-Scott Station, South Pole, Antarctica
	\task \textbf{Period:} 2021 (\textit{operation})
	\task \textbf{Type:} $\nu_{e}$, $\nu_{\mu}$, $\nu_{\tau}$
	\task \textbf{Detection Technique:} Ice Cherenkov  
\end{tasks}
PINGU\cite{aartsen2017pingu} is a 6 Mton Cherenkov detector and is the upgradation of the existing IceCube detector based on a configuration of 40 additional strings, each mounting 96 Digital Optical Modules (DOMs). The detector will follow the model of Deep Core extension and will be located at the centre of the IceCube strings(Figure \ref{Fig.15}). The primary goal of PINGU is the study of neutrino oscillations using atmospheric neutrino flux. The experiment aims at-- 
\begin{enumerate}
	\item determining the neutrino mass hierarchy\cite{clark2016pingu,aartsen2015determining} at 3$\sigma$ median significance within 5 years of operation and also aims at breaking the octant degeneracy of the atmospheric mixing angle $\theta_{23}$\cite{gonzalez2004measuring,barger2012neutrino}.
	\item high precision measurement of the rate of $\nu_{\tau}$ appearance which will test the unitarity of $3\times3$ PMNS neutrino mixing matrix.
	\item improving the sensitivity of searches for low mass dark matter in the Sun.
	\item using neutrino tomography\cite{winter2016atmospheric} to directly probe the compostion of the Earth's core.
	\item improving the IceCube's sensitivity to neutrinos from galactic supernovae.
\end{enumerate}
The experiment has the potential to detect 3000 charge-current $\nu_{\tau}$ appearances every year compared to 180 such interactions observed in 2,806 days of Super-K data\cite{abe2013evidence}. Due to its neutrino path length through the Earth ranging upto 12,700 km, PINGU has an added advantage over the existing and the proposed long-baseline neutrino beam experiments \textit{i.e.} the experiment can observe the oscillation phenomena at energies and baselines an order of magnitude larger than those of LBNEs. Refer Figure \ref{Fig.14} for illustration. For example, the matter effects at the energies and baselines of NO$\nu$A and T2K experiments are relatively weak\cite{agarwalla2013resolving}. So, the sensitivity to $\theta_{23}$ depends considerably on the CPV parameter $\delta_{CP}$\cite{chatterjee2013octant}. However, due its longer baseline length, $\delta_{CP}$ has little impact on PINGU observation. By comparing the $\nu_{\mu}$ survival and transition($\rightarrow\nu_{e}$) probabilities for neutrinos and anti-neutrinos passing through the Earth's core and mantle and observing the resonant matter effect will help to break the octant degeneracy\cite{aartsen2017pingu}.
\section{CURRENT STATUS OF NEUTRINO}
\subsection{UPDATES OF KNOWN AND UNKNOWN OSCILLATION PARAMETERS}
\begin{figure}[h]
	\centering
	\includegraphics[width=4.8in]{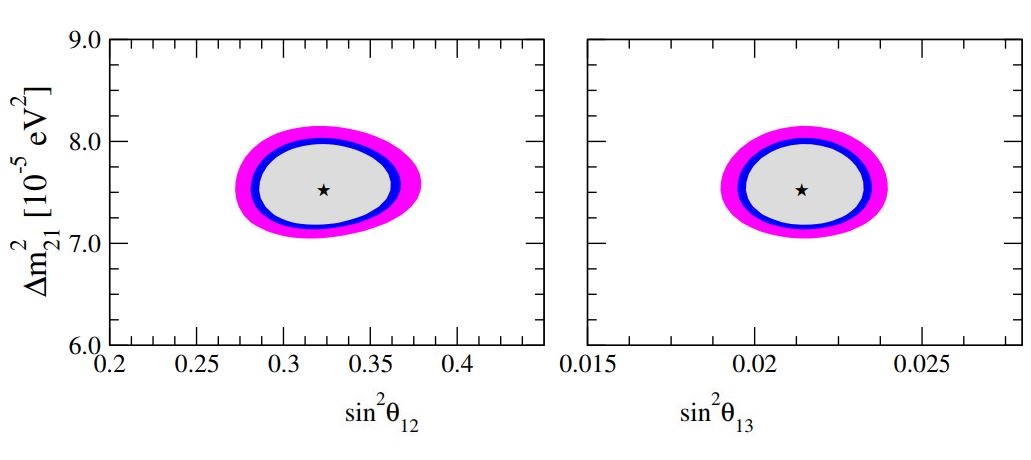}
	\caption{Global Fit of Atmospheric oscillation parameters with 90, 95 and 99\% C.L. The star represents the best fit value from the global analysis\cite{de2017status}.}
	\label{Fig.9}
\end{figure}

Global fit 1(Table \ref{T1}) doesn't include the updated results of Super-Kamiokande phase IV\cite{de2017status} whereas global Fit 2 does\cite{capozzi2017global}. In the global fit 1, there is no significant variations in the solar neutrino oscillation parameters(Figure \ref{Fig.9}).
\begin{table}[t]
	\caption{The neutrino oscillation parameters, according to two different global fits- Global Fit 1\cite{de2017status} \& Global Fit 2\cite{capozzi2017global}.}
	\begin{tabular}{c|lccc}
		\centering
		%\hline\hline
		\textbf{Global Fit 1}&\textbf{Parameters\footnote{$\Delta m_{3\textit{n}}^{2}\equiv\Delta m_{31}^{2}>0$, for Normal Hierarchy, $\Delta m_{31}^{2}$ being the reactor mass squared difference. $\Delta m_{3\textit{n}}^{2}\equiv\Delta m_{32}^{2}<0$, for Inverted Hierarchy, $\Delta m_{32}^{2}$ being the atmospheric mass squared difference.}}& \textbf{Mass Ordering}&\textbf{Best Fit}& \textbf{3$\sigma$}\\\hline\hline
		&$\Delta m_{21}^{2}(\times10^{-5}eV^{2})$&NH,IH&7.56&7.05-8.14\\\cline{2-5}
		&$\theta_{12}$($^o$)&NH,IH&34.5&31.5-38.0\\\cline{2-5}
		&$\lvert\Delta m_{3n}^{2}\rvert(\times10^{-3}eV^{2})$&NH&2.55&2.43-2.67\\
		&&IH&2.49&2.37-2.61\\\cline{2-5}
		&$\theta_{23}$($^o$)&NH&41.0&38.3-52.8\\
		&&IH&50.5&38.5-53.0\\\cline{2-5}
		&$\theta_{13}$($^o$)&NH&8.44&7.9-8.9\\
		&&IH&8.41&7.9-8.9\\\cline{2-5}
		&$\sigma(\pi)$&NH&1.40&0.00-2.00\\
		&&IH&1.44&0.00-0.17 \& 0.79-2.00\\\cline{2-5}
		&$\sigma$($^o$)&NH&252&0-360\\
		&&IH&259&0-31 \& 142-360\\\hline\hline		
		\textbf{Global Fit 2}&$\Delta m_{21}^{2}(\times10^{-5}eV^{2})$&NH,IH&7.37&6.93-7.96\\\cline{2-5}
		&$\sin^{2}\theta_{12}$($\times10^{-1}$)&NH,IH&2.97&2.50-3.54\\\cline{2-5}
		&$\lvert\Delta m_{3n}^{2}\rvert(\times10^{-3}eV^{2})$&NH&2.525&2.411-2.646\\
		&&IH&2.505&2.390-2.624\\\cline{2-5}
		&$\sin^{2}\theta_{23}$($\times10^{-1}$)&NH&4.25&3.81-6.15\\
		&&IH&5.89&3.84-6.36\\\cline{2-5}
		&$\sin^{2}\theta_{13}$($\times10^{-2}$)&NH&2.15&1.90-2.40\\
		&&IH&2.16&1.90-2.42\\\cline{2-5}
		&$\sigma(\pi)$&NH&1.38&0-0.17 \& 0.76-2\\
		&&IH&1.31&0.00-0.15 \& 0.69-2.00\\\cline{2-5}
		&$\sigma$($^o$)&NH&248&0-31 \& 137-360\\
		&&IH&236&0-27 \& 124-360\\\hline				
	\end{tabular}
\label{T1}	
\end{table}
\begin{figure}[h]
	\centering
	\includegraphics[width=4.8in]{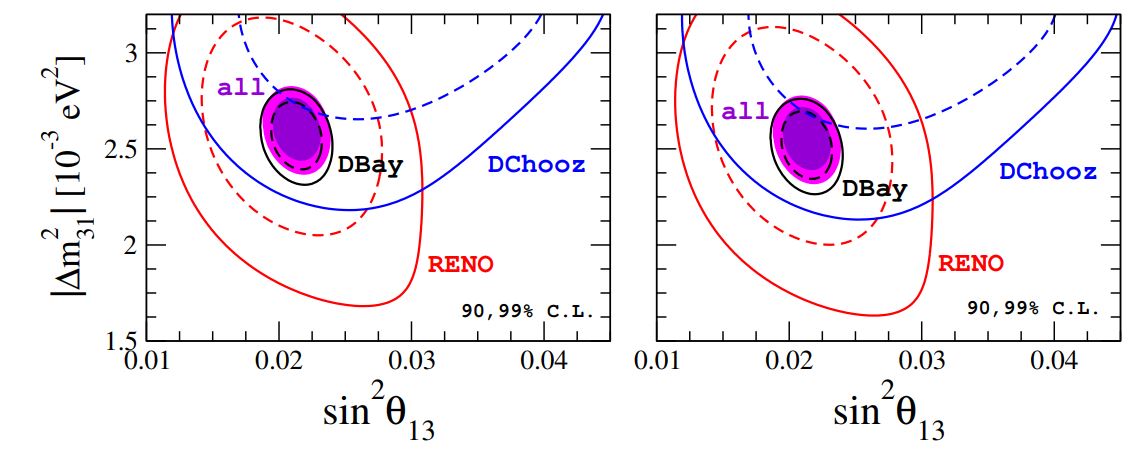}
	\label{(a)}
	\caption{The data representations of the individual \textbf{reactor neutrino experiments} Daya Bay, RENO and Double Chooz and of the combination of the three experiments at the $sin^{2}_{\theta_{13}}-\Delta m_{31}^{2}$ plane for the allowed regions of 90 and 99\% C.L.\cite{de2017status}} 
	\label{Fig.10}
\end{figure}
 Global Fit 1(Table \ref{T1}) includes the recent atmospheric data from ANTARES and IceCube DeepCore, the long-baseline data from T2K, MINOS and No$\nu$A, and the reactor experiments data include from Daya Bay, Double Chooz and RENO; mostly dominated by Daya Bay(Figure \ref{Fig.10}) when compared to the global result\cite{de2017status}. However, none of the papers include the updated results of atmospheric data from IceCube\cite{aartsen2018measurement} and NO$\nu$A as announced in the January 2018 JETP talk "Latest Oscillation Results from NOvA" by A. Radovic\cite{radovic2018latest}.\\
	The \textquotedblleft well-measured" four oscillation parameters are $\Delta m_{21}^{2}$, $\lvert\Delta m_{3\textit{n}}^{2}\rvert$, $\theta_{12}$ and $\theta_{13}$, those are measured to an appreciated accuracy upto a few \% level;  $\theta_{13}$ is measured to 6\% level, the $\theta_{23}$ is affected by an octant ambiguity and measured less accurately at the level of $9.6\%$. \\The remaining unknowns are the sign of  $\Delta m_{3\textit{n}}^{2}$ and the CP violating phase $\delta$. The values of the above parameters guide us to the Mass Hierarchy problem and also set the lower bounds for the absolute neutrino masses. However, the upper bounds can be set only by non-oscillation experiments. The current upper bound for the sum of the mass states with the  cosmological results in combination with the oscillation results is $\sum_{i=1}^{3}m_{i}=0.17$eV at 95\% C.L. with 0.01eV of systematic uncertainities\cite{couchot2017cosmological}.\\Long base-line neutrino oscillation data play an important role in determining  the mass ordering and CP violating phase $\delta$. The current global sensitivity to the CP phase value is dominated by the T2K experiment(Figure \ref{Fig.11}), rejecting the $\delta=\frac{\pi}{2}$ after combining with the other experiments. The results will be more prominent with the operation of DUNE and LBNO which will study matter effects through earth more precisely.
\begin{figure}[h]
	\centering
	\includegraphics[width=4.8in]{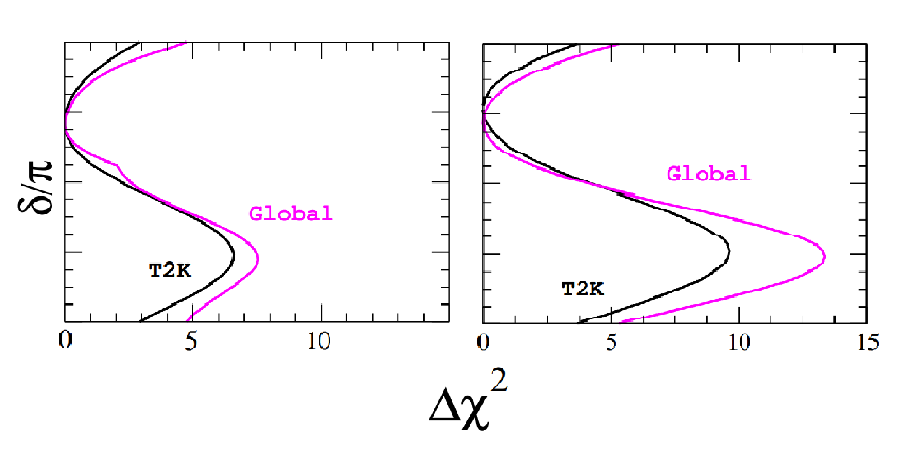}
	\caption{$\Delta\chi^{2}$ observation as a function of $\delta_{CP}$ from T2K and the global fit\cite{de2017status}. The left (right) ones correspond to the normal (inverted) mass hierarchy.}
	\label{Fig.11}
\end{figure}
\begin{figure}[t]
	\centering
	\includegraphics[width=4.2in]{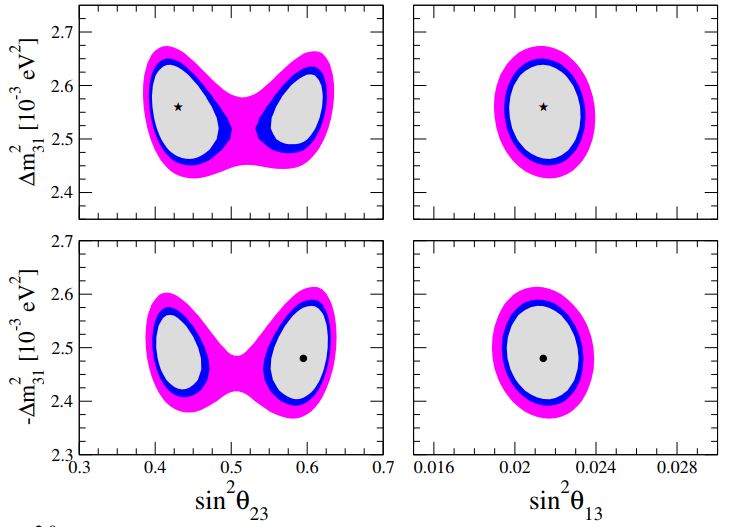}
	\caption{90 and 99\% C.L. allowed regions at the $\Delta m_{31}^{2}-sin^{2}_{\theta_{23}}$ and $\Delta m_{31}^{2}-sin^{2}_{\theta_{13}}$ planes from the global fit of all the oscillation experiments\cite{de2017status}.Upper: normal mass hierarchy. Lower: inverted mass hierarchy. The star represents the best fit value and the black dot shows the local minimum obtained for inverted ordering.}
	\label{Fig.12}
\end{figure}
\begin{figure}
	\includegraphics[width=4.2in]{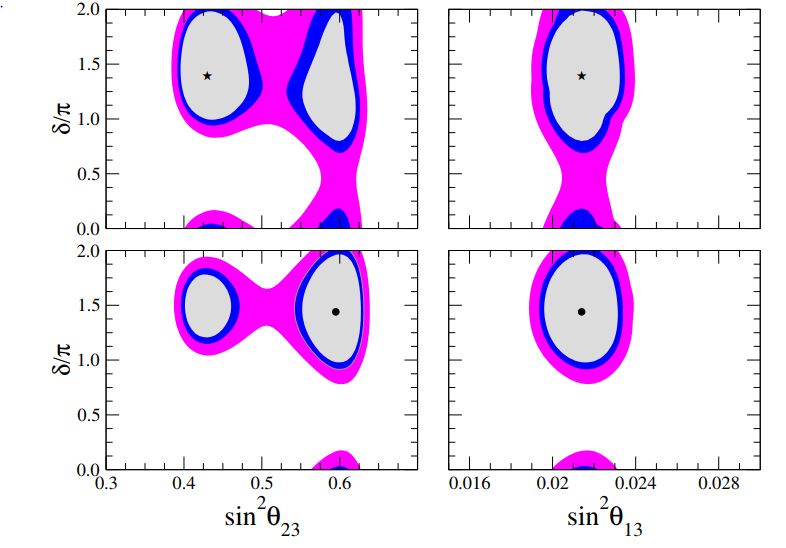}
	\caption{90, 95 and 99\% C.L. allowed regions at the $\delta_{CP}-sin^{2}_{\theta_{23}}$ and $\delta_{CP}-sin^{2}_{\theta_{13}}$ planes from the global fit\cite{de2017status} of all the oscillation experiments. Upper: normal mass hierarchy. Lower: inverted mass hierarchy.  The star and the black dot have the same meaning as the above graphs.}
	\label{Fig.13}
\end{figure}
For NH, $\delta=\frac{\pi}{2}$ is disfavoured at $2.7\sigma$ and even stronger for IH at $3.7\sigma$ w.r.t. the global minimum in the IO and $4\sigma$ w.r.t. the absolute minimum in the NO. It is observed that the current preferred value is $\delta\sim\frac{3\pi}{2}$ for both NO \& IO\cite{de2017status}.\\
The long-baseline data alone indicates non-maximal mixing for $\theta_{23}$ ($<45^{o}$, in the first atmospheric octant). After combining with the reactor data, slight deviation of $\theta_{23}$ to the second octant is favoured in case of IH(Figures \ref{Fig.12} and \ref{Fig.13}). The new results from IceCube and NO$\nu$A may affect the global analysis considerably. The global 3$\nu$ oscillation analyses prefer for NO to $\sim2\sigma$ sensitivity supported by atmospheric data (excluding IceCube DeepCore and ANTARES data), accelarator data and constraints from reactor data.
\subsection{LIMITS ON ABSOLUTE NEUTRINO MASSES}
The absolute neutrino masses is still unknown within the statndard three-neutrino framework. We shall discuss the implications of the oscillation results\cite{capozzi2017global} on the three observables sensitive to the absolute mass scale: \textit{(i)} the effective $\nu_{e}$ mass $m_{\beta}$ probed by beta decay; \textit{(ii)} the sum of $\nu$ masses $\Sigma$ in cosmology, and \textit{(iii)} the effective neutrino mass $m_{\beta\beta}$ in neutrinoless double beta decay ($0\nu\beta\beta$), if neutrinos are Majorana fermions.\\The lower bounds on the absolute neutrino mass can be set by considering the oscillation data of the Global Fit 2(Table \ref{T1})\cite{capozzi2017global}, assuming the lightest $m_{i}$ to be zero. Hence, from equation 2, we get
\begin{equation}
(m_{1},m_{2},m_{3})\geq\left\{
	\begin{array}{lr}
	\bigg( 0,\sqrt{\Delta m^{2}}_{21},\sqrt{|\Delta m^{2}_{31}|+\frac{\Delta m^{2}_{21}}{2}} \bigg) & \text{(NH)},\\\\
	\bigg(\sqrt{|\Delta m^{2}_{32}|-\frac{\Delta m^{2}_{21}}{2}},\sqrt{|\Delta m^{2}_{32}|+\frac{\Delta m^{2}_{21}}{2}},0\bigg)	& \text{(IH)}.
	\end{array}\right.
\end{equation}
whereas the upper bounds can be set by non-oscillaion neutrino experiments like NEMO\cite{pahlka2012nemo}, SuperNEMO\cite{guzowski2015construction}, KATRIN\cite{robertson2013katrin}, KamLAND-Zen\cite{shirai2013kamland} and other $\beta\beta$ and $0\nu\beta\beta$-decay experiments.\\
The three observables can be expressed\cite{capozzi2017global,fogli2006global} as follows:
\paragraph{} The observable effective electron neutrino mass $m_{\beta}$ from $\beta$ spectrum is the square root of the combination of the squared mass states with their respective $\nu_{e}$ admixture, given by
\begin{equation}
m_{\beta}=\sqrt{\sum_{i}|U_{ei}|^{2}m_{i}^{2}}
\end{equation}
\paragraph{} The $0\nu\beta\beta$ decay observable is a linear combination of the $m_{i}$'s associated with the unknown majorana phases $\phi_{i}$, where the Dirac phase $\phi_{1}=0$ by convention, given by
\begin{equation}
m_{\beta\beta}=\Big| \sum_{i}|U_{ei}|^{2}m_{i}e^{i\phi_{i}}\Big|
\end{equation}
\paragraph{} The precision cosmology is sensitive to the sum of the neutrino mass states, given by:
\begin{equation}
\sum_{i=1}^{3}m_{i}=m_{1}+m_{2}+m_{3}
\end{equation}
Taking in consideration the oscillation data in the Global Fit 2(Table \ref{T1}) and substituting in equation 8, the lower bounds of the mass states are obtained as:
\begin{equation}
(m_{1},m_{2},m_{3})\geq\left\{
\begin{array}{lr}
	(0,0.86,5.06) & \text{(NH)},\\
	(4.97,5.04,0) & \text{(IH)}.
\end{array}\right.
\end{equation}
\section{SUMMARY}
We described different oscillation experiments from past, present to future and mentioned the importance of upgrading one to the another, for instance Chooz to Double Chooz, K2K to T2K, then to T2HK; Super-K to Hyper-K, MINOS to NO$\nu$A, Daya Bay to JUNO, NEMO to SuperNEMO, SNO to SNO+, RENO to RENO-50 and IceCube to PINGU. The next decades will be focussed on solving the Mass Hierarchy problem and the CP violation phase, particularly in the Long Baseline Neutrino Experiments (LBNEs). The updated $\theta_{23}$ results from IceCube and NO$\nu$A shows a near-maximal value and its results from the other laboratories will affect significantly. The solar oscillation parameters are well established and non-zero value of $\theta_{13}$ angle upto a precision of $\sim$6\% level is dominated by Daya Bay. The solar and atmospheric mass-square differences $\Delta m^{2}_{21}$ and $\Delta m^{2}_{31}$ will be improved by JUNO in the near future of precision better than 1\%. T2K reports its first result for CP violation search in neutrino oscillations after 6 years of run and rejects the CP conserving hypothesis for $\delta_{CP}=0,\pi$ at a significance of 90\%. The collaboration of T2K and NO$\nu$A for joint analysis of netrino oscillation data by 2021 will give an upperhand picture in the determination of Dirac CP violating phase of the active three-neutrino framework and the mass hierarchy of the three neutrino weak flavour states.
\section*{Disclosure}
We have neither submitted our paper to any other journals nor delivered in any conference.
\section*{Conflict of Interest}
The authors declare that there are no conflicts of interest regarding the publication of this paper.
\begin{acknowledgments}
We acknowledge DST-SERB, Govt. of India for the research project entitled \textbf{\textquotedblleft Neutrino mass ordering, leptonic CP violation and matter-antimatter asymmetry"} vide grant no. \textbf{EMR/2015/001683}. We also thank the authors of the papers from whose various references have been cited.
\end{acknowledgments}
\appendix
\bibliographystyle{unsrt}
\bibliography{apssamp}% Produces the bibliography via BibTeX.

\begin{thebibliography}{100}

\bibitem{reines1996neutrino}
Frederick Reines.
\newblock The neutrino: from poltergeist to particle.
\newblock {\em Reviews of Modern Physics}, 68(2):317, 1996.

\bibitem{cowan1956test}
CL~Cowan, FB~Harrison, LM~Langer, and F~Reines.
\newblock A test of neutrino-antineutrino identity.
\newblock {\em Il Nuovo Cimento (1955-1965)}, 3(3):649--651, 1956.

\bibitem{pauli1930letter}
Wolfgang Pauli.
\newblock Letter to l.
\newblock In {\em Meitner and her colleagues (Open letter to the participants
  of the conference at Tubingen)}, 1930.

\bibitem{danby1962observation}
Gaillard Danby, JM~Gaillard, Konstantin Goulianos, LM~Lederman, N~Mistry,
  M~Schwartz, and J~Steinberger.
\newblock Observation of high-energy neutrino reactions and the existence of
  two kinds of neutrinos.
\newblock {\em Physical Review Letters}, 9(1):36, 1962.

\bibitem{kodama2001observation}
K~Kodama, N~Ushida, C~Andreopoulos, N~Saoulidou, G~Tzanakos, P~Yager, B~Baller,
  D~Boehnlein, W~Freeman, B~Lundberg, et~al.
\newblock Observation of tau neutrino interactions.
\newblock {\em Physics Letters B}, 504(3):218--224, 2001.

\bibitem{maki1962remarks}
Ziro Maki, Masami Nakagawa, and Shoichi Sakata.
\newblock Remarks on the unified model of elementary particles.
\newblock {\em Progress of Theoretical Physics}, 28(5):870--880, 1962.

\bibitem{nakamura2010neutrino}
K~Nakamura and ST~Petcov.
\newblock Neutrino mass, mixing, and oscillations.
\newblock {\em K. Nakamura et al.(Particle Data Group), J. Phys. G}, 37:075021,
  2010.

\bibitem{fogli2002supernova}
Gian~Luigi Fogli, E~Lisi, D~Montanino, and A~Palazzo.
\newblock Supernova neutrino oscillations: A simple analytical approach.
\newblock {\em Physical Review D}, 65(7):073008, 2002.

\bibitem{diwan2016long}
MILIND~V Diwan, V~Galymov, X~Qian, and Andr{\'e} Rubbia.
\newblock Long-baseline neutrino experiments.
\newblock {\em Annual Review of Nuclear and Particle Science}, 66, 2016.

\bibitem{cleveland1998measurement}
Bruce~T Cleveland, Timothy Daily, Raymond Davis~Jr, James~R Distel, Kenneth
  Lande, CK~Lee, Paul~S Wildenhain, and Jack Ullman.
\newblock Measurement of the solar electron neutrino flux with the homestake
  chlorine detector.
\newblock {\em The Astrophysical Journal}, 496(1):505, 1998.

\bibitem{davis1994review}
Raymond Davis.
\newblock A review of the homestake solar neutrino experiment.
\newblock {\em Progress in Particle and Nuclear Physics}, 32:13--32, 1994.

\bibitem{hampel1999gallex}
Wolfgang Hampel, J~Handt, G~Heusser, J~Kiko, T~Kirsten, M~Laubenstein,
  E~Pernicka, W~Rau, M~Wojcik, Yu~Zakharov, et~al.
\newblock Gallex solar neutrino observations: Results for gallex iv.
\newblock {\em Physics Letters B}, 447(1-2):127--133, 1999.

\bibitem{abdurashitov2009measurement}
JN~Abdurashitov, VN~Gavrin, VV~Gorbachev, PP~Gurkina, TV~Ibragimova,
  AV~Kalikhov, NG~Khairnasov, TV~Knodel, IN~Mirmov, AA~Shikhin, et~al.
\newblock Measurement of the solar neutrino capture rate with gallium metal.
  iii. results for the 2002--2007 data-taking period.
\newblock {\em Physical Review C}, 80(1):015807, 2009.

\bibitem{castellani1997solar}
Vittorio Castellani, S~Degl'Innocenti, G~Fiorentini, M~Lissia, and B~Ricci.
\newblock Solar neutrinos: Beyond standard solar models.
\newblock {\em Physics Reports}, 281(5-6):309--398, 1997.

\bibitem{ranucci2016techniques}
Gioacchino Ranucci.
\newblock Techniques and methods for the low-energy neutrino detection.
\newblock {\em The European Physical Journal A}, 52(4):79, 2016.

\bibitem{jelley1955cerenkov}
JV~Jelley.
\newblock Cerenkov radiation and its applications.
\newblock {\em British Journal of Applied Physics}, 6(7):227, 1955.

\bibitem{watson2011discovery}
Alan~A Watson.
\newblock The discovery of cherenkov radiation and its use in the detection of
  extensive air showers.
\newblock {\em Nuclear Physics B-Proceedings Supplements}, 212:13--19, 2011.

\bibitem{abe2013evidence}
K~Abe, Y~Hayato, T~Iida, K~Iyogi, J~Kameda, Y~Koshio, Y~Kozuma, Ll~Marti,
  M~Miura, S~Moriyama, et~al.
\newblock Evidence for the appearance of atmospheric tau neutrinos in
  super-kamiokande.
\newblock {\em Physical review letters}, 110(18):181802, 2013.

\bibitem{adrian2016prototype}
S~Adri{\'a}n-Mart{\'\i}nez, M~Ageron, F~Aharonian, S~Aiello, A~Albert, F~Ameli,
  EG~Anassontzis, GC~Androulakis, M~Anghinolfi, G~Anton, et~al.
\newblock The prototype detection unit of the km3net detector.
\newblock {\em The European Physical Journal C}, 76(2):54, 2016.

\bibitem{aguilarpreprint}
AA~Aguilar-Arevaloe, CE~Andersonp, LM~Bartoszekg, AO~Bazarkom, SJ~Briceg,
  BC~Browng, L~Bugele, J~Caod, L~Coneye, JM~Conrade, et~al.
\newblock The miniboone detector.
\newblock {\em FERMILAB-PUB-08-210, Preprint submitted to Elsevier}, 25 June,
  2008.

\bibitem{avrorin2011gigaton}
A~Avrorin, V~Aynutdinov, I~Belolaptikov, S~Berezhnev, D~Bogorodsky, N~Budnev,
  I~Danilchenko, G~Domogatsky, A~Doroshenko, A~Dyachok, et~al.
\newblock The gigaton volume detector in lake baikal.
\newblock {\em Nuclear Instruments and Methods in Physics Research Section A:
  Accelerators, Spectrometers, Detectors and Associated Equipment},
  639(1):30--32, 2011.

\bibitem{bellerive2016sudbury}
Alain Bellerive, JR~Klein, AB~McDonald, AJ~Noble, AWP Poon, SNO Collaboration,
  et~al.
\newblock The sudbury neutrino observatory.
\newblock {\em Nuclear Physics B}, 908:30--51, 2016.

\bibitem{icecube2001icecube}
IceCube Collaboration et~al.
\newblock Icecube preliminary design document revision 1.24, 2001.

\bibitem{fukuda2003super}
S~Fukuda, Y~Fukuda, T~Hayakawa, E~Ichihara, M~Ishitsuka, Y~Itow, T~Kajita,
  J~Kameda, K~Kaneyuki, S~Kasuga, et~al.
\newblock The super-kamiokande detector.
\newblock {\em Nuclear Instruments and Methods in Physics Research Section A:
  Accelerators, Spectrometers, Detectors and Associated Equipment},
  501(2-3):418--462, 2003.

\bibitem{haranczyk2017icarus}
M~Hara{\'n}czyk.
\newblock The icarus detector: Past, present and future.
\newblock In {\em Journal of Physics: Conference Series}, volume 798, page
  012162. IOP Publishing, 2017.

\bibitem{kindin2015measuring}
VV~Kindin, MB~Amelchakov, NS~Barbashina, AG~Bogdanov, VD~Burtsev, DV~Chernov,
  SS~Khokhlov, VA~Khomyakov, RP~Kokoulin, KG~Kompaniets, et~al.
\newblock Measuring module of the cherenkov water detector nevod.
\newblock In {\em Journal of Physics: Conference Series}, volume 632, page
  012015. IOP Publishing, 2015.

\bibitem{wilkes2005uno}
R~Jeffrey Wilkes.
\newblock Uno.
\newblock {\em arXiv preprint hep-ex/0507097}, 2005.

\bibitem{brooks1979development}
FD~Brooks.
\newblock Development of organic scintillators.
\newblock {\em Nuclear Instruments and Methods}, 162(1-3):477--505, 1979.

\bibitem{birks2013theory}
John~Betteley Birks.
\newblock {\em The Theory and Practice of Scintillation Counting: International
  Series of Monographs in Electronics and Instrumentation}, volume~27.
\newblock Elsevier, 2013.

\bibitem{hamamatsu2007basics}
Photomultiplier~Tubes Hamamatsu and Photomultupliers~Tubes Photonics.
\newblock Basics and applications.
\newblock {\em Hamamatsu Photonics KK, Iwata City}, 2007.

\bibitem{alimonti2009borexino}
G~Alimonti, C~Arpesella, H~Back, M~Balata, D~Bartolomei, A~De~Bellefon,
  G~Bellini, J~Benziger, A~Bevilacqua, D~Bondi, et~al.
\newblock The borexino detector at the laboratori nazionali del gran sasso
  (lngs).
\newblock {\em Nuclear Instruments and Methods in Physics Research Section A:
  Accelerators, Spectrometers, Detectors and Associated Equipment},
  600(3):568--593, 2009.

\bibitem{mckinsey2005neutrino}
Daniel~N McKinsey and KJ~Coakley.
\newblock Neutrino detection with clean.
\newblock {\em Astroparticle Physics}, 22(5-6):355--368, 2005.

\bibitem{an2012side}
FP~An, Q~An, JZ~Bai, AB~Balantekin, HR~Band, W~Beriguete, M~Bishai, S~Blyth,
  RL~Brown, GF~Cao, et~al.
\newblock A side-by-side comparison of daya bay antineutrino detectors.
\newblock {\em Nuclear Instruments and Methods in Physics Research Section A:
  Accelerators, Spectrometers, Detectors and Associated Equipment}, 685:78--97,
  2012.

\bibitem{greiner2007double}
Daniel Greiner, Tobias Lachenmaier, Josef Jochum, and Anatael Cabrera.
\newblock Double chooz detectors design.
\newblock {\em Nuclear Instruments and Methods in Physics Research Section A:
  Accelerators, Spectrometers, Detectors and Associated Equipment},
  581(1-2):139--142, 2007.

\bibitem{eguchi2003first}
K~Eguchi, S~Enomoto, K~Furuno, J~Goldman, H~Hanada, H~Ikeda, K~Ikeda, K~Inoue,
  K~Ishihara, W~Itoh, et~al.
\newblock First results from kamland: evidence for reactor antineutrino
  disappearance.
\newblock {\em Physical Review Letters}, 90(2):021802, 2003.

\bibitem{barabanov1999rare}
IR~Barabanov, VI~Beresnev, VN~Kornoukhov, EA~Yanovich, GT~Zatsepin, NA~Danilov,
  GV~Korpusov, GV~Kostikova, YS~Krylov, and VV~Yakshin.
\newblock Rare-earth loaded liquid scintillator (for lens experiment).
\newblock {\em arXiv preprint physics/9908005}, 1999.

\bibitem{osmanov2011minerva}
Bari Osmanov.
\newblock Minerva detector: description and performance.
\newblock {\em arXiv preprint arXiv:1109.2855}, 2011.

\bibitem{evans2013minos}
JJ~Evans.
\newblock The minos experiment: results and prospects.
\newblock {\em Advances in High Energy Physics}, 2013, 2013.

\bibitem{ahn2010reno}
JK~Ahn, Reno Collaboration, et~al.
\newblock Reno: An experiment for neutrino oscillation parameter $\theta_{13}$
  using reactor neutrinos at yonggwang.
\newblock {\em arXiv:1003.1391}, 2010.

\bibitem{takei2009scibar}
H~Takei.
\newblock Scibar detector for sciboone.
\newblock In {\em Journal of Physics: Conference Series}, volume 160, page
  012034. IOP Publishing, 2009.

\bibitem{walding2007muon}
Joseph Walding.
\newblock The muon range detector at sciboone.
\newblock In {\em AIP Conference Proceedings}, volume 967, pages 289--291. AIP,
  2007.

\bibitem{kamdin2015understanding}
K~Kamdin.
\newblock Understanding the sno+ detector.
\newblock {\em Physics Procedia}, 61:719--723, 2015.

\bibitem{labare2017solid}
Mathieu Labare.
\newblock Solid detector technology.
\newblock In {\em Journal of Physics: Conference Series}, volume 888, page
  012180. IOP Publishing, 2017.

\bibitem{buck2017scintillation}
Christian Buck, Manfred Lindner, and Christian Roca.
\newblock Scintillation light production, propagation and detection in the
  stereo reactor antineutrino experiment.
\newblock In {\em Journal of Physics: Conference Series}, volume 888, page
  012101. IOP Publishing, 2017.

\bibitem{he2015jiangmen}
Miao He, JUNO collaboration, et~al.
\newblock Jiangmen underground neutrino observatory.
\newblock {\em Nuclear and Particle Physics Proceedings}, 265:111--113, 2015.

\bibitem{bolton2005braidwood}
T~Bolton.
\newblock The braidwood reactor anitneutrino experiment.
\newblock {\em Nuclear Physics B-Proceedings Supplements}, 149:166--169, 2005.

\bibitem{kuze2005kaska}
Masahiro Kuze, Kaska Collaboration, et~al.
\newblock The kaska project--a japanese medium-baseline reactor-neutrino
  oscillation experiment to measure the mixing angle $\theta_{13}$.
\newblock {\em Nuclear Physics B-Proceedings Supplements}, 149:160--162, 2005.

\bibitem{dornelas2016front}
TI~Dornelas, FTH Ara{\'u}jo, AS~Cerqueira, JA~Costa, and RA~N{\'o}brega.
\newblock Front-end design and characterization for the $\nu$-angra nuclear
  reactor monitoring detector.
\newblock {\em Journal of Instrumentation}, 11(07):P07018, 2016.

\bibitem{kumar2017invited}
A~Kumar, AM~Vinod Kumar, Abhik Jash, Ajit~K Mohanty, Aleena Chacko, Ali Ajmi,
  Ambar Ghosal, Amina Khatun, Amitava Raychaudhuri, Amol Dighe, et~al.
\newblock Invited review: Physics potential of the ical detector at the
  india-based neutrino observatory (ino).
\newblock {\em Pramana}, 88(5):79, 2017.

\bibitem{nygren1974proposal}
David~R Nygren.
\newblock Proposal to investigate the feasibility of a novel concept in
  particle detection.
\newblock 1974.

\bibitem{rubbia1977liquid}
Carlo Rubbia.
\newblock The liquid argon time projection chamber: a new concept for neutrino
  detectors.
\newblock Technical report, 1977.

\bibitem{miyajima1974average}
M~Miyajima, T~Takahashi, S~Konno, T~Hamada, S~Kubota, H~Shibamura, and T~Doke.
\newblock Average energy expended per ion pair in liquid argon.
\newblock {\em Physical Review A}, 9(3):1438, 1974.

\bibitem{hofmann1976production}
W~Hofmann, U~Klein, M~Schulz, J~Spengler, and D~Wegener.
\newblock Production and transport of conduction electrons in a liquid argon
  ionization chamber.
\newblock {\em Nuclear Instruments and Methods}, 135(1):151--156, 1976.

\bibitem{acciarri2016long}
R~Acciarri, MA~Acero, M~Adamowski, C~Adams, P~Adamson, S~Adhikari, Z~Ahmad,
  CH~Albright, T~Alion, E~Amador, et~al.
\newblock Long-baseline neutrino facility (lbnf) and deep underground neutrino
  experiment (dune) conceptual design report, volume 4 the dune detectors at
  (lbnf).
\newblock {\em arXiv:1601.02984}, 2016.

\bibitem{lbne2012long}
LBNE collaboration et~al.
\newblock The long baseline neutrino experiment (lbne) water cherenkov detector
  (wcd) conceptual design report (cdr).
\newblock {\em arXiv:1204.2295}, 2012.

\bibitem{jelley2009sudbury}
Nick Jelley, Arthur~B McDonald, and RG~Hamish Robertson.
\newblock The sudbury neutrino observatory.
\newblock {\em Annual Review of Nuclear and Particle Science}, 59:431--465,
  2009.

\bibitem{mcdonald2016sudbury}
Arthur~B McDonald.
\newblock The sudbury neutrino observatory: Observation of flavor change for
  solar neutrinos.
\newblock {\em Annalen der Physik}, 528(6):469--480, 2016.

\bibitem{de2001experimentalist}
Antonella De~Santo.
\newblock An experimentalist's view of neutrino oscillations.
\newblock {\em International Journal of Modern Physics A}, 16(25):4085--4151,
  2001.

\bibitem{suzuki2009history}
Atsuto Suzuki and Masatoshi Koshiba.
\newblock History of neutrino telescope/astronomy.
\newblock {\em Experimental Astronomy}, 25(1-3):209--224, 2009.

\bibitem{adamson2008measurement}
P~Adamson, C~Andreopoulos, KE~Arms, R~Armstrong, DJ~Auty, DS~Ayres, B~Baller,
  PD~Barnes~Jr, G~Barr, WL~Barrett, et~al.
\newblock Measurement of neutrino oscillations with the minos detectors in the
  numi beam.
\newblock {\em Physical Review Letters}, 101(13):131802, 2008.

\bibitem{adamson2011measurement}
P~Adamson, C~Andreopoulos, R~Armstrong, DJ~Auty, DS~Ayres, C~Backhouse, G~Barr,
  M~Bishai, A~Blake, GJ~Bock, et~al.
\newblock Measurement of the neutrino mass splitting and flavor mixing by
  minos.
\newblock {\em Physical Review Letters}, 106(18):181801, 2011.

\bibitem{sousa2015first}
Alexandre~B Sousa, MINOS, and MINOS+ Collaborations).
\newblock First minos+ data and new results from minos.
\newblock In {\em AIP Conference Proceedings}, volume 1666, page 110004. AIP
  Publishing, 2015.

\bibitem{whitehead2016neutrino}
Leigh~H Whitehead.
\newblock Neutrino oscillations with minos and minos+.
\newblock {\em Nuclear Physics B}, 908:130--150, 2016.

\bibitem{backhouse2015results}
Christopher Backhouse.
\newblock Results from minos and no$\nu$a.
\newblock In {\em Journal of Physics: Conference Series}, volume 598, page
  012004. IOP Publishing, 2015.

\bibitem{tzanankos2011minos+}
G~Tzanankos, A~Weber, K~Lang, CO~Escobar, J~Evans, E~Falk, SG~Wojcicki,
  P~Vahle, M~Marshak, J~Nelson, et~al.
\newblock Minos+: a proposal to fnal to run minos with the medium energy numi
  beam.
\newblock Technical report, 2011.

\bibitem{holin2015results}
Anna Holin, MINOS+ Collaborations, et~al.
\newblock Results from the minos experiment and new minos+ data.
\newblock {\em arXiv:1507.08564}, 2015.

\bibitem{apollonio1998initial}
M~Apollonio, A~Baldini, C~Bemporad, E~Caffau, F~Cei, Y~Declais, H~De~Kerret,
  B~Dieterle, A~Etenko, J~George, et~al.
\newblock Initial results from the chooz long baseline reactor neutrino
  oscillation experiment.
\newblock {\em Physics Letters B}, 420(3-4):397--404, 1998.

\bibitem{cao2013detection}
Jun Cao.
\newblock Detection methods at reactor neutrino experiments.
\newblock {\em Nuclear Instruments and Methods in Physics Research Section A:
  Accelerators, Spectrometers, Detectors and Associated Equipment}, 732:9--15,
  2013.

\bibitem{mikaelyan2000chooz}
L~Mikaelyan.
\newblock Chooz, palo verde, krasnoyarsk.
\newblock {\em Nuclear Physics B-Proceedings Supplements}, 87(1-3):284--287,
  2000.

\bibitem{apollonio2003search}
M~Apollonio, A~Baldini, C~Bemporad, E~Caffau, F~Cei, Y~Declais, H~De~Kerret,
  B~Dieterle, A~Etenko, L~Foresti, et~al.
\newblock Search for neutrino oscillations on a long base-line at the chooz
  nuclear power station.
\newblock {\em The European Physical Journal C-Particles and Fields},
  27(3):331--374, 2003.

\bibitem{pahlka2012nemo}
RB~Pahlka, NEMO-3 Collaboration, et~al.
\newblock The nemo-3 experiment.
\newblock {\em Nuclear Physics B-Proceedings Supplements}, 229:491, 2012.

\bibitem{arnold2016measurement}
R~Arnold, C~Augier, AM~Bakalyarov, JD~Baker, AS~Barabash,
  A~Basharina-Freshville, S~Blondel, S~Blot, M~Bongrand, V~Brudanin, et~al.
\newblock Measurement of the double-beta decay half-life and search for the
  neutrinoless double-beta decay of ca-48 with the nemo-3 detector.
\newblock {\em Physical Review D}, 93(11):112008, 2016.

\bibitem{saakyan2013two}
Ruben Saakyan.
\newblock Two-neutrino double-beta decay.
\newblock {\em Annual Review of Nuclear and Particle Science}, 63:503--529,
  2013.

\bibitem{gomez2016latest}
H~G{\'o}mez et~al.
\newblock Latest results of nemo-3 experiment and present status of supernemo.
\newblock {\em Nuclear and Particle Physics Proceedings}, 273:1765--1770, 2016.

\bibitem{waters2017latest}
David Waters et~al.
\newblock Latest results from nemo-3 \& status of the supernemo experiment.
\newblock In {\em Journal of Physics: Conference Series}, volume 888, page
  012033. IOP Publishing, 2017.

\bibitem{rubbia2011underground}
C~Rubbia, M~Antonello, P~Aprili, B~Baibussinov, M~Baldo Ceolin, L~Barze,
  P~Benetti, E~Calligarich, N~Canci, F~Carbonara, et~al.
\newblock Underground operation of the $icarus$ $t600$ $lar-tpc$: first
  results.
\newblock {\em Journal of Instrumentation}, 6(07):P07011, 2011.

\bibitem{menegolli2016some}
A~Menegolli.
\newblock Some recent results from the icarus experiment.
\newblock {\em Nucl. Part. Phys. Proc.}, 273:1891--1896, 2016.

\bibitem{ianni2011neutrino}
A~Ianni.
\newblock Neutrino physics with borexino.
\newblock {\em Progress in Particle and Nuclear Physics}, 66(2):405--411, 2011.

\bibitem{arpesella2008first}
C~Arpesella, G~Bellini, J~Benziger, S~Bonetti, B~Caccianiga, F~Calaprice,
  F~Dalnoki-Veress, D~D'Angelo, H~de~Kerret, A~Derbin, et~al.
\newblock First real time detection of 7be solar neutrinos by borexino.
\newblock {\em Physics Letters B}, 658(4):101--108, 2008.

\bibitem{bellerive2004review}
Alain Bellerive.
\newblock Review of solar neutrino experiments.
\newblock {\em International Journal of Modern Physics A}, 19(08):1167--1179,
  2004.

\bibitem{bellini2011precision}
Gianpaolo Bellini, J~Benziger, D~Bick, S~Bonetti, G~Bonfini, M~Buizza Avanzini,
  B~Caccianiga, L~Cadonati, F~Calaprice, C~Carraro, et~al.
\newblock Precision measurement of the $^{7}be$ solar neutrino interaction rate
  in borexino.
\newblock {\em Physical Review Letters}, 107(14):141302, 2011.

\bibitem{bellini2012first}
G~Bellini, J~Benziger, D~Bick, S~Bonetti, G~Bonfini, D~Bravo, M~Buizza
  Avanzini, B~Caccianiga, L~Cadonati, F~Calaprice, et~al.
\newblock First evidence of $pep$ solar neutrinos by direct detection in
  borexino.
\newblock {\em Physical Review Letters}, 108(5):051302, 2012.

\bibitem{bellini2012absence}
Gianpaolo Bellini, J~Benziger, D~Bick, S~Bonetti, G~Bonfini, M~Buizza Avanzini,
  B~Caccianiga, L~Cadonati, F~Calaprice, C~Carraro, et~al.
\newblock Absence of a day-night asymmetry in the $^{7}be$ solar neutrino rate
  in borexino.
\newblock {\em Physics Letters B}, 707(1):22--26, 2012.

\bibitem{bellini2014neutrinos}
G~Bellini, J~Benziger, D~Bick, G~Bonfini, D~Bravo, B~Caccianiga, L~Cadonati,
  F~Calaprice, A~Caminata, P~Cavalcante, et~al.
\newblock Neutrinos from the primary proton--proton fusion process in the sun.
\newblock {\em Nature}, 512(7515):383, 2014.

\bibitem{bellini2016impact}
Gianpaolo Bellini.
\newblock The impact of borexino on the solar and neutrino physics.
\newblock {\em Nuclear Physics B}, 908:178--198, 2016.

\bibitem{palomares2011double}
Carmen Palomares, Double~CHOOZ Collaboration, et~al.
\newblock Double-chooz neutrino experiment.
\newblock In {\em Journal of Physics: Conference Series}, volume 335, page
  012055. IOP Publishing, 2011.

\bibitem{carr2016new}
R~Carr, S~Lutch, and P~Novella.
\newblock New results from the double chooz experiment.
\newblock {\em Nuclear and particle physics proceedings}, 273:2648--2650, 2016.

\bibitem{abe2012indication}
Y~Abe, Christoph Aberle, T~Akiri, JC~Dos~Anjos, F~Ardellier, AF~Barbosa,
  A~Baxter, M~Bergevin, A~Bernstein, TJC Bezerra, et~al.
\newblock Indication of reactor $\nu$ e disappearance in the double chooz
  experiment.
\newblock {\em Physical Review Letters}, 108(13):131801, 2012.

\bibitem{abe2016solar}
K~Abe, Y~Haga, Y~Hayato, M~Ikeda, K~Iyogi, J~Kameda, Y~Kishimoto, Ll~Marti,
  M~Miura, S~Moriyama, et~al.
\newblock Solar neutrino measurements in super-kamiokande-$iv$.
\newblock {\em Physical Review D}, 94(5):052010, 2016.

\bibitem{farzan2017neutrino}
Y~Farzan and M~Tortola.
\newblock Neutrino oscillations and non-standard interactions.
\newblock {\em arXiv preprint arXiv:1710.09360}, 2017.

\bibitem{abe2017search}
K~Abe, Y~Haga, Y~Hayato, M~Ikeda, K~Iyogi, J~Kameda, Y~Kishimoto, M~Miura,
  S~Moriyama, M~Nakahata, et~al.
\newblock Search for proton decay via $p\rightarrow e^{+}\pi^{0}$ and
  $p\rightarrow\mu^{+}\pi^{0}$ in 0.31 megaton{\textperiodcentered} years
  exposure of the super-kamiokande water cherenkov detector.
\newblock {\em Physical Review D}, 95(1):012004, 2017.

\bibitem{abe2011t2k}
K~Abe, N~Abgrall, H~Aihara, Y~Ajima, JB~Albert, D~Allan, P-A Amaudruz,
  C~Andreopoulos, B~Andrieu, MD~Anerella, et~al.
\newblock The $t2k$ experiment.
\newblock {\em Nuclear Instruments and Methods in Physics Research Section A:
  Accelerators, Spectrometers, Detectors and Associated Equipment},
  659(1):106--135, 2011.

\bibitem{abe2013measurement}
Kou Abe, J~Adam, H~Aihara, T~Akiri, C~Andreopoulos, S~Aoki, A~Ariga, T~Ariga,
  S~Assylbekov, D~Autiero, et~al.
\newblock Measurement of neutrino oscillation parameters from muon neutrino
  disappearance with an off-axis beam.
\newblock {\em Physical review letters}, 111(21):211803, 2013.

\bibitem{abe2014observation}
K~Abe, J~Adam, H~Aihara, T~Akiri, C~Andreopoulos, S~Aoki, Akitaka Ariga, Tomoko
  Ariga, S~Assylbekov, D~Autiero, et~al.
\newblock Observation of electron neutrino appearance in a muon neutrino beam.
\newblock {\em Physical review letters}, 112(6):061802, 2014.

\bibitem{abe2017combined}
K~Abe, J~Amey, C~Andreopoulos, M~Antonova, S~Aoki, A~Ariga, D~Autiero, S~Ban,
  M~Barbi, GJ~Barker, et~al.
\newblock Combined analysis of neutrino and antineutrino oscillations at t2k.
\newblock {\em Physical review letters}, 118(15):151801, 2017.

\bibitem{karle2010icecube}
Albrecht Karle et~al.
\newblock Icecube.
\newblock {\em arXiv:1003.5715}, 2010.

\bibitem{aartsen2018measurement}
MG~Aartsen, M~Ackermann, J~Adams, JA~Aguilar, M~Ahlers, M~Ahrens, I~Al~Samarai,
  D~Altmann, K~Andeen, T~Anderson, et~al.
\newblock Measurement of atmospheric neutrino oscillations at $6-56 gev$ with
  icecube deepcore.
\newblock {\em Physical Review Letters}, 120(7):071801, 2018.

\bibitem{jediny2017nova}
Filip Jediny et~al.
\newblock No$\nu$a latest results.
\newblock {\em PoS, SISSA}, page 025, 2017.

\bibitem{radovic2018latest}
A~Radovic.
\newblock Latest oscillation results from no$\nu$a.
\newblock In {\em Joint Experimental-Theoretical Physics Seminar, Fermilab,
  USA}, 2018.

\bibitem{kim2016measurement}
Soo-Bong Kim, RENO Collaboration, et~al.
\newblock Measurement of neutrino mixing angle $\theta_{13}$ and mass
  difference $\delta m_{ee}^2$ from reactor antineutrino disappearance in the
  reno experiment.
\newblock {\em Nuclear Physics B}, 908:94--115, 2016.

\bibitem{ahn2012observation}
JK~Ahn, S~Chebotaryov, JH~Choi, S~Choi, W~Choi, Y~Choi, HI~Jang, JS~Jang,
  EJ~Jeon, IS~Jeong, et~al.
\newblock Observation of reactor electron antineutrinos disappearance in the
  reno experiment.
\newblock {\em Physical Review Letters}, 108(19):191802, 2012.

\bibitem{joo2012status}
Kyung~Kwang Joo, RENO Collaboration, et~al.
\newblock Status of the reno reactor neutrino experiment.
\newblock {\em Nuclear Physics B-Proceedings Supplements}, 229:97--100, 2012.

\bibitem{seo2017talk}
H~Seo.
\newblock Talk given on behalf of the reno collaboration.
\newblock {\em EPS conference on High Energy Physics, Venice, Italy}, 511,
  July, 2017.

\bibitem{an2017improved}
Feng~Peng An, AB~Balantekin, HR~Band, M~Bishai, S~Blyth, D~Cao, GF~Cao, J~Cao,
  WR~Cen, YL~Chan, et~al.
\newblock Improved measurement of the reactor antineutrino flux and spectrum at
  daya bay.
\newblock {\em Chinese Physics C}, 41(1):013002, 2017.

\bibitem{cao2016overview}
Jun Cao and Kam-Biu Luk.
\newblock An overview of the daya bay reactor neutrino experiment.
\newblock {\em Nuclear Physics B}, 908:62--73, 2016.

\bibitem{kim2013reactor}
Soo-Bong Kim, Thierry Lasserre, and Yifang Wang.
\newblock Reactor neutrinos.
\newblock {\em Advances in High Energy Physics}, 2013.

\bibitem{an2012observation}
FP~An, JZ~Bai, AB~Balantekin, HR~Band, D~Beavis, W~Beriguete, M~Bishai,
  S~Blyth, K~Boddy, RL~Brown, et~al.
\newblock Observation of electron-antineutrino disappearance at daya bay.
\newblock {\em Physical Review Letters}, 108(17):171803, 2012.

\bibitem{an2017measurement}
Feng~Peng An, AB~Balantekin, HR~Band, M~Bishai, S~Blyth, D~Cao, GF~Cao, J~Cao,
  WR~Cen, YL~Chan, et~al.
\newblock Measurement of electron antineutrino oscillation based on 1230 days
  of operation of the daya bay experiment.
\newblock {\em Physical Review D}, 95(7):072006, 2017.

\bibitem{andringa2016current}
S~Andringa, E~Arushanova, S~Asahi, M~Askins, DJ~Auty, AR~Back, Z~Barnard,
  N~Barros, EW~Beier, A~Bialek, et~al.
\newblock Current status and future prospects of the sno.
\newblock {\em Advances in High Energy Physics}, 2016.

\bibitem{maneira2016status}
J~Maneira.
\newblock Status and prospects of the sno+ experiment.
\newblock In {\em Journal of Physics: Conference Series}, volume 718, page
  062040. IOP Publishing, 2016.

\bibitem{maneira2013sno+}
Jos{\'e} Maneira, Sno+ Collaboration, et~al.
\newblock The sno+ experiment: status and overview.
\newblock In {\em Journal of Physics: Conference Series}, volume 447, page
  012065. IOP Publishing, 2013.

\bibitem{chen2008sno+}
Mark~C Chen et~al.
\newblock The sno+ experiment.
\newblock {\em arXiv:0810.3694}, 2008.

\bibitem{abe2011letter}
K~Abe, T~Abe, H~Aihara, Y~Fukuda, Y~Hayato, K~Huang, AK~Ichikawa, M~Ikeda,
  K~Inoue, H~Ishino, et~al.
\newblock Letter of intent: The hyper-kamiokande experiment-- detector design
  and physics potential.
\newblock {\em arXiv:1109.3262}, 2011.

\bibitem{cremonesi2015sensitivity}
Linda Cremonesi.
\newblock Sensitivity to the neutrino oscillation parameters in the
  hyper-kamiokande experiment.
\newblock In {\em Journal of Physics: Conference Series}, volume 598, page
  012017. IOP Publishing, 2015.

\bibitem{di2017hyper}
Francesca Di~Lodovico, Hyper-Kamiokande Collaboration, et~al.
\newblock The hyper-kamiokande experiment.
\newblock In {\em Journal of Physics: Conference Series}, volume 888, page
  012020. IOP Publishing, 2017.

\bibitem{group2014long}
Hyper-Kamiokande~Working Group, K~Abe, H~Aihara, C~Andreopoulos, I~Anghel,
  A~Ariga, T~Ariga, R~Asfandiyarov, M~Askins, JJ~Back, et~al.
\newblock A long baseline neutrino oscillation experiment using j-parc neutrino
  beam and hyper-kamiokande.
\newblock {\em arXiv:1412.4673}, 2014.

\bibitem{ballett2017sensitivities}
Peter Ballett, Stephen~F King, Silvia Pascoli, Nick~W Prouse, and TseChun Wang.
\newblock Sensitivities and synergies of dune and t2hk.
\newblock {\em Physical Review D}, 96(3):033003, 2017.

\bibitem{raut2017matter}
Sushant~K Raut.
\newblock Matter effects at the t2hk and t2hkk experiments.
\newblock {\em Physical Review D}, 96(7):075029, 2017.

\bibitem{brugiere2017jiangmen}
Timoth{\'e}e Brugi{\`e}re.
\newblock The jiangmen underground neutrino observatory experiment.
\newblock {\em Nuclear Instruments and Methods in Physics Research Section A:
  Accelerators, Spectrometers, Detectors and Associated Equipment},
  845:326--329, 2017.

\bibitem{an2016neutrino}
Fengpeng An, Guangpeng An, Qi~An, Vito Antonelli, Eric Baussan, John Beacom,
  Leonid Bezrukov, Simon Blyth, Riccardo Brugnera, Margherita~Buizza Avanzini,
  et~al.
\newblock Neutrino physics with juno.
\newblock {\em Journal of Physics G: Nuclear and Particle Physics},
  43(3):030401, 2016.

\bibitem{kim2015review}
Soo-Bong Kim.
\newblock Review of reactor neutrino experiments.
\newblock In {\em Proceedings of the 2nd International Symposium on Science at
  J-PARC-- Unlocking the Mysteries of Life, Matter and the Universe}, page
  023005, 2015.

\bibitem{djurcic2013review}
Zelimir Djurcic.
\newblock Review of reactor antineutrino experiments.
\newblock In {\em Journal of Physics: Conference Series}, volume 408, page
  012008. IOP Publishing, 2013.

\bibitem{datar2017india}
Vivek~M Datar and Naba~K Mondal.
\newblock India-based neutrino observatory.
\newblock {\em CURRENT SCIENCE}, 113(4):701--706, 2017.

\bibitem{bhattacharya2014error}
Kolahal Bhattacharya, Arnab~K Pal, Gobinda Majumder, and Naba~K Mondal.
\newblock Error propagation of the track model and track fitting strategy for
  the iron calorimeter detector in india-based neutrino observatory.
\newblock {\em Computer Physics Communications}, 185(12):3259--3268, 2014.

\bibitem{khatun2018can}
Amina Khatun, Tarak Thakore, and Sanjib~Kumar Agarwalla.
\newblock Can ino be sensitive to flavor-dependent long-range forces?
\newblock {\em arXiv:1801.00949}, 2018.

\bibitem{bhattacharya2006india}
Sudeb Bhattacharya, INO Collaboration, et~al.
\newblock India-based neutrino observatory- the present status.
\newblock {\em Progress in Particle and Nuclear Physics}, 57(1):299--308, 2006.

\bibitem{adams2013long}
Corey Adams, David Adams, Tarek Akiri, Tyler Alion, Kris Anderson, Costas
  Andreopoulos, Mike Andrews, Ioana Anghel, Jo{\~a}o Carlos Costa~dos Anjos,
  Maddalena Antonello, et~al.
\newblock The long-baseline neutrino experiment: exploring fundamental
  symmetries of the universe.
\newblock {\em arXiv:1307.7335}, 2013.

\bibitem{aartsen2017pingu}
MG~Aartsen, K~Abraham, M~Ackermann, J~Adams, JA~Aguilar, M~Ahlers, M~Ahrens,
  D~Altmann, K~Andeen, T~Anderson, et~al.
\newblock Pingu: a vision for neutrino and particle physics at the south pole.
\newblock {\em Journal of Physics G: Nuclear and Particle Physics},
  44(5):054006, 2017.

\bibitem{vilela2015supernemo}
Cristovao Vilela, NEMO collaboration, et~al.
\newblock The supernemo neutrinoless double beta decay experiment.
\newblock In {\em Journal of Physics: Conference Series}, volume 598, page
  012034. IOP Publishing, 2015.

\bibitem{guzowski2015construction}
P~Guzowski.
\newblock Construction of the tracker for the supernemo experiment.
\newblock In {\em Journal of Physics: Conference Series}, volume 598, page
  012020. IOP Publishing, 2015.

\bibitem{zaborov2018km3net}
Dmitry Zaborov et~al.
\newblock The km3net neutrino telescope and the potential of a neutrino beam
  from russia to the mediterranean sea.
\newblock {\em arXiv:1803.08017}, 2018.

\bibitem{collaboration2016km3net}
KM3NeT Collaboration.
\newblock Km3net 2.0 letter of intent for arca and orca.
\newblock {\em J. Phys. G: Nucl. Part. Phys}, 43:084001, 2016.

\bibitem{bruijn2016km3net}
R~Bruijn and Daan van Eijk.
\newblock The km3net multi-pmt digital optical module.
\newblock In {\em The 34th International Cosmic Ray Conference}, volume 236,
  page 1157. SISSA Medialab, 2016.

\bibitem{katz2014orca}
Ulrich~F Katz.
\newblock The orca option for km3net.
\newblock {\em arXiv:1402.1022}, 2014.

\bibitem{kouchner2016km3net}
Antoine Kouchner.
\newblock Km3net-orca: measuring the neutrino mass ordering in the
  mediterranean.
\newblock In {\em Journal of Physics: Conference Series}, volume 718, page
  062030. IOP Publishing, 2016.

\bibitem{wolfenstein1978neutrino}
Lincoln Wolfenstein.
\newblock Neutrino oscillations in matter.
\newblock {\em Physical Review D}, 17(9):2369, 1978.

\bibitem{mikheyev1985sp}
SP~Mikheyev and A.~Yu. Smirnov.
\newblock Resonance amplification of oscillations in matter and spectroscopy of
  solar neutrinos, yad. fiz. 42, 1441-1448 (1985).
\newblock {\em Sov. J. Nucl. Phys.}, 42:913--917, 1985.

\bibitem{migliozzi2016high}
Pasquale Migliozzi, KM3NeT Collaboration, et~al.
\newblock High energy neutrino detection with km3net.
\newblock In {\em Journal of Physics: Conference Series}, volume 718, page
  052024. IOP Publishing, 2016.

\bibitem{brunner2013measurement}
J{\"u}rgen Brunner.
\newblock Measurement of neutrino oscillations with neutrino telescopes.
\newblock {\em Advances in High Energy Physics}, 2013.

\bibitem{sakharov1967violation}
Andrej~Dmitrievich Sakharov.
\newblock Violation of cp invariance, c asymmetry, and baryon asymmetry of the
  universe.
\newblock {\em JETP lett.}, 5:24--27, 1967.

\bibitem{гаркушаисследование}
F.N. Novoskoltsev \& A.A.~Sokolov V.I.~Garkusha.
\newblock Neutrino oscillation research using the u-70 accelerator complex (in
  russian).
\newblock {\em IHEP Preprint 2015-5}, 2015.

\bibitem{U-70}
N.~E.~Tyurin et~al.
\newblock {\em Facility for intense hadron beams}.
\newblock News and Problems of Fundamental Physics 2 (9), 2010.

\bibitem{stahl2012expression}
A~Stahl, A~Jipa, P~Kuusiniemi, A~Finch, D~Yilmaz, J~Ilic, L~Periale,
  S~S{\"o}ldner-Rembold, B~Andrieu, G~Galvanin, et~al.
\newblock Expression of interest for a very long baseline neutrino oscillation
  experiment (lbno).
\newblock Technical report, 2012.

\bibitem{rubbia2013laguna}
Andr{\'e} Rubbia.
\newblock Laguna-lbno: Design of an underground neutrino observatory coupled to
  long baseline neutrino beams from cern.
\newblock In {\em Journal of Physics: Conference Series}, volume 408, page
  012006. IOP Publishing, 2013.

\bibitem{murphy2015glacier}
Sebastien Murphy.
\newblock Glacier for lbno: physics motivation and r\&d results.
\newblock {\em Physics Procedia}, 61:560--567, 2015.

\bibitem{agarwalla2014mass}
Sanjib~Kumar Agarwalla, L~Agostino, M~Aittola, A~Alekou, B~Andrieu, D~Angus,
  F~Antoniou, A~Ariga, T~Ariga, R~Asfandiyarov, et~al.
\newblock The mass-hierarchy and cp-violation discovery reach of the lbno
  long-baseline neutrino experiment.
\newblock {\em Journal of High Energy Physics}, 2014(5):94, 2014.

\bibitem{avanzini2015laguna}
Margherita~Buizza Avanzini.
\newblock The laguna-lbno project.
\newblock {\em Physics Procedia}, 61:524--533, 2015.

\bibitem{clark2016pingu}
Ken Clark.
\newblock Pingu and the neutrino mass hierarchy.
\newblock {\em Nucl. Part. Phys. Proc.}, 273:1870--1875, 2016.

\bibitem{aartsen2015determining}
MG~Aartsen, M~Ackermann, J~Adams, JA~Aguilar, M~Ahlers, Maryon Ahrens,
  D~Altmann, T~Anderson, C~Arguelles, TC~Arlen, et~al.
\newblock Determining neutrino oscillation parameters from atmospheric muon
  neutrino disappearance with three years of icecube deepcore data.
\newblock {\em Physical Review D}, 91(7):072004, 2015.

\bibitem{gonzalez2004measuring}
MC~Gonzalez-Garcia, M~Maltoni, and A~Yu Smirnov.
\newblock Measuring the deviation of the 2-3 lepton mixing from maximal with
  atmospheric neutrinos.
\newblock {\em Physical Review D}, 70(9):093005, 2004.

\bibitem{barger2012neutrino}
Vernon Barger, Raj Gandhi, Pomita Ghoshal, Srubabati Goswami, Danny Marfatia,
  Suprabh Prakash, Sushant~K Raut, and S~Uma Sankar.
\newblock Neutrino mass hierarchy and octant determination with atmospheric
  neutrinos.
\newblock {\em Physical review letters}, 109(9):091801, 2012.

\bibitem{winter2016atmospheric}
Walter Winter.
\newblock Atmospheric neutrino oscillations for earth tomography.
\newblock {\em Nuclear Physics B}, 908:250--267, 2016.

\bibitem{agarwalla2013resolving}
Sanjib~Kumar Agarwalla, Suprabh Prakash, and S~Uma Sankar.
\newblock Resolving the octant of $\theta_{23}$ with t2k and no$\nu$a.
\newblock {\em Journal of High Energy Physics}, 2013(7):131, 2013.

\bibitem{chatterjee2013octant}
Animesh Chatterjee, Pomita Ghoshal, Srubabati Goswami, and Sushant~K Raut.
\newblock Octant sensitivity for large $\theta_{13}$ in atmospheric and
  long-baseline neutrino experiments.
\newblock {\em Journal of High Energy Physics}, 2013(6):10, 2013.

\bibitem{de2017status}
PF~de~Salas, DV~Forero, CA~Ternes, M~Tortola, and JWF Valle.
\newblock Status of neutrino oscillations 2017.
\newblock {\em arXiv:1708.01186}, 2017.

\bibitem{capozzi2017global}
Francesco Capozzi, Eleonora Di~Valentino, Eligio Lisi, Antonio Marrone,
  Alessandro Melchiorri, and Antonio Palazzo.
\newblock Global constraints on absolute neutrino masses and their ordering.
\newblock {\em Physical Review D}, 95(9):096014, 2017.

\bibitem{couchot2017cosmological}
F~Couchot, S~Henrot-Versill{\'e}, O~Perdereau, S~Plaszczynski, B~Rouill{\'e}
  d’Orfeuil, M~Spinelli, and M~Tristram.
\newblock Cosmological constraints on the neutrino mass including systematic
  uncertainties.
\newblock {\em Astronomy \& Astrophysics}, 606:A104, 2017.

\bibitem{robertson2013katrin}
RG~Robertson.
\newblock Katrin: an experiment to determine the neutrino mass from the beta
  decay of tritium.
\newblock {\em arXiv:1307.5486}, 2013.

\bibitem{shirai2013kamland}
Junpei Shirai.
\newblock Kamland-zen: status and future.
\newblock {\em Nuclear Physics B-Proceedings Supplements}, 237:28--30, 2013.

\bibitem{fogli2006global}
GL~Fogli, E~Lisi, A~Marrone, and A~Palazzo.
\newblock Global analysis of three-flavor neutrino masses and mixings.
\newblock {\em Progress in Particle and Nuclear Physics}, 57(2):742--795, 2006.

\end{thebibliography}


\begin{thebibliography}{161}%
\makeatletter
\providecommand \@ifxundefined [1]{%
 \@ifx{#1\undefined}
}%
\providecommand \@ifnum [1]{%
 \ifnum #1\expandafter \@firstoftwo
 \else \expandafter \@secondoftwo
 \fi
}%
\providecommand \@ifx [1]{%
 \ifx #1\expandafter \@firstoftwo
 \else \expandafter \@secondoftwo
 \fi
}%
\providecommand \natexlab [1]{#1}%
\providecommand \enquote  [1]{``#1''}%
\providecommand \bibnamefont  [1]{#1}%
\providecommand \bibfnamefont [1]{#1}%
\providecommand \citenamefont [1]{#1}%
\providecommand \href@noop [0]{\@secondoftwo}%
\providecommand \href [0]{\begingroup \@sanitize@url \@href}%
\providecommand \@href[1]{\@@startlink{#1}\@@href}%
\providecommand \@@href[1]{\endgroup#1\@@endlink}%
\providecommand \@sanitize@url [0]{\catcode `\\12\catcode `\$12\catcode
  `\&12\catcode `\#12\catcode `\^12\catcode `\_12\catcode `\%12\relax}%
\providecommand \@@startlink[1]{}%
\providecommand \@@endlink[0]{}%
\providecommand \url  [0]{\begingroup\@sanitize@url \@url }%
\providecommand \@url [1]{\endgroup\@href {#1}{\urlprefix }}%
\providecommand \urlprefix  [0]{URL }%
\providecommand \Eprint [0]{\href }%
\providecommand \doibase [0]{http://dx.doi.org/}%
\providecommand \selectlanguage [0]{\@gobble}%
\providecommand \bibinfo  [0]{\@secondoftwo}%
\providecommand \bibfield  [0]{\@secondoftwo}%
\providecommand \translation [1]{[#1]}%
\providecommand \BibitemOpen [0]{}%
\providecommand \bibitemStop [0]{}%
\providecommand \bibitemNoStop [0]{.\EOS\space}%
\providecommand \EOS [0]{\spacefactor3000\relax}%
\providecommand \BibitemShut  [1]{\csname bibitem#1\endcsname}%
\let\auto@bib@innerbib\@empty
%</preamble>
\bibitem [{Note1()}]{Note1}%
  \BibitemOpen
  \bibinfo {note} {Automatically placing footnotes into the bibliography
  requires using BibTeX to compile the bibliography.}\BibitemShut {Stop}%
\bibitem [{\citenamefont {Cowan}\ \emph {et~al.}(1956)\citenamefont {Cowan},
  \citenamefont {Harrison}, \citenamefont {Langer},\ and\ \citenamefont
  {Reines}}]{cowan1956test}%
  \BibitemOpen
  \bibfield  {author} {\bibinfo {author} {\bibfnamefont {C.}~\bibnamefont
  {Cowan}}, \bibinfo {author} {\bibfnamefont {F.}~\bibnamefont {Harrison}},
  \bibinfo {author} {\bibfnamefont {L.}~\bibnamefont {Langer}}, \ and\ \bibinfo
  {author} {\bibfnamefont {F.}~\bibnamefont {Reines}},\ }\href@noop {}
  {\bibfield  {journal} {\bibinfo  {journal} {Il Nuovo Cimento (1955-1965)}\
  }\textbf {\bibinfo {volume} {3}},\ \bibinfo {pages} {649} (\bibinfo {year}
  {1956})}\BibitemShut {NoStop}%
\bibitem [{\citenamefont {Pauli}(1930)}]{pauli1930letter}%
  \BibitemOpen
  \bibfield  {author} {\bibinfo {author} {\bibfnamefont {W.}~\bibnamefont
  {Pauli}},\ }in\ \href@noop {} {\emph {\bibinfo {booktitle} {Meitner and her
  colleagues (Open letter to the participants of the conference at
  Tubingen)}}}\ (\bibinfo {year} {1930})\BibitemShut {NoStop}%
\bibitem [{\citenamefont {Diwan}\ \emph {et~al.}(2016)\citenamefont {Diwan},
  \citenamefont {Galymov}, \citenamefont {Qian},\ and\ \citenamefont
  {Rubbia}}]{diwan2016long}%
  \BibitemOpen
  \bibfield  {author} {\bibinfo {author} {\bibfnamefont {M.~V.}\ \bibnamefont
  {Diwan}}, \bibinfo {author} {\bibfnamefont {V.}~\bibnamefont {Galymov}},
  \bibinfo {author} {\bibfnamefont {X.}~\bibnamefont {Qian}}, \ and\ \bibinfo
  {author} {\bibfnamefont {A.}~\bibnamefont {Rubbia}},\ }\href@noop {}
  {\bibfield  {journal} {\bibinfo  {journal} {Annual Review of Nuclear and
  Particle Science}\ }\textbf {\bibinfo {volume} {66}} (\bibinfo {year}
  {2016})}\BibitemShut {NoStop}%
\bibitem [{\citenamefont {Bellerive}(2004)}]{bellerive2004review}%
  \BibitemOpen
  \bibfield  {author} {\bibinfo {author} {\bibfnamefont {A.}~\bibnamefont
  {Bellerive}},\ }\href@noop {} {\bibfield  {journal} {\bibinfo  {journal}
  {International Journal of Modern Physics A}\ }\textbf {\bibinfo {volume}
  {19}},\ \bibinfo {pages} {1167} (\bibinfo {year} {2004})}\BibitemShut
  {NoStop}%
\bibitem [{\citenamefont {Cleveland}\ \emph {et~al.}(1998)\citenamefont
  {Cleveland}, \citenamefont {Daily}, \citenamefont {Davis~Jr}, \citenamefont
  {Distel}, \citenamefont {Lande}, \citenamefont {Lee}, \citenamefont
  {Wildenhain},\ and\ \citenamefont {Ullman}}]{cleveland1998measurement}%
  \BibitemOpen
  \bibfield  {author} {\bibinfo {author} {\bibfnamefont {B.~T.}\ \bibnamefont
  {Cleveland}}, \bibinfo {author} {\bibfnamefont {T.}~\bibnamefont {Daily}},
  \bibinfo {author} {\bibfnamefont {R.}~\bibnamefont {Davis~Jr}}, \bibinfo
  {author} {\bibfnamefont {J.~R.}\ \bibnamefont {Distel}}, \bibinfo {author}
  {\bibfnamefont {K.}~\bibnamefont {Lande}}, \bibinfo {author} {\bibfnamefont
  {C.}~\bibnamefont {Lee}}, \bibinfo {author} {\bibfnamefont {P.~S.}\
  \bibnamefont {Wildenhain}}, \ and\ \bibinfo {author} {\bibfnamefont
  {J.}~\bibnamefont {Ullman}},\ }\href@noop {} {\bibfield  {journal} {\bibinfo
  {journal} {The Astrophysical Journal}\ }\textbf {\bibinfo {volume} {496}},\
  \bibinfo {pages} {505} (\bibinfo {year} {1998})}\BibitemShut {NoStop}%
\bibitem [{\citenamefont {Hampel}\ \emph {et~al.}(1999)\citenamefont {Hampel},
  \citenamefont {Handt}, \citenamefont {Heusser}, \citenamefont {Kiko},
  \citenamefont {Kirsten}, \citenamefont {Laubenstein}, \citenamefont
  {Pernicka}, \citenamefont {Rau}, \citenamefont {Wojcik}, \citenamefont
  {Zakharov} \emph {et~al.}}]{hampel1999gallex}%
  \BibitemOpen
  \bibfield  {author} {\bibinfo {author} {\bibfnamefont {W.}~\bibnamefont
  {Hampel}}, \bibinfo {author} {\bibfnamefont {J.}~\bibnamefont {Handt}},
  \bibinfo {author} {\bibfnamefont {G.}~\bibnamefont {Heusser}}, \bibinfo
  {author} {\bibfnamefont {J.}~\bibnamefont {Kiko}}, \bibinfo {author}
  {\bibfnamefont {T.}~\bibnamefont {Kirsten}}, \bibinfo {author} {\bibfnamefont
  {M.}~\bibnamefont {Laubenstein}}, \bibinfo {author} {\bibfnamefont
  {E.}~\bibnamefont {Pernicka}}, \bibinfo {author} {\bibfnamefont
  {W.}~\bibnamefont {Rau}}, \bibinfo {author} {\bibfnamefont {M.}~\bibnamefont
  {Wojcik}}, \bibinfo {author} {\bibfnamefont {Y.}~\bibnamefont {Zakharov}},
  \emph {et~al.},\ }\href@noop {} {\bibfield  {journal} {\bibinfo  {journal}
  {Physics Letters B}\ }\textbf {\bibinfo {volume} {447}},\ \bibinfo {pages}
  {127} (\bibinfo {year} {1999})}\BibitemShut {NoStop}%
\bibitem [{\citenamefont {Abdurashitov}\ \emph {et~al.}(2009)\citenamefont
  {Abdurashitov}, \citenamefont {Gavrin}, \citenamefont {Gorbachev},
  \citenamefont {Gurkina}, \citenamefont {Ibragimova}, \citenamefont
  {Kalikhov}, \citenamefont {Khairnasov}, \citenamefont {Knodel}, \citenamefont
  {Mirmov}, \citenamefont {Shikhin} \emph
  {et~al.}}]{abdurashitov2009measurement}%
  \BibitemOpen
  \bibfield  {author} {\bibinfo {author} {\bibfnamefont {J.}~\bibnamefont
  {Abdurashitov}}, \bibinfo {author} {\bibfnamefont {V.}~\bibnamefont
  {Gavrin}}, \bibinfo {author} {\bibfnamefont {V.}~\bibnamefont {Gorbachev}},
  \bibinfo {author} {\bibfnamefont {P.}~\bibnamefont {Gurkina}}, \bibinfo
  {author} {\bibfnamefont {T.}~\bibnamefont {Ibragimova}}, \bibinfo {author}
  {\bibfnamefont {A.}~\bibnamefont {Kalikhov}}, \bibinfo {author}
  {\bibfnamefont {N.}~\bibnamefont {Khairnasov}}, \bibinfo {author}
  {\bibfnamefont {T.}~\bibnamefont {Knodel}}, \bibinfo {author} {\bibfnamefont
  {I.}~\bibnamefont {Mirmov}}, \bibinfo {author} {\bibfnamefont
  {A.}~\bibnamefont {Shikhin}},  \emph {et~al.},\ }\href@noop {} {\bibfield
  {journal} {\bibinfo  {journal} {Physical Review C}\ }\textbf {\bibinfo
  {volume} {80}},\ \bibinfo {pages} {015807} (\bibinfo {year}
  {2009})}\BibitemShut {NoStop}%
\bibitem [{\citenamefont {Ranucci}(2016)}]{ranucci2016techniques}%
  \BibitemOpen
  \bibfield  {author} {\bibinfo {author} {\bibfnamefont {G.}~\bibnamefont
  {Ranucci}},\ }\href@noop {} {\bibfield  {journal} {\bibinfo  {journal} {The
  European Physical Journal A}\ }\textbf {\bibinfo {volume} {52}},\ \bibinfo
  {pages} {79} (\bibinfo {year} {2016})}\BibitemShut {NoStop}%
\bibitem [{\citenamefont {Fukuda}\ \emph {et~al.}(2003)\citenamefont {Fukuda},
  \citenamefont {Fukuda}, \citenamefont {Hayakawa}, \citenamefont {Ichihara},
  \citenamefont {Ishitsuka}, \citenamefont {Itow}, \citenamefont {Kajita},
  \citenamefont {Kameda}, \citenamefont {Kaneyuki}, \citenamefont {Kasuga}
  \emph {et~al.}}]{fukuda2003super}%
  \BibitemOpen
  \bibfield  {author} {\bibinfo {author} {\bibfnamefont {S.}~\bibnamefont
  {Fukuda}}, \bibinfo {author} {\bibfnamefont {Y.}~\bibnamefont {Fukuda}},
  \bibinfo {author} {\bibfnamefont {T.}~\bibnamefont {Hayakawa}}, \bibinfo
  {author} {\bibfnamefont {E.}~\bibnamefont {Ichihara}}, \bibinfo {author}
  {\bibfnamefont {M.}~\bibnamefont {Ishitsuka}}, \bibinfo {author}
  {\bibfnamefont {Y.}~\bibnamefont {Itow}}, \bibinfo {author} {\bibfnamefont
  {T.}~\bibnamefont {Kajita}}, \bibinfo {author} {\bibfnamefont
  {J.}~\bibnamefont {Kameda}}, \bibinfo {author} {\bibfnamefont
  {K.}~\bibnamefont {Kaneyuki}}, \bibinfo {author} {\bibfnamefont
  {S.}~\bibnamefont {Kasuga}},  \emph {et~al.},\ }\href@noop {} {\bibfield
  {journal} {\bibinfo  {journal} {Nuclear Instruments and Methods in Physics
  Research Section A: Accelerators, Spectrometers, Detectors and Associated
  Equipment}\ }\textbf {\bibinfo {volume} {501}},\ \bibinfo {pages} {418}
  (\bibinfo {year} {2003})}\BibitemShut {NoStop}%
\bibitem [{\citenamefont {Jelley}\ \emph {et~al.}(2009)\citenamefont {Jelley},
  \citenamefont {McDonald},\ and\ \citenamefont
  {Robertson}}]{jelley2009sudbury}%
  \BibitemOpen
  \bibfield  {author} {\bibinfo {author} {\bibfnamefont {N.}~\bibnamefont
  {Jelley}}, \bibinfo {author} {\bibfnamefont {A.~B.}\ \bibnamefont
  {McDonald}}, \ and\ \bibinfo {author} {\bibfnamefont {R.~H.}\ \bibnamefont
  {Robertson}},\ }\href@noop {} {\bibfield  {journal} {\bibinfo  {journal}
  {Annual Review of Nuclear and Particle Science}\ }\textbf {\bibinfo {volume}
  {59}},\ \bibinfo {pages} {431} (\bibinfo {year} {2009})}\BibitemShut
  {NoStop}%
\bibitem [{\citenamefont {Abe}\ \emph {et~al.}(2011{\natexlab{a}})\citenamefont
  {Abe}, \citenamefont {Abe}, \citenamefont {Aihara}, \citenamefont {Fukuda},
  \citenamefont {Hayato}, \citenamefont {Huang}, \citenamefont {Ichikawa},
  \citenamefont {Ikeda}, \citenamefont {Inoue}, \citenamefont {Ishino} \emph
  {et~al.}}]{abe2011letter}%
  \BibitemOpen
  \bibfield  {author} {\bibinfo {author} {\bibfnamefont {K.}~\bibnamefont
  {Abe}}, \bibinfo {author} {\bibfnamefont {T.}~\bibnamefont {Abe}}, \bibinfo
  {author} {\bibfnamefont {H.}~\bibnamefont {Aihara}}, \bibinfo {author}
  {\bibfnamefont {Y.}~\bibnamefont {Fukuda}}, \bibinfo {author} {\bibfnamefont
  {Y.}~\bibnamefont {Hayato}}, \bibinfo {author} {\bibfnamefont
  {K.}~\bibnamefont {Huang}}, \bibinfo {author} {\bibfnamefont
  {A.}~\bibnamefont {Ichikawa}}, \bibinfo {author} {\bibfnamefont
  {M.}~\bibnamefont {Ikeda}}, \bibinfo {author} {\bibfnamefont
  {K.}~\bibnamefont {Inoue}}, \bibinfo {author} {\bibfnamefont
  {H.}~\bibnamefont {Ishino}},  \emph {et~al.},\ }\href@noop {} {\bibfield
  {journal} {\bibinfo  {journal} {arXiv preprint arXiv:1109.3262}\ } (\bibinfo
  {year} {2011}{\natexlab{a}})}\BibitemShut {NoStop}%
\bibitem [{\citenamefont {Karle}\ \emph {et~al.}(2010)\citenamefont {Karle}
  \emph {et~al.}}]{karle2010icecube}%
  \BibitemOpen
  \bibfield  {author} {\bibinfo {author} {\bibfnamefont {A.}~\bibnamefont
  {Karle}} \emph {et~al.},\ }\href@noop {} {\bibfield  {journal} {\bibinfo
  {journal} {arXiv preprint arXiv:1003.5715}\ } (\bibinfo {year}
  {2010})}\BibitemShut {NoStop}%
\bibitem [{\citenamefont {Collaboration}\ \emph {et~al.}(2001)\citenamefont
  {Collaboration} \emph {et~al.}}]{icecube2001icecube}%
  \BibitemOpen
  \bibfield  {author} {\bibinfo {author} {\bibfnamefont {I.}~\bibnamefont
  {Collaboration}} \emph {et~al.},\ }\href@noop {} {\enquote {\bibinfo {title}
  {Icecube preliminary design document revision 1.24},}\ } (\bibinfo {year}
  {2001})\BibitemShut {NoStop}%
\bibitem [{\citenamefont {Cao}\ and\ \citenamefont
  {Luk}(2016)}]{cao2016overview}%
  \BibitemOpen
  \bibfield  {author} {\bibinfo {author} {\bibfnamefont {J.}~\bibnamefont
  {Cao}}\ and\ \bibinfo {author} {\bibfnamefont {K.-B.}\ \bibnamefont {Luk}},\
  }\href@noop {} {\bibfield  {journal} {\bibinfo  {journal} {Nuclear Physics
  B}\ }\textbf {\bibinfo {volume} {908}},\ \bibinfo {pages} {62} (\bibinfo
  {year} {2016})}\BibitemShut {NoStop}%
\bibitem [{\citenamefont {An}\ \emph {et~al.}(2012)\citenamefont {An},
  \citenamefont {An}, \citenamefont {Bai}, \citenamefont {Balantekin},
  \citenamefont {Band}, \citenamefont {Beriguete}, \citenamefont {Bishai},
  \citenamefont {Blyth}, \citenamefont {Brown}, \citenamefont {Cao} \emph
  {et~al.}}]{an2012side}%
  \BibitemOpen
  \bibfield  {author} {\bibinfo {author} {\bibfnamefont {F.}~\bibnamefont
  {An}}, \bibinfo {author} {\bibfnamefont {Q.}~\bibnamefont {An}}, \bibinfo
  {author} {\bibfnamefont {J.}~\bibnamefont {Bai}}, \bibinfo {author}
  {\bibfnamefont {A.}~\bibnamefont {Balantekin}}, \bibinfo {author}
  {\bibfnamefont {H.}~\bibnamefont {Band}}, \bibinfo {author} {\bibfnamefont
  {W.}~\bibnamefont {Beriguete}}, \bibinfo {author} {\bibfnamefont
  {M.}~\bibnamefont {Bishai}}, \bibinfo {author} {\bibfnamefont
  {S.}~\bibnamefont {Blyth}}, \bibinfo {author} {\bibfnamefont
  {R.}~\bibnamefont {Brown}}, \bibinfo {author} {\bibfnamefont
  {G.}~\bibnamefont {Cao}},  \emph {et~al.},\ }\href@noop {} {\bibfield
  {journal} {\bibinfo  {journal} {Nuclear Instruments and Methods in Physics
  Research Section A: Accelerators, Spectrometers, Detectors and Associated
  Equipment}\ }\textbf {\bibinfo {volume} {685}},\ \bibinfo {pages} {78}
  (\bibinfo {year} {2012})}\BibitemShut {NoStop}%
\bibitem [{\citenamefont {Eguchi}\ \emph {et~al.}(2003)\citenamefont {Eguchi},
  \citenamefont {Enomoto}, \citenamefont {Furuno}, \citenamefont {Goldman},
  \citenamefont {Hanada}, \citenamefont {Ikeda}, \citenamefont {Ikeda},
  \citenamefont {Inoue}, \citenamefont {Ishihara}, \citenamefont {Itoh} \emph
  {et~al.}}]{eguchi2003first}%
  \BibitemOpen
  \bibfield  {author} {\bibinfo {author} {\bibfnamefont {K.}~\bibnamefont
  {Eguchi}}, \bibinfo {author} {\bibfnamefont {S.}~\bibnamefont {Enomoto}},
  \bibinfo {author} {\bibfnamefont {K.}~\bibnamefont {Furuno}}, \bibinfo
  {author} {\bibfnamefont {J.}~\bibnamefont {Goldman}}, \bibinfo {author}
  {\bibfnamefont {H.}~\bibnamefont {Hanada}}, \bibinfo {author} {\bibfnamefont
  {H.}~\bibnamefont {Ikeda}}, \bibinfo {author} {\bibfnamefont
  {K.}~\bibnamefont {Ikeda}}, \bibinfo {author} {\bibfnamefont
  {K.}~\bibnamefont {Inoue}}, \bibinfo {author} {\bibfnamefont
  {K.}~\bibnamefont {Ishihara}}, \bibinfo {author} {\bibfnamefont
  {W.}~\bibnamefont {Itoh}},  \emph {et~al.},\ }\href@noop {} {\bibfield
  {journal} {\bibinfo  {journal} {Physical Review Letters}\ }\textbf {\bibinfo
  {volume} {90}},\ \bibinfo {pages} {021802} (\bibinfo {year}
  {2003})}\BibitemShut {NoStop}%
\bibitem [{\citenamefont {Kajita}(2010)}]{kajita2010atmospheric}%
  \BibitemOpen
  \bibfield  {author} {\bibinfo {author} {\bibfnamefont {T.}~\bibnamefont
  {Kajita}},\ }\href@noop {} {\bibfield  {journal} {\bibinfo  {journal}
  {Proceedings of the Japan Academy, Series B}\ }\textbf {\bibinfo {volume}
  {86}},\ \bibinfo {pages} {303} (\bibinfo {year} {2010})}\BibitemShut
  {NoStop}%
\bibitem [{\citenamefont {Bellerive}\ \emph {et~al.}(2016)\citenamefont
  {Bellerive}, \citenamefont {Klein}, \citenamefont {McDonald}, \citenamefont
  {Noble}, \citenamefont {Poon}, \citenamefont {Collaboration} \emph
  {et~al.}}]{bellerive2016sudbury}%
  \BibitemOpen
  \bibfield  {author} {\bibinfo {author} {\bibfnamefont {A.}~\bibnamefont
  {Bellerive}}, \bibinfo {author} {\bibfnamefont {J.}~\bibnamefont {Klein}},
  \bibinfo {author} {\bibfnamefont {A.}~\bibnamefont {McDonald}}, \bibinfo
  {author} {\bibfnamefont {A.}~\bibnamefont {Noble}}, \bibinfo {author}
  {\bibfnamefont {A.}~\bibnamefont {Poon}}, \bibinfo {author} {\bibfnamefont
  {S.}~\bibnamefont {Collaboration}},  \emph {et~al.},\ }\href@noop {}
  {\bibfield  {journal} {\bibinfo  {journal} {Nuclear Physics B}\ }\textbf
  {\bibinfo {volume} {908}},\ \bibinfo {pages} {30} (\bibinfo {year}
  {2016})}\BibitemShut {NoStop}%
\bibitem [{\citenamefont {Albright}\ \emph {et~al.}(2004)\citenamefont
  {Albright}, \citenamefont {Barger}, \citenamefont {Beacom}, \citenamefont
  {Brice}, \citenamefont {Gomez-Cadenas}, \citenamefont {Goodman},
  \citenamefont {Harris}, \citenamefont {Huber}, \citenamefont {Jansson},
  \citenamefont {Lindner} \emph {et~al.}}]{albright2004neutrino}%
  \BibitemOpen
  \bibfield  {author} {\bibinfo {author} {\bibfnamefont {C.}~\bibnamefont
  {Albright}}, \bibinfo {author} {\bibfnamefont {V.}~\bibnamefont {Barger}},
  \bibinfo {author} {\bibfnamefont {J.}~\bibnamefont {Beacom}}, \bibinfo
  {author} {\bibfnamefont {S.}~\bibnamefont {Brice}}, \bibinfo {author}
  {\bibfnamefont {J.}~\bibnamefont {Gomez-Cadenas}}, \bibinfo {author}
  {\bibfnamefont {M.}~\bibnamefont {Goodman}}, \bibinfo {author} {\bibfnamefont
  {D.}~\bibnamefont {Harris}}, \bibinfo {author} {\bibfnamefont
  {P.}~\bibnamefont {Huber}}, \bibinfo {author} {\bibfnamefont
  {A.}~\bibnamefont {Jansson}}, \bibinfo {author} {\bibfnamefont
  {M.}~\bibnamefont {Lindner}},  \emph {et~al.},\ }\href@noop {} {\bibfield
  {journal} {\bibinfo  {journal} {arXiv preprint physics/0411123}\ } (\bibinfo
  {year} {2004})}\BibitemShut {NoStop}%
\bibitem [{\citenamefont {Zucchelli}(2002)}]{zucchelli2002novel}%
  \BibitemOpen
  \bibfield  {author} {\bibinfo {author} {\bibfnamefont {P.}~\bibnamefont
  {Zucchelli}},\ }\href@noop {} {\bibfield  {journal} {\bibinfo  {journal}
  {Physics Letters B}\ }\textbf {\bibinfo {volume} {532}},\ \bibinfo {pages}
  {166} (\bibinfo {year} {2002})}\BibitemShut {NoStop}%
\bibitem [{\citenamefont {Ianni}(2017)}]{ianni2017solar}%
  \BibitemOpen
  \bibfield  {author} {\bibinfo {author} {\bibfnamefont {A.}~\bibnamefont
  {Ianni}},\ }\href@noop {} {\bibfield  {journal} {\bibinfo  {journal}
  {Progress in Particle and Nuclear Physics}\ }\textbf {\bibinfo {volume}
  {94}},\ \bibinfo {pages} {257} (\bibinfo {year} {2017})}\BibitemShut
  {NoStop}%
\bibitem [{\citenamefont {Davis}(1994)}]{davis1994review}%
  \BibitemOpen
  \bibfield  {author} {\bibinfo {author} {\bibfnamefont {R.}~\bibnamefont
  {Davis}},\ }\href@noop {} {\bibfield  {journal} {\bibinfo  {journal}
  {Progress in Particle and Nuclear Physics}\ }\textbf {\bibinfo {volume}
  {32}},\ \bibinfo {pages} {13} (\bibinfo {year} {1994})}\BibitemShut {NoStop}%
\bibitem [{\citenamefont {Leitner}\ \emph {et~al.}(2017)\citenamefont {Leitner}
  \emph {et~al.}}]{leitner2017recent}%
  \BibitemOpen
  \bibfield  {author} {\bibinfo {author} {\bibfnamefont {R.}~\bibnamefont
  {Leitner}} \emph {et~al.},\ }\href@noop {} {\bibfield  {journal} {\bibinfo
  {journal} {Nuclear and particle physics proceedings}\ }\textbf {\bibinfo
  {volume} {285}},\ \bibinfo {pages} {32} (\bibinfo {year} {2017})}\BibitemShut
  {NoStop}%
\bibitem [{\citenamefont {Ianni}(2014)}]{ianni2014solar}%
  \BibitemOpen
  \bibfield  {author} {\bibinfo {author} {\bibfnamefont {A.}~\bibnamefont
  {Ianni}},\ }\href@noop {} {\bibfield  {journal} {\bibinfo  {journal} {Physics
  of the Dark Universe}\ }\textbf {\bibinfo {volume} {4}},\ \bibinfo {pages}
  {44} (\bibinfo {year} {2014})}\BibitemShut {NoStop}%
\bibitem [{\citenamefont {Suekane}\ and\ \citenamefont
  {de~Castro~Bezerra}(2016)}]{suekane2016double}%
  \BibitemOpen
  \bibfield  {author} {\bibinfo {author} {\bibfnamefont {F.}~\bibnamefont
  {Suekane}}\ and\ \bibinfo {author} {\bibfnamefont {T.~J.}\ \bibnamefont
  {de~Castro~Bezerra}},\ }\href@noop {} {\bibfield  {journal} {\bibinfo
  {journal} {Nuclear Physics B}\ }\textbf {\bibinfo {volume} {908}},\ \bibinfo
  {pages} {74} (\bibinfo {year} {2016})}\BibitemShut {NoStop}%
\bibitem [{\citenamefont {Kim}\ \emph {et~al.}(2016)\citenamefont {Kim},
  \citenamefont {Collaboration} \emph {et~al.}}]{kim2016measurement}%
  \BibitemOpen
  \bibfield  {author} {\bibinfo {author} {\bibfnamefont {S.-B.}\ \bibnamefont
  {Kim}}, \bibinfo {author} {\bibfnamefont {R.}~\bibnamefont {Collaboration}},
  \emph {et~al.},\ }\href@noop {} {\bibfield  {journal} {\bibinfo  {journal}
  {Nuclear Physics B}\ }\textbf {\bibinfo {volume} {908}},\ \bibinfo {pages}
  {94} (\bibinfo {year} {2016})}\BibitemShut {NoStop}%
\bibitem [{\citenamefont {Palomares}\ \emph {et~al.}(2011)\citenamefont
  {Palomares}, \citenamefont {Collaboration} \emph
  {et~al.}}]{palomares2011double}%
  \BibitemOpen
  \bibfield  {author} {\bibinfo {author} {\bibfnamefont {C.}~\bibnamefont
  {Palomares}}, \bibinfo {author} {\bibfnamefont {D.~C.}\ \bibnamefont
  {Collaboration}},  \emph {et~al.},\ }in\ \href@noop {} {\emph {\bibinfo
  {booktitle} {Journal of Physics: Conference Series}}},\ Vol.\ \bibinfo
  {volume} {335}\ (\bibinfo {organization} {IOP Publishing},\ \bibinfo {year}
  {2011})\ p.\ \bibinfo {pages} {012055}\BibitemShut {NoStop}%
\bibitem [{\citenamefont {Ahn}\ \emph {et~al.}(2012)\citenamefont {Ahn},
  \citenamefont {Chebotaryov}, \citenamefont {Choi}, \citenamefont {Choi},
  \citenamefont {Choi}, \citenamefont {Choi}, \citenamefont {Jang},
  \citenamefont {Jang}, \citenamefont {Jeon}, \citenamefont {Jeong} \emph
  {et~al.}}]{ahn2012observation}%
  \BibitemOpen
  \bibfield  {author} {\bibinfo {author} {\bibfnamefont {J.}~\bibnamefont
  {Ahn}}, \bibinfo {author} {\bibfnamefont {S.}~\bibnamefont {Chebotaryov}},
  \bibinfo {author} {\bibfnamefont {J.}~\bibnamefont {Choi}}, \bibinfo {author}
  {\bibfnamefont {S.}~\bibnamefont {Choi}}, \bibinfo {author} {\bibfnamefont
  {W.}~\bibnamefont {Choi}}, \bibinfo {author} {\bibfnamefont {Y.}~\bibnamefont
  {Choi}}, \bibinfo {author} {\bibfnamefont {H.}~\bibnamefont {Jang}}, \bibinfo
  {author} {\bibfnamefont {J.}~\bibnamefont {Jang}}, \bibinfo {author}
  {\bibfnamefont {E.}~\bibnamefont {Jeon}}, \bibinfo {author} {\bibfnamefont
  {I.}~\bibnamefont {Jeong}},  \emph {et~al.},\ }\href@noop {} {\bibfield
  {journal} {\bibinfo  {journal} {Physical Review Letters}\ }\textbf {\bibinfo
  {volume} {108}},\ \bibinfo {pages} {191802} (\bibinfo {year}
  {2012})}\BibitemShut {NoStop}%
\bibitem [{\citenamefont {Dwyer}(2015)}]{dwyer2015antineutrinos}%
  \BibitemOpen
  \bibfield  {author} {\bibinfo {author} {\bibfnamefont {D.}~\bibnamefont
  {Dwyer}},\ }\href@noop {} {\bibfield  {journal} {\bibinfo  {journal} {New
  Journal of Physics}\ }\textbf {\bibinfo {volume} {17}},\ \bibinfo {pages}
  {025003} (\bibinfo {year} {2015})}\BibitemShut {NoStop}%
\bibitem [{\citenamefont {Djurcic}(2013)}]{djurcic2013review}%
  \BibitemOpen
  \bibfield  {author} {\bibinfo {author} {\bibfnamefont {Z.}~\bibnamefont
  {Djurcic}},\ }in\ \href@noop {} {\emph {\bibinfo {booktitle} {Journal of
  Physics: Conference Series}}},\ Vol.\ \bibinfo {volume} {408}\ (\bibinfo
  {organization} {IOP Publishing},\ \bibinfo {year} {2013})\ p.\ \bibinfo
  {pages} {012008}\BibitemShut {NoStop}%
\bibitem [{\citenamefont {Datar}\ and\ \citenamefont
  {Mondal}(2017)}]{datar2017india}%
  \BibitemOpen
  \bibfield  {author} {\bibinfo {author} {\bibfnamefont {V.~M.}\ \bibnamefont
  {Datar}}\ and\ \bibinfo {author} {\bibfnamefont {N.~K.}\ \bibnamefont
  {Mondal}},\ }\href@noop {} {\bibfield  {journal} {\bibinfo  {journal}
  {CURRENT SCIENCE}\ }\textbf {\bibinfo {volume} {113}},\ \bibinfo {pages}
  {701} (\bibinfo {year} {2017})}\BibitemShut {NoStop}%
\bibitem [{\citenamefont {Kumar}\ \emph {et~al.}(2017)\citenamefont {Kumar},
  \citenamefont {Kumar}, \citenamefont {Jash}, \citenamefont {Mohanty},
  \citenamefont {Chacko}, \citenamefont {Ajmi}, \citenamefont {Ghosal},
  \citenamefont {Khatun}, \citenamefont {Raychaudhuri}, \citenamefont {Dighe}
  \emph {et~al.}}]{kumar2017invited}%
  \BibitemOpen
  \bibfield  {author} {\bibinfo {author} {\bibfnamefont {A.}~\bibnamefont
  {Kumar}}, \bibinfo {author} {\bibfnamefont {A.~V.}\ \bibnamefont {Kumar}},
  \bibinfo {author} {\bibfnamefont {A.}~\bibnamefont {Jash}}, \bibinfo {author}
  {\bibfnamefont {A.~K.}\ \bibnamefont {Mohanty}}, \bibinfo {author}
  {\bibfnamefont {A.}~\bibnamefont {Chacko}}, \bibinfo {author} {\bibfnamefont
  {A.}~\bibnamefont {Ajmi}}, \bibinfo {author} {\bibfnamefont {A.}~\bibnamefont
  {Ghosal}}, \bibinfo {author} {\bibfnamefont {A.}~\bibnamefont {Khatun}},
  \bibinfo {author} {\bibfnamefont {A.}~\bibnamefont {Raychaudhuri}}, \bibinfo
  {author} {\bibfnamefont {A.}~\bibnamefont {Dighe}},  \emph {et~al.},\
  }\href@noop {} {\bibfield  {journal} {\bibinfo  {journal} {Pramana}\ }\textbf
  {\bibinfo {volume} {88}},\ \bibinfo {pages} {79} (\bibinfo {year}
  {2017})}\BibitemShut {NoStop}%
\bibitem [{\citenamefont {Abe}\ \emph {et~al.}(2016)\citenamefont {Abe},
  \citenamefont {Haga}, \citenamefont {Hayato}, \citenamefont {Ikeda},
  \citenamefont {Iyogi}, \citenamefont {Kameda}, \citenamefont {Kishimoto},
  \citenamefont {Marti}, \citenamefont {Miura}, \citenamefont {Moriyama} \emph
  {et~al.}}]{abe2016solar}%
  \BibitemOpen
  \bibfield  {author} {\bibinfo {author} {\bibfnamefont {K.}~\bibnamefont
  {Abe}}, \bibinfo {author} {\bibfnamefont {Y.}~\bibnamefont {Haga}}, \bibinfo
  {author} {\bibfnamefont {Y.}~\bibnamefont {Hayato}}, \bibinfo {author}
  {\bibfnamefont {M.}~\bibnamefont {Ikeda}}, \bibinfo {author} {\bibfnamefont
  {K.}~\bibnamefont {Iyogi}}, \bibinfo {author} {\bibfnamefont
  {J.}~\bibnamefont {Kameda}}, \bibinfo {author} {\bibfnamefont
  {Y.}~\bibnamefont {Kishimoto}}, \bibinfo {author} {\bibfnamefont
  {L.}~\bibnamefont {Marti}}, \bibinfo {author} {\bibfnamefont
  {M.}~\bibnamefont {Miura}}, \bibinfo {author} {\bibfnamefont
  {S.}~\bibnamefont {Moriyama}},  \emph {et~al.},\ }\href@noop {} {\bibfield
  {journal} {\bibinfo  {journal} {Physical Review D}\ }\textbf {\bibinfo
  {volume} {94}},\ \bibinfo {pages} {052010} (\bibinfo {year}
  {2016})}\BibitemShut {NoStop}%
\bibitem [{\citenamefont {Abe}\ \emph {et~al.}(2017)\citenamefont {Abe},
  \citenamefont {Haga}, \citenamefont {Hayato}, \citenamefont {Ikeda},
  \citenamefont {Iyogi}, \citenamefont {Kameda}, \citenamefont {Kishimoto},
  \citenamefont {Miura}, \citenamefont {Moriyama}, \citenamefont {Nakahata}
  \emph {et~al.}}]{abe2017search}%
  \BibitemOpen
  \bibfield  {author} {\bibinfo {author} {\bibfnamefont {K.}~\bibnamefont
  {Abe}}, \bibinfo {author} {\bibfnamefont {Y.}~\bibnamefont {Haga}}, \bibinfo
  {author} {\bibfnamefont {Y.}~\bibnamefont {Hayato}}, \bibinfo {author}
  {\bibfnamefont {M.}~\bibnamefont {Ikeda}}, \bibinfo {author} {\bibfnamefont
  {K.}~\bibnamefont {Iyogi}}, \bibinfo {author} {\bibfnamefont
  {J.}~\bibnamefont {Kameda}}, \bibinfo {author} {\bibfnamefont
  {Y.}~\bibnamefont {Kishimoto}}, \bibinfo {author} {\bibfnamefont
  {M.}~\bibnamefont {Miura}}, \bibinfo {author} {\bibfnamefont
  {S.}~\bibnamefont {Moriyama}}, \bibinfo {author} {\bibfnamefont
  {M.}~\bibnamefont {Nakahata}},  \emph {et~al.},\ }\href@noop {} {\bibfield
  {journal} {\bibinfo  {journal} {Physical Review D}\ }\textbf {\bibinfo
  {volume} {95}},\ \bibinfo {pages} {012004} (\bibinfo {year}
  {2017})}\BibitemShut {NoStop}%
\bibitem [{\citenamefont {Bellini}(2016)}]{bellini2016impact}%
  \BibitemOpen
  \bibfield  {author} {\bibinfo {author} {\bibfnamefont {G.}~\bibnamefont
  {Bellini}},\ }\href@noop {} {\bibfield  {journal} {\bibinfo  {journal}
  {Nuclear Physics B}\ }\textbf {\bibinfo {volume} {908}},\ \bibinfo {pages}
  {178} (\bibinfo {year} {2016})}\BibitemShut {NoStop}%
\bibitem [{\citenamefont {Bellini}\ \emph {et~al.}(2011)\citenamefont
  {Bellini}, \citenamefont {Benziger}, \citenamefont {Bick}, \citenamefont
  {Bonetti}, \citenamefont {Bonfini}, \citenamefont {Avanzini}, \citenamefont
  {Caccianiga}, \citenamefont {Cadonati}, \citenamefont {Calaprice},
  \citenamefont {Carraro} \emph {et~al.}}]{bellini2011precision}%
  \BibitemOpen
  \bibfield  {author} {\bibinfo {author} {\bibfnamefont {G.}~\bibnamefont
  {Bellini}}, \bibinfo {author} {\bibfnamefont {J.}~\bibnamefont {Benziger}},
  \bibinfo {author} {\bibfnamefont {D.}~\bibnamefont {Bick}}, \bibinfo {author}
  {\bibfnamefont {S.}~\bibnamefont {Bonetti}}, \bibinfo {author} {\bibfnamefont
  {G.}~\bibnamefont {Bonfini}}, \bibinfo {author} {\bibfnamefont {M.~B.}\
  \bibnamefont {Avanzini}}, \bibinfo {author} {\bibfnamefont {B.}~\bibnamefont
  {Caccianiga}}, \bibinfo {author} {\bibfnamefont {L.}~\bibnamefont
  {Cadonati}}, \bibinfo {author} {\bibfnamefont {F.}~\bibnamefont {Calaprice}},
  \bibinfo {author} {\bibfnamefont {C.}~\bibnamefont {Carraro}},  \emph
  {et~al.},\ }\href@noop {} {\bibfield  {journal} {\bibinfo  {journal}
  {Physical Review Letters}\ }\textbf {\bibinfo {volume} {107}},\ \bibinfo
  {pages} {141302} (\bibinfo {year} {2011})}\BibitemShut {NoStop}%
\bibitem [{\citenamefont {Bellini}\ \emph
  {et~al.}(2012{\natexlab{a}})\citenamefont {Bellini}, \citenamefont
  {Benziger}, \citenamefont {Bick}, \citenamefont {Bonetti}, \citenamefont
  {Bonfini}, \citenamefont {Avanzini}, \citenamefont {Caccianiga},
  \citenamefont {Cadonati}, \citenamefont {Calaprice}, \citenamefont {Carraro}
  \emph {et~al.}}]{bellini2012absence}%
  \BibitemOpen
  \bibfield  {author} {\bibinfo {author} {\bibfnamefont {G.}~\bibnamefont
  {Bellini}}, \bibinfo {author} {\bibfnamefont {J.}~\bibnamefont {Benziger}},
  \bibinfo {author} {\bibfnamefont {D.}~\bibnamefont {Bick}}, \bibinfo {author}
  {\bibfnamefont {S.}~\bibnamefont {Bonetti}}, \bibinfo {author} {\bibfnamefont
  {G.}~\bibnamefont {Bonfini}}, \bibinfo {author} {\bibfnamefont {M.~B.}\
  \bibnamefont {Avanzini}}, \bibinfo {author} {\bibfnamefont {B.}~\bibnamefont
  {Caccianiga}}, \bibinfo {author} {\bibfnamefont {L.}~\bibnamefont
  {Cadonati}}, \bibinfo {author} {\bibfnamefont {F.}~\bibnamefont {Calaprice}},
  \bibinfo {author} {\bibfnamefont {C.}~\bibnamefont {Carraro}},  \emph
  {et~al.},\ }\href@noop {} {\bibfield  {journal} {\bibinfo  {journal} {Physics
  Letters B}\ }\textbf {\bibinfo {volume} {707}},\ \bibinfo {pages} {22}
  (\bibinfo {year} {2012}{\natexlab{a}})}\BibitemShut {NoStop}%
\bibitem [{\citenamefont {Raut}(2017)}]{raut2017matter}%
  \BibitemOpen
  \bibfield  {author} {\bibinfo {author} {\bibfnamefont {S.~K.}\ \bibnamefont
  {Raut}},\ }\href@noop {} {\bibfield  {journal} {\bibinfo  {journal} {Physical
  Review D}\ }\textbf {\bibinfo {volume} {96}},\ \bibinfo {pages} {075029}
  (\bibinfo {year} {2017})}\BibitemShut {NoStop}%
\bibitem [{\citenamefont {He}\ \emph {et~al.}(2015)\citenamefont {He},
  \citenamefont {collaboration} \emph {et~al.}}]{he2015jiangmen}%
  \BibitemOpen
  \bibfield  {author} {\bibinfo {author} {\bibfnamefont {M.}~\bibnamefont
  {He}}, \bibinfo {author} {\bibfnamefont {J.}~\bibnamefont {collaboration}},
  \emph {et~al.},\ }\href@noop {} {\bibfield  {journal} {\bibinfo  {journal}
  {Nuclear and Particle Physics Proceedings}\ }\textbf {\bibinfo {volume}
  {265}},\ \bibinfo {pages} {111} (\bibinfo {year} {2015})}\BibitemShut
  {NoStop}%
\bibitem [{\citenamefont {Brugi{\`e}re}(2017)}]{brugiere2017jiangmen}%
  \BibitemOpen
  \bibfield  {author} {\bibinfo {author} {\bibfnamefont {T.}~\bibnamefont
  {Brugi{\`e}re}},\ }\href@noop {} {\bibfield  {journal} {\bibinfo  {journal}
  {Nuclear Instruments and Methods in Physics Research Section A: Accelerators,
  Spectrometers, Detectors and Associated Equipment}\ }\textbf {\bibinfo
  {volume} {845}},\ \bibinfo {pages} {326} (\bibinfo {year}
  {2017})}\BibitemShut {NoStop}%
\bibitem [{\citenamefont {Evans}(2013)}]{evans2013minos}%
  \BibitemOpen
  \bibfield  {author} {\bibinfo {author} {\bibfnamefont {J.}~\bibnamefont
  {Evans}},\ }\href@noop {} {\bibfield  {journal} {\bibinfo  {journal}
  {Advances in High Energy Physics}\ }\textbf {\bibinfo {volume} {2013}}
  (\bibinfo {year} {2013})}\BibitemShut {NoStop}%
\bibitem [{\citenamefont {Group}\ \emph {et~al.}(2014)\citenamefont {Group},
  \citenamefont {Abe}, \citenamefont {Aihara}, \citenamefont {Andreopoulos},
  \citenamefont {Anghel}, \citenamefont {Ariga}, \citenamefont {Ariga},
  \citenamefont {Asfandiyarov}, \citenamefont {Askins}, \citenamefont {Back}
  \emph {et~al.}}]{group2014long}%
  \BibitemOpen
  \bibfield  {author} {\bibinfo {author} {\bibfnamefont {H.-K.~W.}\
  \bibnamefont {Group}}, \bibinfo {author} {\bibfnamefont {K.}~\bibnamefont
  {Abe}}, \bibinfo {author} {\bibfnamefont {H.}~\bibnamefont {Aihara}},
  \bibinfo {author} {\bibfnamefont {C.}~\bibnamefont {Andreopoulos}}, \bibinfo
  {author} {\bibfnamefont {I.}~\bibnamefont {Anghel}}, \bibinfo {author}
  {\bibfnamefont {A.}~\bibnamefont {Ariga}}, \bibinfo {author} {\bibfnamefont
  {T.}~\bibnamefont {Ariga}}, \bibinfo {author} {\bibfnamefont
  {R.}~\bibnamefont {Asfandiyarov}}, \bibinfo {author} {\bibfnamefont
  {M.}~\bibnamefont {Askins}}, \bibinfo {author} {\bibfnamefont
  {J.}~\bibnamefont {Back}},  \emph {et~al.},\ }\href@noop {} {\bibfield
  {journal} {\bibinfo  {journal} {arXiv preprint arXiv:1412.4673}\ } (\bibinfo
  {year} {2014})}\BibitemShut {NoStop}%
\bibitem [{\citenamefont {Ballett}\ \emph {et~al.}(2017)\citenamefont
  {Ballett}, \citenamefont {King}, \citenamefont {Pascoli}, \citenamefont
  {Prouse},\ and\ \citenamefont {Wang}}]{ballett2017sensitivities}%
  \BibitemOpen
  \bibfield  {author} {\bibinfo {author} {\bibfnamefont {P.}~\bibnamefont
  {Ballett}}, \bibinfo {author} {\bibfnamefont {S.~F.}\ \bibnamefont {King}},
  \bibinfo {author} {\bibfnamefont {S.}~\bibnamefont {Pascoli}}, \bibinfo
  {author} {\bibfnamefont {N.~W.}\ \bibnamefont {Prouse}}, \ and\ \bibinfo
  {author} {\bibfnamefont {T.}~\bibnamefont {Wang}},\ }\href@noop {} {\bibfield
   {journal} {\bibinfo  {journal} {Physical Review D}\ }\textbf {\bibinfo
  {volume} {96}},\ \bibinfo {pages} {033003} (\bibinfo {year}
  {2017})}\BibitemShut {NoStop}%
\bibitem [{\citenamefont {Jelley}(1955)}]{jelley1955cerenkov}%
  \BibitemOpen
  \bibfield  {author} {\bibinfo {author} {\bibfnamefont {J.}~\bibnamefont
  {Jelley}},\ }\href@noop {} {\bibfield  {journal} {\bibinfo  {journal}
  {British Journal of Applied Physics}\ }\textbf {\bibinfo {volume} {6}},\
  \bibinfo {pages} {227} (\bibinfo {year} {1955})}\BibitemShut {NoStop}%
\bibitem [{\citenamefont {Lasserre}(2006)}]{lasserre2006double}%
  \BibitemOpen
  \bibfield  {author} {\bibinfo {author} {\bibfnamefont {M.~G.~T.}\
  \bibnamefont {Lasserre}},\ }\href@noop {} {\bibfield  {journal} {\bibinfo
  {journal} {arXiv preprint hep-ex/0606025}\ } (\bibinfo {year}
  {2006})}\BibitemShut {NoStop}%
\bibitem [{\citenamefont {Bellini}\ \emph
  {et~al.}(2012{\natexlab{b}})\citenamefont {Bellini}, \citenamefont
  {Benziger}, \citenamefont {Bick}, \citenamefont {Bonetti}, \citenamefont
  {Bonfini}, \citenamefont {Bravo}, \citenamefont {Avanzini}, \citenamefont
  {Caccianiga}, \citenamefont {Cadonati}, \citenamefont {Calaprice} \emph
  {et~al.}}]{bellini2012first}%
  \BibitemOpen
  \bibfield  {author} {\bibinfo {author} {\bibfnamefont {G.}~\bibnamefont
  {Bellini}}, \bibinfo {author} {\bibfnamefont {J.}~\bibnamefont {Benziger}},
  \bibinfo {author} {\bibfnamefont {D.}~\bibnamefont {Bick}}, \bibinfo {author}
  {\bibfnamefont {S.}~\bibnamefont {Bonetti}}, \bibinfo {author} {\bibfnamefont
  {G.}~\bibnamefont {Bonfini}}, \bibinfo {author} {\bibfnamefont
  {D.}~\bibnamefont {Bravo}}, \bibinfo {author} {\bibfnamefont {M.~B.}\
  \bibnamefont {Avanzini}}, \bibinfo {author} {\bibfnamefont {B.}~\bibnamefont
  {Caccianiga}}, \bibinfo {author} {\bibfnamefont {L.}~\bibnamefont
  {Cadonati}}, \bibinfo {author} {\bibfnamefont {F.}~\bibnamefont {Calaprice}},
   \emph {et~al.},\ }\href@noop {} {\bibfield  {journal} {\bibinfo  {journal}
  {Physical Review Letters}\ }\textbf {\bibinfo {volume} {108}},\ \bibinfo
  {pages} {051302} (\bibinfo {year} {2012}{\natexlab{b}})}\BibitemShut
  {NoStop}%
\bibitem [{\citenamefont {Bellini}\ \emph {et~al.}(2014)\citenamefont
  {Bellini}, \citenamefont {Benziger}, \citenamefont {Bick}, \citenamefont
  {Bonfini}, \citenamefont {Bravo}, \citenamefont {Caccianiga}, \citenamefont
  {Cadonati}, \citenamefont {Calaprice}, \citenamefont {Caminata},
  \citenamefont {Cavalcante} \emph {et~al.}}]{bellini2014neutrinos}%
  \BibitemOpen
  \bibfield  {author} {\bibinfo {author} {\bibfnamefont {G.}~\bibnamefont
  {Bellini}}, \bibinfo {author} {\bibfnamefont {J.}~\bibnamefont {Benziger}},
  \bibinfo {author} {\bibfnamefont {D.}~\bibnamefont {Bick}}, \bibinfo {author}
  {\bibfnamefont {G.}~\bibnamefont {Bonfini}}, \bibinfo {author} {\bibfnamefont
  {D.}~\bibnamefont {Bravo}}, \bibinfo {author} {\bibfnamefont
  {B.}~\bibnamefont {Caccianiga}}, \bibinfo {author} {\bibfnamefont
  {L.}~\bibnamefont {Cadonati}}, \bibinfo {author} {\bibfnamefont
  {F.}~\bibnamefont {Calaprice}}, \bibinfo {author} {\bibfnamefont
  {A.}~\bibnamefont {Caminata}}, \bibinfo {author} {\bibfnamefont
  {P.}~\bibnamefont {Cavalcante}},  \emph {et~al.},\ }\href@noop {} {\bibfield
  {journal} {\bibinfo  {journal} {Nature}\ }\textbf {\bibinfo {volume} {512}},\
  \bibinfo {pages} {383} (\bibinfo {year} {2014})}\BibitemShut {NoStop}%
\bibitem [{\citenamefont {Sousa}\ \emph {et~al.}(2015)\citenamefont {Sousa},
  \citenamefont {MINOS},\ and\ \citenamefont
  {Collaborations)}}]{sousa2015first}%
  \BibitemOpen
  \bibfield  {author} {\bibinfo {author} {\bibfnamefont {A.~B.}\ \bibnamefont
  {Sousa}}, \bibinfo {author} {\bibnamefont {MINOS}}, \ and\ \bibinfo {author}
  {\bibfnamefont {M.}~\bibnamefont {Collaborations)}},\ }in\ \href@noop {}
  {\emph {\bibinfo {booktitle} {AIP Conference Proceedings}}},\ Vol.\ \bibinfo
  {volume} {1666}\ (\bibinfo {organization} {AIP Publishing},\ \bibinfo {year}
  {2015})\ p.\ \bibinfo {pages} {110004}\BibitemShut {NoStop}%
\bibitem [{\citenamefont {Whitehead}(2016)}]{whitehead2016neutrino}%
  \BibitemOpen
  \bibfield  {author} {\bibinfo {author} {\bibfnamefont {L.~H.}\ \bibnamefont
  {Whitehead}},\ }\href@noop {} {\bibfield  {journal} {\bibinfo  {journal}
  {Nuclear Physics B}\ }\textbf {\bibinfo {volume} {908}},\ \bibinfo {pages}
  {130} (\bibinfo {year} {2016})}\BibitemShut {NoStop}%
\bibitem [{\citenamefont {Maneira}\ \emph {et~al.}(2013)\citenamefont
  {Maneira}, \citenamefont {Collaboration} \emph {et~al.}}]{maneira2013sno+}%
  \BibitemOpen
  \bibfield  {author} {\bibinfo {author} {\bibfnamefont {J.}~\bibnamefont
  {Maneira}}, \bibinfo {author} {\bibfnamefont {S.}~\bibnamefont
  {Collaboration}},  \emph {et~al.},\ }in\ \href@noop {} {\emph {\bibinfo
  {booktitle} {Journal of Physics: Conference Series}}},\ Vol.\ \bibinfo
  {volume} {447}\ (\bibinfo {organization} {IOP Publishing},\ \bibinfo {year}
  {2013})\ p.\ \bibinfo {pages} {012065}\BibitemShut {NoStop}%
\bibitem [{\citenamefont {Maneira}(2016)}]{maneira2016status}%
  \BibitemOpen
  \bibfield  {author} {\bibinfo {author} {\bibfnamefont {J.}~\bibnamefont
  {Maneira}},\ }in\ \href@noop {} {\emph {\bibinfo {booktitle} {Journal of
  Physics: Conference Series}}},\ Vol.\ \bibinfo {volume} {718}\ (\bibinfo
  {organization} {IOP Publishing},\ \bibinfo {year} {2016})\ p.\ \bibinfo
  {pages} {062040}\BibitemShut {NoStop}%
\bibitem [{\citenamefont {Andringa}\ \emph {et~al.}(2016)\citenamefont
  {Andringa}, \citenamefont {Arushanova}, \citenamefont {Asahi}, \citenamefont
  {Askins}, \citenamefont {Auty}, \citenamefont {Back}, \citenamefont
  {Barnard}, \citenamefont {Barros}, \citenamefont {Beier}, \citenamefont
  {Bialek} \emph {et~al.}}]{andringa2016current}%
  \BibitemOpen
  \bibfield  {author} {\bibinfo {author} {\bibfnamefont {S.}~\bibnamefont
  {Andringa}}, \bibinfo {author} {\bibfnamefont {E.}~\bibnamefont
  {Arushanova}}, \bibinfo {author} {\bibfnamefont {S.}~\bibnamefont {Asahi}},
  \bibinfo {author} {\bibfnamefont {M.}~\bibnamefont {Askins}}, \bibinfo
  {author} {\bibfnamefont {D.}~\bibnamefont {Auty}}, \bibinfo {author}
  {\bibfnamefont {A.}~\bibnamefont {Back}}, \bibinfo {author} {\bibfnamefont
  {Z.}~\bibnamefont {Barnard}}, \bibinfo {author} {\bibfnamefont
  {N.}~\bibnamefont {Barros}}, \bibinfo {author} {\bibfnamefont
  {E.}~\bibnamefont {Beier}}, \bibinfo {author} {\bibfnamefont
  {A.}~\bibnamefont {Bialek}},  \emph {et~al.},\ }\href@noop {} {\bibfield
  {journal} {\bibinfo  {journal} {Advances in High Energy Physics}\ }\textbf
  {\bibinfo {volume} {2016}} (\bibinfo {year} {2016})}\BibitemShut {NoStop}%
\bibitem [{\citenamefont {Chen}\ \emph {et~al.}(2008)\citenamefont {Chen} \emph
  {et~al.}}]{chen2008sno+}%
  \BibitemOpen
  \bibfield  {author} {\bibinfo {author} {\bibfnamefont {M.~C.}\ \bibnamefont
  {Chen}} \emph {et~al.},\ }\href@noop {} {\bibfield  {journal} {\bibinfo
  {journal} {arXiv preprint arXiv:0810.3694}\ } (\bibinfo {year}
  {2008})}\BibitemShut {NoStop}%
\bibitem [{\citenamefont {Kamdin}(2015)}]{kamdin2015understanding}%
  \BibitemOpen
  \bibfield  {author} {\bibinfo {author} {\bibfnamefont {K.}~\bibnamefont
  {Kamdin}},\ }\href@noop {} {\bibfield  {journal} {\bibinfo  {journal}
  {Physics Procedia}\ }\textbf {\bibinfo {volume} {61}},\ \bibinfo {pages}
  {719} (\bibinfo {year} {2015})}\BibitemShut {NoStop}%
\bibitem [{\citenamefont {Ahn}\ \emph {et~al.}(2010)\citenamefont {Ahn},
  \citenamefont {Collaboration} \emph {et~al.}}]{ahn2010reno}%
  \BibitemOpen
  \bibfield  {author} {\bibinfo {author} {\bibfnamefont {J.}~\bibnamefont
  {Ahn}}, \bibinfo {author} {\bibfnamefont {R.}~\bibnamefont {Collaboration}},
  \emph {et~al.},\ }\href@noop {} {\bibfield  {journal} {\bibinfo  {journal}
  {arXiv preprint arXiv:1003.1391}\ } (\bibinfo {year} {2010})}\BibitemShut
  {NoStop}%
\bibitem [{\citenamefont {Pahlka}\ \emph {et~al.}(2012)\citenamefont {Pahlka},
  \citenamefont {Collaboration} \emph {et~al.}}]{pahlka2012nemo}%
  \BibitemOpen
  \bibfield  {author} {\bibinfo {author} {\bibfnamefont {R.}~\bibnamefont
  {Pahlka}}, \bibinfo {author} {\bibfnamefont {N.-.}\ \bibnamefont
  {Collaboration}},  \emph {et~al.},\ }\href@noop {} {\bibfield  {journal}
  {\bibinfo  {journal} {Nuclear Physics B-Proceedings Supplements}\ }\textbf
  {\bibinfo {volume} {229}},\ \bibinfo {pages} {491} (\bibinfo {year}
  {2012})}\BibitemShut {NoStop}%
\bibitem [{\citenamefont {Guzowski}(2015)}]{guzowski2015construction}%
  \BibitemOpen
  \bibfield  {author} {\bibinfo {author} {\bibfnamefont {P.}~\bibnamefont
  {Guzowski}},\ }in\ \href@noop {} {\emph {\bibinfo {booktitle} {Journal of
  Physics: Conference Series}}},\ Vol.\ \bibinfo {volume} {598}\ (\bibinfo
  {organization} {IOP Publishing},\ \bibinfo {year} {2015})\ p.\ \bibinfo
  {pages} {012020}\BibitemShut {NoStop}%
\bibitem [{\citenamefont {Vilela}\ \emph {et~al.}(2015)\citenamefont {Vilela},
  \citenamefont {collaboration} \emph {et~al.}}]{vilela2015supernemo}%
  \BibitemOpen
  \bibfield  {author} {\bibinfo {author} {\bibfnamefont {C.}~\bibnamefont
  {Vilela}}, \bibinfo {author} {\bibfnamefont {N.}~\bibnamefont
  {collaboration}},  \emph {et~al.},\ }in\ \href@noop {} {\emph {\bibinfo
  {booktitle} {Journal of Physics: Conference Series}}},\ Vol.\ \bibinfo
  {volume} {598}\ (\bibinfo {organization} {IOP Publishing},\ \bibinfo {year}
  {2015})\ p.\ \bibinfo {pages} {012034}\BibitemShut {NoStop}%
\bibitem [{\citenamefont {G{\'o}mez}\ \emph {et~al.}(2016)\citenamefont
  {G{\'o}mez} \emph {et~al.}}]{gomez2016latest}%
  \BibitemOpen
  \bibfield  {author} {\bibinfo {author} {\bibfnamefont {H.}~\bibnamefont
  {G{\'o}mez}} \emph {et~al.},\ }\href@noop {} {\bibfield  {journal} {\bibinfo
  {journal} {Nuclear and Particle Physics Proceedings}\ }\textbf {\bibinfo
  {volume} {273}},\ \bibinfo {pages} {1765} (\bibinfo {year}
  {2016})}\BibitemShut {NoStop}%
\bibitem [{\citenamefont {Waters}\ \emph {et~al.}(2017)\citenamefont {Waters}
  \emph {et~al.}}]{waters2017latest}%
  \BibitemOpen
  \bibfield  {author} {\bibinfo {author} {\bibfnamefont {D.}~\bibnamefont
  {Waters}} \emph {et~al.},\ }in\ \href@noop {} {\emph {\bibinfo {booktitle}
  {Journal of Physics: Conference Series}}},\ Vol.\ \bibinfo {volume} {888}\
  (\bibinfo {organization} {IOP Publishing},\ \bibinfo {year} {2017})\ p.\
  \bibinfo {pages} {012033}\BibitemShut {NoStop}%
\bibitem [{\citenamefont {Arnold}\ \emph {et~al.}(2016)\citenamefont {Arnold},
  \citenamefont {Augier}, \citenamefont {Bakalyarov}, \citenamefont {Baker},
  \citenamefont {Barabash}, \citenamefont {Basharina-Freshville}, \citenamefont
  {Blondel}, \citenamefont {Blot}, \citenamefont {Bongrand}, \citenamefont
  {Brudanin} \emph {et~al.}}]{arnold2016measurement}%
  \BibitemOpen
  \bibfield  {author} {\bibinfo {author} {\bibfnamefont {R.}~\bibnamefont
  {Arnold}}, \bibinfo {author} {\bibfnamefont {C.}~\bibnamefont {Augier}},
  \bibinfo {author} {\bibfnamefont {A.}~\bibnamefont {Bakalyarov}}, \bibinfo
  {author} {\bibfnamefont {J.}~\bibnamefont {Baker}}, \bibinfo {author}
  {\bibfnamefont {A.}~\bibnamefont {Barabash}}, \bibinfo {author}
  {\bibfnamefont {A.}~\bibnamefont {Basharina-Freshville}}, \bibinfo {author}
  {\bibfnamefont {S.}~\bibnamefont {Blondel}}, \bibinfo {author} {\bibfnamefont
  {S.}~\bibnamefont {Blot}}, \bibinfo {author} {\bibfnamefont {M.}~\bibnamefont
  {Bongrand}}, \bibinfo {author} {\bibfnamefont {V.}~\bibnamefont {Brudanin}},
  \emph {et~al.},\ }\href@noop {} {\bibfield  {journal} {\bibinfo  {journal}
  {Physical Review D}\ }\textbf {\bibinfo {volume} {93}},\ \bibinfo {pages}
  {112008} (\bibinfo {year} {2016})}\BibitemShut {NoStop}%
\bibitem [{\citenamefont {Saakyan}(2013)}]{saakyan2013two}%
  \BibitemOpen
  \bibfield  {author} {\bibinfo {author} {\bibfnamefont {R.}~\bibnamefont
  {Saakyan}},\ }\href@noop {} {\bibfield  {journal} {\bibinfo  {journal}
  {Annual Review of Nuclear and Particle Science}\ }\textbf {\bibinfo {volume}
  {63}},\ \bibinfo {pages} {503} (\bibinfo {year} {2013})}\BibitemShut
  {NoStop}%
\bibitem [{\citenamefont {Greiner}\ \emph {et~al.}(2007)\citenamefont
  {Greiner}, \citenamefont {Lachenmaier}, \citenamefont {Jochum},\ and\
  \citenamefont {Cabrera}}]{greiner2007double}%
  \BibitemOpen
  \bibfield  {author} {\bibinfo {author} {\bibfnamefont {D.}~\bibnamefont
  {Greiner}}, \bibinfo {author} {\bibfnamefont {T.}~\bibnamefont
  {Lachenmaier}}, \bibinfo {author} {\bibfnamefont {J.}~\bibnamefont {Jochum}},
  \ and\ \bibinfo {author} {\bibfnamefont {A.}~\bibnamefont {Cabrera}},\
  }\href@noop {} {\bibfield  {journal} {\bibinfo  {journal} {Nuclear
  Instruments and Methods in Physics Research Section A: Accelerators,
  Spectrometers, Detectors and Associated Equipment}\ }\textbf {\bibinfo
  {volume} {581}},\ \bibinfo {pages} {139} (\bibinfo {year}
  {2007})}\BibitemShut {NoStop}%
\bibitem [{\citenamefont {Apollonio}\ \emph {et~al.}(1999)\citenamefont
  {Apollonio}, \citenamefont {Baldini}, \citenamefont {Bemporad}, \citenamefont
  {Caffau}, \citenamefont {Cei}, \citenamefont {Declais}, \citenamefont
  {De~Kerret}, \citenamefont {Dieterle}, \citenamefont {Etenko}, \citenamefont
  {Foresti} \emph {et~al.}}]{apollonio1999determination}%
  \BibitemOpen
  \bibfield  {author} {\bibinfo {author} {\bibfnamefont {M.}~\bibnamefont
  {Apollonio}}, \bibinfo {author} {\bibfnamefont {A.}~\bibnamefont {Baldini}},
  \bibinfo {author} {\bibfnamefont {C.}~\bibnamefont {Bemporad}}, \bibinfo
  {author} {\bibfnamefont {E.}~\bibnamefont {Caffau}}, \bibinfo {author}
  {\bibfnamefont {F.}~\bibnamefont {Cei}}, \bibinfo {author} {\bibfnamefont
  {Y.}~\bibnamefont {Declais}}, \bibinfo {author} {\bibfnamefont
  {H.}~\bibnamefont {De~Kerret}}, \bibinfo {author} {\bibfnamefont
  {B.}~\bibnamefont {Dieterle}}, \bibinfo {author} {\bibfnamefont
  {A.}~\bibnamefont {Etenko}}, \bibinfo {author} {\bibfnamefont
  {L.}~\bibnamefont {Foresti}},  \emph {et~al.},\ }\href@noop {} {\bibfield
  {journal} {\bibinfo  {journal} {Physical Review D}\ }\textbf {\bibinfo
  {volume} {61}},\ \bibinfo {pages} {012001} (\bibinfo {year}
  {1999})}\BibitemShut {NoStop}%
\bibitem [{\citenamefont {Abe}\ \emph {et~al.}(2012)\citenamefont {Abe},
  \citenamefont {Aberle}, \citenamefont {Akiri}, \citenamefont {Dos~Anjos},
  \citenamefont {Ardellier}, \citenamefont {Barbosa}, \citenamefont {Baxter},
  \citenamefont {Bergevin}, \citenamefont {Bernstein}, \citenamefont {Bezerra}
  \emph {et~al.}}]{abe2012indication}%
  \BibitemOpen
  \bibfield  {author} {\bibinfo {author} {\bibfnamefont {Y.}~\bibnamefont
  {Abe}}, \bibinfo {author} {\bibfnamefont {C.}~\bibnamefont {Aberle}},
  \bibinfo {author} {\bibfnamefont {T.}~\bibnamefont {Akiri}}, \bibinfo
  {author} {\bibfnamefont {J.}~\bibnamefont {Dos~Anjos}}, \bibinfo {author}
  {\bibfnamefont {F.}~\bibnamefont {Ardellier}}, \bibinfo {author}
  {\bibfnamefont {A.}~\bibnamefont {Barbosa}}, \bibinfo {author} {\bibfnamefont
  {A.}~\bibnamefont {Baxter}}, \bibinfo {author} {\bibfnamefont
  {M.}~\bibnamefont {Bergevin}}, \bibinfo {author} {\bibfnamefont
  {A.}~\bibnamefont {Bernstein}}, \bibinfo {author} {\bibfnamefont
  {T.}~\bibnamefont {Bezerra}},  \emph {et~al.},\ }\href@noop {} {\bibfield
  {journal} {\bibinfo  {journal} {Physical Review Letters}\ }\textbf {\bibinfo
  {volume} {108}},\ \bibinfo {pages} {131801} (\bibinfo {year}
  {2012})}\BibitemShut {NoStop}%
\bibitem [{\citenamefont {Acciarri}\ \emph {et~al.}(2016)\citenamefont
  {Acciarri}, \citenamefont {Acero}, \citenamefont {Adamowski}, \citenamefont
  {Adams}, \citenamefont {Adamson}, \citenamefont {Adhikari}, \citenamefont
  {Ahmad}, \citenamefont {Albright}, \citenamefont {Alion}, \citenamefont
  {Amador} \emph {et~al.}}]{acciarri2016long}%
  \BibitemOpen
  \bibfield  {author} {\bibinfo {author} {\bibfnamefont {R.}~\bibnamefont
  {Acciarri}}, \bibinfo {author} {\bibfnamefont {M.}~\bibnamefont {Acero}},
  \bibinfo {author} {\bibfnamefont {M.}~\bibnamefont {Adamowski}}, \bibinfo
  {author} {\bibfnamefont {C.}~\bibnamefont {Adams}}, \bibinfo {author}
  {\bibfnamefont {P.}~\bibnamefont {Adamson}}, \bibinfo {author} {\bibfnamefont
  {S.}~\bibnamefont {Adhikari}}, \bibinfo {author} {\bibfnamefont
  {Z.}~\bibnamefont {Ahmad}}, \bibinfo {author} {\bibfnamefont
  {C.}~\bibnamefont {Albright}}, \bibinfo {author} {\bibfnamefont
  {T.}~\bibnamefont {Alion}}, \bibinfo {author} {\bibfnamefont
  {E.}~\bibnamefont {Amador}},  \emph {et~al.},\ }\href@noop {} {\bibfield
  {journal} {\bibinfo  {journal} {arXiv preprint arXiv:1601.02984}\ } (\bibinfo
  {year} {2016})}\BibitemShut {NoStop}%
\bibitem [{\citenamefont {Adams}\ \emph {et~al.}(2013)\citenamefont {Adams},
  \citenamefont {Adams}, \citenamefont {Akiri}, \citenamefont {Alion},
  \citenamefont {Anderson}, \citenamefont {Andreopoulos}, \citenamefont
  {Andrews}, \citenamefont {Anghel}, \citenamefont {Anjos}, \citenamefont
  {Antonello} \emph {et~al.}}]{adams2013long}%
  \BibitemOpen
  \bibfield  {author} {\bibinfo {author} {\bibfnamefont {C.}~\bibnamefont
  {Adams}}, \bibinfo {author} {\bibfnamefont {D.}~\bibnamefont {Adams}},
  \bibinfo {author} {\bibfnamefont {T.}~\bibnamefont {Akiri}}, \bibinfo
  {author} {\bibfnamefont {T.}~\bibnamefont {Alion}}, \bibinfo {author}
  {\bibfnamefont {K.}~\bibnamefont {Anderson}}, \bibinfo {author}
  {\bibfnamefont {C.}~\bibnamefont {Andreopoulos}}, \bibinfo {author}
  {\bibfnamefont {M.}~\bibnamefont {Andrews}}, \bibinfo {author} {\bibfnamefont
  {I.}~\bibnamefont {Anghel}}, \bibinfo {author} {\bibfnamefont {J.~C. C.~d.}\
  \bibnamefont {Anjos}}, \bibinfo {author} {\bibfnamefont {M.}~\bibnamefont
  {Antonello}},  \emph {et~al.},\ }\href@noop {} {\bibfield  {journal}
  {\bibinfo  {journal} {arXiv preprint arXiv:1307.7335}\ } (\bibinfo {year}
  {2013})}\BibitemShut {NoStop}%
\bibitem [{\citenamefont {Carr}\ \emph {et~al.}(2016)\citenamefont {Carr},
  \citenamefont {Lutch},\ and\ \citenamefont {Novella}}]{carr2016new}%
  \BibitemOpen
  \bibfield  {author} {\bibinfo {author} {\bibfnamefont {R.}~\bibnamefont
  {Carr}}, \bibinfo {author} {\bibfnamefont {S.}~\bibnamefont {Lutch}}, \ and\
  \bibinfo {author} {\bibfnamefont {P.}~\bibnamefont {Novella}},\ }\href@noop
  {} {\bibfield  {journal} {\bibinfo  {journal} {Nuclear and particle physics
  proceedings}\ }\textbf {\bibinfo {volume} {273}},\ \bibinfo {pages} {2648}
  (\bibinfo {year} {2016})}\BibitemShut {NoStop}%
\bibitem [{\citenamefont {de~Salas}\ \emph {et~al.}(2017)\citenamefont
  {de~Salas}, \citenamefont {Forero}, \citenamefont {Ternes}, \citenamefont
  {Tortola},\ and\ \citenamefont {Valle}}]{de2017status}%
  \BibitemOpen
  \bibfield  {author} {\bibinfo {author} {\bibfnamefont {P.}~\bibnamefont
  {de~Salas}}, \bibinfo {author} {\bibfnamefont {D.}~\bibnamefont {Forero}},
  \bibinfo {author} {\bibfnamefont {C.}~\bibnamefont {Ternes}}, \bibinfo
  {author} {\bibfnamefont {M.}~\bibnamefont {Tortola}}, \ and\ \bibinfo
  {author} {\bibfnamefont {J.}~\bibnamefont {Valle}},\ }\href@noop {}
  {\bibfield  {journal} {\bibinfo  {journal} {arXiv preprint arXiv:1708.01186}\
  } (\bibinfo {year} {2017})}\BibitemShut {NoStop}%
\bibitem [{\citenamefont {Capozzi}\ \emph {et~al.}(2017)\citenamefont
  {Capozzi}, \citenamefont {Di~Valentino}, \citenamefont {Lisi}, \citenamefont
  {Marrone}, \citenamefont {Melchiorri},\ and\ \citenamefont
  {Palazzo}}]{capozzi2017global}%
  \BibitemOpen
  \bibfield  {author} {\bibinfo {author} {\bibfnamefont {F.}~\bibnamefont
  {Capozzi}}, \bibinfo {author} {\bibfnamefont {E.}~\bibnamefont
  {Di~Valentino}}, \bibinfo {author} {\bibfnamefont {E.}~\bibnamefont {Lisi}},
  \bibinfo {author} {\bibfnamefont {A.}~\bibnamefont {Marrone}}, \bibinfo
  {author} {\bibfnamefont {A.}~\bibnamefont {Melchiorri}}, \ and\ \bibinfo
  {author} {\bibfnamefont {A.}~\bibnamefont {Palazzo}},\ }\href@noop {}
  {\bibfield  {journal} {\bibinfo  {journal} {Physical Review D}\ }\textbf
  {\bibinfo {volume} {95}},\ \bibinfo {pages} {096014} (\bibinfo {year}
  {2017})}\BibitemShut {NoStop}%
\bibitem [{\citenamefont {Castellani}\ \emph {et~al.}(1997)\citenamefont
  {Castellani}, \citenamefont {Degl'Innocenti}, \citenamefont {Fiorentini},
  \citenamefont {Lissia},\ and\ \citenamefont {Ricci}}]{castellani1997solar}%
  \BibitemOpen
  \bibfield  {author} {\bibinfo {author} {\bibfnamefont {V.}~\bibnamefont
  {Castellani}}, \bibinfo {author} {\bibfnamefont {S.}~\bibnamefont
  {Degl'Innocenti}}, \bibinfo {author} {\bibfnamefont {G.}~\bibnamefont
  {Fiorentini}}, \bibinfo {author} {\bibfnamefont {M.}~\bibnamefont {Lissia}},
  \ and\ \bibinfo {author} {\bibfnamefont {B.}~\bibnamefont {Ricci}},\
  }\href@noop {} {\bibfield  {journal} {\bibinfo  {journal} {Physics Reports}\
  }\textbf {\bibinfo {volume} {281}},\ \bibinfo {pages} {309} (\bibinfo {year}
  {1997})}\BibitemShut {NoStop}%
\bibitem [{\citenamefont {McDonald}(2016)}]{mcdonald2016sudbury}%
  \BibitemOpen
  \bibfield  {author} {\bibinfo {author} {\bibfnamefont {A.~B.}\ \bibnamefont
  {McDonald}},\ }\href@noop {} {\bibfield  {journal} {\bibinfo  {journal}
  {Annalen der Physik}\ }\textbf {\bibinfo {volume} {528}},\ \bibinfo {pages}
  {469} (\bibinfo {year} {2016})}\BibitemShut {NoStop}%
\bibitem [{\citenamefont {Seo}\ \emph {et~al.}(2016)\citenamefont {Seo},
  \citenamefont {Collaboration} \emph {et~al.}}]{seo2016recent}%
  \BibitemOpen
  \bibfield  {author} {\bibinfo {author} {\bibfnamefont {H.}~\bibnamefont
  {Seo}}, \bibinfo {author} {\bibfnamefont {R.}~\bibnamefont {Collaboration}},
  \emph {et~al.},\ }in\ \href@noop {} {\emph {\bibinfo {booktitle} {Journal of
  Physics: Conference Series}}},\ Vol.\ \bibinfo {volume} {718}\ (\bibinfo
  {organization} {IOP Publishing},\ \bibinfo {year} {2016})\ p.\ \bibinfo
  {pages} {062053}\BibitemShut {NoStop}%
\bibitem [{\citenamefont {Arpesella}\ \emph {et~al.}(2008)\citenamefont
  {Arpesella}, \citenamefont {Bellini}, \citenamefont {Benziger}, \citenamefont
  {Bonetti}, \citenamefont {Caccianiga}, \citenamefont {Calaprice},
  \citenamefont {Dalnoki-Veress}, \citenamefont {D'Angelo}, \citenamefont
  {de~Kerret}, \citenamefont {Derbin} \emph {et~al.}}]{arpesella2008first}%
  \BibitemOpen
  \bibfield  {author} {\bibinfo {author} {\bibfnamefont {C.}~\bibnamefont
  {Arpesella}}, \bibinfo {author} {\bibfnamefont {G.}~\bibnamefont {Bellini}},
  \bibinfo {author} {\bibfnamefont {J.}~\bibnamefont {Benziger}}, \bibinfo
  {author} {\bibfnamefont {S.}~\bibnamefont {Bonetti}}, \bibinfo {author}
  {\bibfnamefont {B.}~\bibnamefont {Caccianiga}}, \bibinfo {author}
  {\bibfnamefont {F.}~\bibnamefont {Calaprice}}, \bibinfo {author}
  {\bibfnamefont {F.}~\bibnamefont {Dalnoki-Veress}}, \bibinfo {author}
  {\bibfnamefont {D.}~\bibnamefont {D'Angelo}}, \bibinfo {author}
  {\bibfnamefont {H.}~\bibnamefont {de~Kerret}}, \bibinfo {author}
  {\bibfnamefont {A.}~\bibnamefont {Derbin}},  \emph {et~al.},\ }\href@noop {}
  {\bibfield  {journal} {\bibinfo  {journal} {Physics Letters B}\ }\textbf
  {\bibinfo {volume} {658}},\ \bibinfo {pages} {101} (\bibinfo {year}
  {2008})}\BibitemShut {NoStop}%
\bibitem [{\citenamefont {Ranucci}\ \emph {et~al.}(2016)\citenamefont
  {Ranucci}, \citenamefont {Agostini}, \citenamefont {Appel}, \citenamefont
  {Bellini}, \citenamefont {Benziger}, \citenamefont {Bick}, \citenamefont
  {Bonfini}, \citenamefont {Bravo}, \citenamefont {Caccianiga}, \citenamefont
  {Calaprice} \emph {et~al.}}]{ranucci2016overview}%
  \BibitemOpen
  \bibfield  {author} {\bibinfo {author} {\bibfnamefont {G.}~\bibnamefont
  {Ranucci}}, \bibinfo {author} {\bibfnamefont {M.}~\bibnamefont {Agostini}},
  \bibinfo {author} {\bibfnamefont {S.}~\bibnamefont {Appel}}, \bibinfo
  {author} {\bibfnamefont {G.}~\bibnamefont {Bellini}}, \bibinfo {author}
  {\bibfnamefont {J.}~\bibnamefont {Benziger}}, \bibinfo {author}
  {\bibfnamefont {D.}~\bibnamefont {Bick}}, \bibinfo {author} {\bibfnamefont
  {G.}~\bibnamefont {Bonfini}}, \bibinfo {author} {\bibfnamefont
  {D.}~\bibnamefont {Bravo}}, \bibinfo {author} {\bibfnamefont
  {B.}~\bibnamefont {Caccianiga}}, \bibinfo {author} {\bibfnamefont
  {F.}~\bibnamefont {Calaprice}},  \emph {et~al.},\ }in\ \href@noop {} {\emph
  {\bibinfo {booktitle} {Journal of Physics: Conference Series}}},\ Vol.\
  \bibinfo {volume} {675}\ (\bibinfo {organization} {IOP Publishing},\ \bibinfo
  {year} {2016})\ p.\ \bibinfo {pages} {012036}\BibitemShut {NoStop}%
\bibitem [{\citenamefont {Morf{\'\i}n}\ \emph {et~al.}(2012)\citenamefont
  {Morf{\'\i}n}, \citenamefont {Nieves},\ and\ \citenamefont
  {Sobczyk}}]{morfin2012recent}%
  \BibitemOpen
  \bibfield  {author} {\bibinfo {author} {\bibfnamefont {J.~G.}\ \bibnamefont
  {Morf{\'\i}n}}, \bibinfo {author} {\bibfnamefont {J.}~\bibnamefont {Nieves}},
  \ and\ \bibinfo {author} {\bibfnamefont {J.~T.}\ \bibnamefont {Sobczyk}},\
  }\href@noop {} {\bibfield  {journal} {\bibinfo  {journal} {Advances in High
  Energy Physics}\ }\textbf {\bibinfo {volume} {2012}} (\bibinfo {year}
  {2012})}\BibitemShut {NoStop}%
\bibitem [{\citenamefont {Joo}\ \emph {et~al.}(2012)\citenamefont {Joo},
  \citenamefont {Collaboration} \emph {et~al.}}]{joo2012status}%
  \BibitemOpen
  \bibfield  {author} {\bibinfo {author} {\bibfnamefont {K.~K.}\ \bibnamefont
  {Joo}}, \bibinfo {author} {\bibfnamefont {R.}~\bibnamefont {Collaboration}},
  \emph {et~al.},\ }\href@noop {} {\bibfield  {journal} {\bibinfo  {journal}
  {Nuclear Physics B-Proceedings Supplements}\ }\textbf {\bibinfo {volume}
  {229}},\ \bibinfo {pages} {97} (\bibinfo {year} {2012})}\BibitemShut
  {NoStop}%
\bibitem [{\citenamefont {Mueller}\ \emph {et~al.}(2011)\citenamefont {Mueller}
  \emph {et~al.}}]{mueller2011double}%
  \BibitemOpen
  \bibfield  {author} {\bibinfo {author} {\bibfnamefont {T.~A.}\ \bibnamefont
  {Mueller}} \emph {et~al.},\ }in\ \href@noop {} {\emph {\bibinfo {booktitle}
  {Journal of Physics: Conference Series}}},\ Vol.\ \bibinfo {volume} {312}\
  (\bibinfo {organization} {IOP Publishing},\ \bibinfo {year} {2011})\ p.\
  \bibinfo {pages} {072013}\BibitemShut {NoStop}%
\bibitem [{\citenamefont {Lozza}(2014)}]{lozza2014neutrinoless}%
  \BibitemOpen
  \bibfield  {author} {\bibinfo {author} {\bibfnamefont {V.}~\bibnamefont
  {Lozza}},\ }in\ \href@noop {} {\emph {\bibinfo {booktitle} {EPJ Web of
  Conferences}}},\ Vol.~\bibinfo {volume} {65}\ (\bibinfo {organization} {EDP
  Sciences},\ \bibinfo {year} {2014})\ p.\ \bibinfo {pages} {01003}\BibitemShut
  {NoStop}%
\bibitem [{\citenamefont {Bhattacharya}\ \emph {et~al.}(2006)\citenamefont
  {Bhattacharya}, \citenamefont {Collaboration} \emph
  {et~al.}}]{bhattacharya2006india}%
  \BibitemOpen
  \bibfield  {author} {\bibinfo {author} {\bibfnamefont {S.}~\bibnamefont
  {Bhattacharya}}, \bibinfo {author} {\bibfnamefont {I.}~\bibnamefont
  {Collaboration}},  \emph {et~al.},\ }\href@noop {} {\bibfield  {journal}
  {\bibinfo  {journal} {Progress in Particle and Nuclear Physics}\ }\textbf
  {\bibinfo {volume} {57}},\ \bibinfo {pages} {299} (\bibinfo {year}
  {2006})}\BibitemShut {NoStop}%
\bibitem [{\citenamefont {Gomez-Cadenas}\ and\ \citenamefont
  {Harris}(2002)}]{gomez2002physics}%
  \BibitemOpen
  \bibfield  {author} {\bibinfo {author} {\bibfnamefont {J.}~\bibnamefont
  {Gomez-Cadenas}}\ and\ \bibinfo {author} {\bibfnamefont {D.~A.}\ \bibnamefont
  {Harris}},\ }\href@noop {} {\bibfield  {journal} {\bibinfo  {journal} {Annual
  Review of Nuclear and Particle Science}\ }\textbf {\bibinfo {volume} {52}},\
  \bibinfo {pages} {253} (\bibinfo {year} {2002})}\BibitemShut {NoStop}%
\bibitem [{\citenamefont {Calaprice}\ \emph {et~al.}(2012)\citenamefont
  {Calaprice}, \citenamefont {Galbiati}, \citenamefont {Wright},\ and\
  \citenamefont {Ianni}}]{calaprice2012results}%
  \BibitemOpen
  \bibfield  {author} {\bibinfo {author} {\bibfnamefont {F.}~\bibnamefont
  {Calaprice}}, \bibinfo {author} {\bibfnamefont {C.}~\bibnamefont {Galbiati}},
  \bibinfo {author} {\bibfnamefont {A.}~\bibnamefont {Wright}}, \ and\ \bibinfo
  {author} {\bibfnamefont {A.}~\bibnamefont {Ianni}},\ }\href@noop {}
  {\bibfield  {journal} {\bibinfo  {journal} {Annual Review of Nuclear and
  Particle Science}\ }\textbf {\bibinfo {volume} {62}},\ \bibinfo {pages} {315}
  (\bibinfo {year} {2012})}\BibitemShut {NoStop}%
\bibitem [{\citenamefont {An}\ \emph {et~al.}(2016)\citenamefont {An},
  \citenamefont {An}, \citenamefont {An}, \citenamefont {Antonelli},
  \citenamefont {Baussan}, \citenamefont {Beacom}, \citenamefont {Bezrukov},
  \citenamefont {Blyth}, \citenamefont {Brugnera}, \citenamefont {Avanzini}
  \emph {et~al.}}]{an2016neutrino}%
  \BibitemOpen
  \bibfield  {author} {\bibinfo {author} {\bibfnamefont {F.}~\bibnamefont
  {An}}, \bibinfo {author} {\bibfnamefont {G.}~\bibnamefont {An}}, \bibinfo
  {author} {\bibfnamefont {Q.}~\bibnamefont {An}}, \bibinfo {author}
  {\bibfnamefont {V.}~\bibnamefont {Antonelli}}, \bibinfo {author}
  {\bibfnamefont {E.}~\bibnamefont {Baussan}}, \bibinfo {author} {\bibfnamefont
  {J.}~\bibnamefont {Beacom}}, \bibinfo {author} {\bibfnamefont
  {L.}~\bibnamefont {Bezrukov}}, \bibinfo {author} {\bibfnamefont
  {S.}~\bibnamefont {Blyth}}, \bibinfo {author} {\bibfnamefont
  {R.}~\bibnamefont {Brugnera}}, \bibinfo {author} {\bibfnamefont {M.~B.}\
  \bibnamefont {Avanzini}},  \emph {et~al.},\ }\href@noop {} {\bibfield
  {journal} {\bibinfo  {journal} {Journal of Physics G: Nuclear and Particle
  Physics}\ }\textbf {\bibinfo {volume} {43}},\ \bibinfo {pages} {030401}
  (\bibinfo {year} {2016})}\BibitemShut {NoStop}%
\bibitem [{\citenamefont {Kobayashi}(2001)}]{kobayashi2001present}%
  \BibitemOpen
  \bibfield  {author} {\bibinfo {author} {\bibfnamefont {T.}~\bibnamefont
  {Kobayashi}},\ }in\ \href@noop {} {\emph {\bibinfo {booktitle} {Neutrino
  Oscillations And Their Origin}}}\ (\bibinfo  {publisher} {World Scientific},\
  \bibinfo {year} {2001})\ pp.\ \bibinfo {pages} {152--161}\BibitemShut
  {NoStop}%
\bibitem [{\citenamefont {Das}\ \emph {et~al.}(2017)\citenamefont {Das},
  \citenamefont {Maalampi}, \citenamefont {Pulido},\ and\ \citenamefont
  {Vihonen}}]{das2017optimizing}%
  \BibitemOpen
  \bibfield  {author} {\bibinfo {author} {\bibfnamefont {C.~R.}\ \bibnamefont
  {Das}}, \bibinfo {author} {\bibfnamefont {J.}~\bibnamefont {Maalampi}},
  \bibinfo {author} {\bibfnamefont {J.}~\bibnamefont {Pulido}}, \ and\ \bibinfo
  {author} {\bibfnamefont {S.}~\bibnamefont {Vihonen}},\ }in\ \href@noop {}
  {\emph {\bibinfo {booktitle} {Journal of Physics: Conference Series}}},\
  Vol.\ \bibinfo {volume} {888}\ (\bibinfo {organization} {IOP Publishing},\
  \bibinfo {year} {2017})\ p.\ \bibinfo {pages} {012219}\BibitemShut {NoStop}%
\bibitem [{\citenamefont {Cao}(2013)}]{cao2013detection}%
  \BibitemOpen
  \bibfield  {author} {\bibinfo {author} {\bibfnamefont {J.}~\bibnamefont
  {Cao}},\ }\href@noop {} {\bibfield  {journal} {\bibinfo  {journal} {Nuclear
  Instruments and Methods in Physics Research Section A: Accelerators,
  Spectrometers, Detectors and Associated Equipment}\ }\textbf {\bibinfo
  {volume} {732}},\ \bibinfo {pages} {9} (\bibinfo {year} {2013})}\BibitemShut
  {NoStop}%
\bibitem [{\citenamefont {Ianni}(2011)}]{ianni2011neutrino}%
  \BibitemOpen
  \bibfield  {author} {\bibinfo {author} {\bibfnamefont {A.}~\bibnamefont
  {Ianni}},\ }\href@noop {} {\bibfield  {journal} {\bibinfo  {journal}
  {Progress in Particle and Nuclear Physics}\ }\textbf {\bibinfo {volume}
  {66}},\ \bibinfo {pages} {405} (\bibinfo {year} {2011})}\BibitemShut
  {NoStop}%
\bibitem [{\citenamefont {Cremonesi}(2015)}]{cremonesi2015sensitivity}%
  \BibitemOpen
  \bibfield  {author} {\bibinfo {author} {\bibfnamefont {L.}~\bibnamefont
  {Cremonesi}},\ }in\ \href@noop {} {\emph {\bibinfo {booktitle} {Journal of
  Physics: Conference Series}}},\ Vol.\ \bibinfo {volume} {598}\ (\bibinfo
  {organization} {IOP Publishing},\ \bibinfo {year} {2015})\ p.\ \bibinfo
  {pages} {012017}\BibitemShut {NoStop}%
\bibitem [{\citenamefont {Di~Lodovico}\ \emph {et~al.}(2017)\citenamefont
  {Di~Lodovico}, \citenamefont {Collaboration} \emph {et~al.}}]{di2017hyper}%
  \BibitemOpen
  \bibfield  {author} {\bibinfo {author} {\bibfnamefont {F.}~\bibnamefont
  {Di~Lodovico}}, \bibinfo {author} {\bibfnamefont {H.-K.}\ \bibnamefont
  {Collaboration}},  \emph {et~al.},\ }in\ \href@noop {} {\emph {\bibinfo
  {booktitle} {Journal of Physics: Conference Series}}},\ Vol.\ \bibinfo
  {volume} {888}\ (\bibinfo {organization} {IOP Publishing},\ \bibinfo {year}
  {2017})\ p.\ \bibinfo {pages} {012020}\BibitemShut {NoStop}%
\bibitem [{\citenamefont {Kim}\ \emph {et~al.}(2013)\citenamefont {Kim},
  \citenamefont {Lasserre},\ and\ \citenamefont {Wang}}]{kim2013reactor}%
  \BibitemOpen
  \bibfield  {author} {\bibinfo {author} {\bibfnamefont {S.-B.}\ \bibnamefont
  {Kim}}, \bibinfo {author} {\bibfnamefont {T.}~\bibnamefont {Lasserre}}, \
  and\ \bibinfo {author} {\bibfnamefont {Y.}~\bibnamefont {Wang}},\ }\href@noop
  {} {\bibfield  {journal} {\bibinfo  {journal} {Advances in High Energy
  Physics}\ }\textbf {\bibinfo {volume} {2013}} (\bibinfo {year}
  {2013})}\BibitemShut {NoStop}%
\bibitem [{\citenamefont {Timmons}(2017)}]{timmons2017search}%
  \BibitemOpen
  \bibfield  {author} {\bibinfo {author} {\bibfnamefont {A.~M.}\ \bibnamefont
  {Timmons}},\ }\href@noop {} {\emph {\bibinfo {title} {Search for Sterile
  Neutrinos with the MINOS Long-Baseline Experiment}}}\ (\bibinfo  {publisher}
  {Springer},\ \bibinfo {year} {2017})\BibitemShut {NoStop}%
\bibitem [{\citenamefont {Aartsen}\ \emph {et~al.}(2016)\citenamefont
  {Aartsen}, \citenamefont {Abraham}, \citenamefont {Ackermann}, \citenamefont
  {Adams}, \citenamefont {Aguilar}, \citenamefont {Ahlers}, \citenamefont
  {Ahrens}, \citenamefont {Altmann}, \citenamefont {Andeen}, \citenamefont
  {Anderson} \emph {et~al.}}]{aartsen2016searches}%
  \BibitemOpen
  \bibfield  {author} {\bibinfo {author} {\bibfnamefont {M.}~\bibnamefont
  {Aartsen}}, \bibinfo {author} {\bibfnamefont {K.}~\bibnamefont {Abraham}},
  \bibinfo {author} {\bibfnamefont {M.}~\bibnamefont {Ackermann}}, \bibinfo
  {author} {\bibfnamefont {J.}~\bibnamefont {Adams}}, \bibinfo {author}
  {\bibfnamefont {J.}~\bibnamefont {Aguilar}}, \bibinfo {author} {\bibfnamefont
  {M.}~\bibnamefont {Ahlers}}, \bibinfo {author} {\bibfnamefont
  {M.}~\bibnamefont {Ahrens}}, \bibinfo {author} {\bibfnamefont
  {D.}~\bibnamefont {Altmann}}, \bibinfo {author} {\bibfnamefont
  {K.}~\bibnamefont {Andeen}}, \bibinfo {author} {\bibfnamefont
  {T.}~\bibnamefont {Anderson}},  \emph {et~al.},\ }\href@noop {} {\bibfield
  {journal} {\bibinfo  {journal} {Physical review letters}\ }\textbf {\bibinfo
  {volume} {117}},\ \bibinfo {pages} {071801} (\bibinfo {year}
  {2016})}\BibitemShut {NoStop}%
\bibitem [{\citenamefont {Aartsen}\ \emph {et~al.}(2018)\citenamefont
  {Aartsen}, \citenamefont {Ackermann}, \citenamefont {Adams}, \citenamefont
  {Aguilar}, \citenamefont {Ahlers}, \citenamefont {Ahrens}, \citenamefont
  {Al~Samarai}, \citenamefont {Altmann}, \citenamefont {Andeen}, \citenamefont
  {Anderson} \emph {et~al.}}]{aartsen2018measurement}%
  \BibitemOpen
  \bibfield  {author} {\bibinfo {author} {\bibfnamefont {M.}~\bibnamefont
  {Aartsen}}, \bibinfo {author} {\bibfnamefont {M.}~\bibnamefont {Ackermann}},
  \bibinfo {author} {\bibfnamefont {J.}~\bibnamefont {Adams}}, \bibinfo
  {author} {\bibfnamefont {J.}~\bibnamefont {Aguilar}}, \bibinfo {author}
  {\bibfnamefont {M.}~\bibnamefont {Ahlers}}, \bibinfo {author} {\bibfnamefont
  {M.}~\bibnamefont {Ahrens}}, \bibinfo {author} {\bibfnamefont
  {I.}~\bibnamefont {Al~Samarai}}, \bibinfo {author} {\bibfnamefont
  {D.}~\bibnamefont {Altmann}}, \bibinfo {author} {\bibfnamefont
  {K.}~\bibnamefont {Andeen}}, \bibinfo {author} {\bibfnamefont
  {T.}~\bibnamefont {Anderson}},  \emph {et~al.},\ }\href@noop {} {\bibfield
  {journal} {\bibinfo  {journal} {Physical Review Letters}\ }\textbf {\bibinfo
  {volume} {120}},\ \bibinfo {pages} {071801} (\bibinfo {year}
  {2018})}\BibitemShut {NoStop}%
\bibitem [{\citenamefont {Watson}(2011)}]{watson2011discovery}%
  \BibitemOpen
  \bibfield  {author} {\bibinfo {author} {\bibfnamefont {A.~A.}\ \bibnamefont
  {Watson}},\ }\href@noop {} {\bibfield  {journal} {\bibinfo  {journal}
  {Nuclear Physics B-Proceedings Supplements}\ }\textbf {\bibinfo {volume}
  {212}},\ \bibinfo {pages} {13} (\bibinfo {year} {2011})}\BibitemShut
  {NoStop}%
\bibitem [{\citenamefont {Birks}(2013)}]{birks2013theory}%
  \BibitemOpen
  \bibfield  {author} {\bibinfo {author} {\bibfnamefont {J.~B.}\ \bibnamefont
  {Birks}},\ }\href@noop {} {\emph {\bibinfo {title} {The Theory and Practice
  of Scintillation Counting: International Series of Monographs in Electronics
  and Instrumentation}}},\ Vol.~\bibinfo {volume} {27}\ (\bibinfo  {publisher}
  {Elsevier},\ \bibinfo {year} {2013})\BibitemShut {NoStop}%
\bibitem [{\citenamefont {Abe}\ \emph {et~al.}(2013)\citenamefont {Abe},
  \citenamefont {Hayato}, \citenamefont {Iida}, \citenamefont {Iyogi},
  \citenamefont {Kameda}, \citenamefont {Koshio}, \citenamefont {Kozuma},
  \citenamefont {Marti}, \citenamefont {Miura}, \citenamefont {Moriyama} \emph
  {et~al.}}]{abe2013evidence}%
  \BibitemOpen
  \bibfield  {author} {\bibinfo {author} {\bibfnamefont {K.}~\bibnamefont
  {Abe}}, \bibinfo {author} {\bibfnamefont {Y.}~\bibnamefont {Hayato}},
  \bibinfo {author} {\bibfnamefont {T.}~\bibnamefont {Iida}}, \bibinfo {author}
  {\bibfnamefont {K.}~\bibnamefont {Iyogi}}, \bibinfo {author} {\bibfnamefont
  {J.}~\bibnamefont {Kameda}}, \bibinfo {author} {\bibfnamefont
  {Y.}~\bibnamefont {Koshio}}, \bibinfo {author} {\bibfnamefont
  {Y.}~\bibnamefont {Kozuma}}, \bibinfo {author} {\bibfnamefont
  {L.}~\bibnamefont {Marti}}, \bibinfo {author} {\bibfnamefont
  {M.}~\bibnamefont {Miura}}, \bibinfo {author} {\bibfnamefont
  {S.}~\bibnamefont {Moriyama}},  \emph {et~al.},\ }\href@noop {} {\bibfield
  {journal} {\bibinfo  {journal} {Physical review letters}\ }\textbf {\bibinfo
  {volume} {110}},\ \bibinfo {pages} {181802} (\bibinfo {year}
  {2013})}\BibitemShut {NoStop}%
\bibitem [{\citenamefont {De~Santo}(2001)}]{de2001experimentalist}%
  \BibitemOpen
  \bibfield  {author} {\bibinfo {author} {\bibfnamefont {A.}~\bibnamefont
  {De~Santo}},\ }\href@noop {} {\bibfield  {journal} {\bibinfo  {journal}
  {International Journal of Modern Physics A}\ }\textbf {\bibinfo {volume}
  {16}},\ \bibinfo {pages} {4085} (\bibinfo {year} {2001})}\BibitemShut
  {NoStop}%
\bibitem [{\citenamefont {Suzuki}\ and\ \citenamefont
  {Koshiba}(2009)}]{suzuki2009history}%
  \BibitemOpen
  \bibfield  {author} {\bibinfo {author} {\bibfnamefont {A.}~\bibnamefont
  {Suzuki}}\ and\ \bibinfo {author} {\bibfnamefont {M.}~\bibnamefont
  {Koshiba}},\ }\href@noop {} {\bibfield  {journal} {\bibinfo  {journal}
  {Experimental Astronomy}\ }\textbf {\bibinfo {volume} {25}},\ \bibinfo
  {pages} {209} (\bibinfo {year} {2009})}\BibitemShut {NoStop}%
\bibitem [{\citenamefont {Mikaelyan}(2000)}]{mikaelyan2000chooz}%
  \BibitemOpen
  \bibfield  {author} {\bibinfo {author} {\bibfnamefont {L.}~\bibnamefont
  {Mikaelyan}},\ }\href@noop {} {\bibfield  {journal} {\bibinfo  {journal}
  {Nuclear Physics B-Proceedings Supplements}\ }\textbf {\bibinfo {volume}
  {87}},\ \bibinfo {pages} {284} (\bibinfo {year} {2000})}\BibitemShut
  {NoStop}%
\bibitem [{\citenamefont {Apollonio}\ \emph {et~al.}(1998)\citenamefont
  {Apollonio}, \citenamefont {Baldini}, \citenamefont {Bemporad}, \citenamefont
  {Caffau}, \citenamefont {Cei}, \citenamefont {Declais}, \citenamefont
  {De~Kerret}, \citenamefont {Dieterle}, \citenamefont {Etenko}, \citenamefont
  {George} \emph {et~al.}}]{apollonio1998initial}%
  \BibitemOpen
  \bibfield  {author} {\bibinfo {author} {\bibfnamefont {M.}~\bibnamefont
  {Apollonio}}, \bibinfo {author} {\bibfnamefont {A.}~\bibnamefont {Baldini}},
  \bibinfo {author} {\bibfnamefont {C.}~\bibnamefont {Bemporad}}, \bibinfo
  {author} {\bibfnamefont {E.}~\bibnamefont {Caffau}}, \bibinfo {author}
  {\bibfnamefont {F.}~\bibnamefont {Cei}}, \bibinfo {author} {\bibfnamefont
  {Y.}~\bibnamefont {Declais}}, \bibinfo {author} {\bibfnamefont
  {H.}~\bibnamefont {De~Kerret}}, \bibinfo {author} {\bibfnamefont
  {B.}~\bibnamefont {Dieterle}}, \bibinfo {author} {\bibfnamefont
  {A.}~\bibnamefont {Etenko}}, \bibinfo {author} {\bibfnamefont
  {J.}~\bibnamefont {George}},  \emph {et~al.},\ }\href@noop {} {\bibfield
  {journal} {\bibinfo  {journal} {Physics Letters B}\ }\textbf {\bibinfo
  {volume} {420}},\ \bibinfo {pages} {397} (\bibinfo {year}
  {1998})}\BibitemShut {NoStop}%
\bibitem [{\citenamefont {Brugnera}(2011)}]{brugnera2011neutrino}%
  \BibitemOpen
  \bibfield  {author} {\bibinfo {author} {\bibfnamefont {R.}~\bibnamefont
  {Brugnera}},\ }\href@noop {} {\bibfield  {journal} {\bibinfo  {journal}
  {International Journal of Modern Physics A}\ }\textbf {\bibinfo {volume}
  {26}},\ \bibinfo {pages} {4901} (\bibinfo {year} {2011})}\BibitemShut
  {NoStop}%
\bibitem [{\citenamefont {Farzan}\ and\ \citenamefont
  {Tortola}(2017)}]{farzan2017neutrino}%
  \BibitemOpen
  \bibfield  {author} {\bibinfo {author} {\bibfnamefont {Y.}~\bibnamefont
  {Farzan}}\ and\ \bibinfo {author} {\bibfnamefont {M.}~\bibnamefont
  {Tortola}},\ }\href@noop {} {\bibfield  {journal} {\bibinfo  {journal} {arXiv
  preprint arXiv:1710.09360}\ } (\bibinfo {year} {2017})}\BibitemShut {NoStop}%
\bibitem [{\citenamefont {Abe}\ \emph {et~al.}(2011{\natexlab{b}})\citenamefont
  {Abe}, \citenamefont {Abgrall}, \citenamefont {Aihara}, \citenamefont
  {Ajima}, \citenamefont {Albert}, \citenamefont {Allan}, \citenamefont
  {Amaudruz}, \citenamefont {Andreopoulos}, \citenamefont {Andrieu},
  \citenamefont {Anerella} \emph {et~al.}}]{abe2011t2k}%
  \BibitemOpen
  \bibfield  {author} {\bibinfo {author} {\bibfnamefont {K.}~\bibnamefont
  {Abe}}, \bibinfo {author} {\bibfnamefont {N.}~\bibnamefont {Abgrall}},
  \bibinfo {author} {\bibfnamefont {H.}~\bibnamefont {Aihara}}, \bibinfo
  {author} {\bibfnamefont {Y.}~\bibnamefont {Ajima}}, \bibinfo {author}
  {\bibfnamefont {J.}~\bibnamefont {Albert}}, \bibinfo {author} {\bibfnamefont
  {D.}~\bibnamefont {Allan}}, \bibinfo {author} {\bibfnamefont {P.-A.}\
  \bibnamefont {Amaudruz}}, \bibinfo {author} {\bibfnamefont {C.}~\bibnamefont
  {Andreopoulos}}, \bibinfo {author} {\bibfnamefont {B.}~\bibnamefont
  {Andrieu}}, \bibinfo {author} {\bibfnamefont {M.}~\bibnamefont {Anerella}},
  \emph {et~al.},\ }\href@noop {} {\bibfield  {journal} {\bibinfo  {journal}
  {Nuclear Instruments and Methods in Physics Research Section A: Accelerators,
  Spectrometers, Detectors and Associated Equipment}\ }\textbf {\bibinfo
  {volume} {659}},\ \bibinfo {pages} {106} (\bibinfo {year}
  {2011}{\natexlab{b}})}\BibitemShut {NoStop}%
\bibitem [{\citenamefont {Backhouse}(2015)}]{backhouse2015results}%
  \BibitemOpen
  \bibfield  {author} {\bibinfo {author} {\bibfnamefont {C.}~\bibnamefont
  {Backhouse}},\ }in\ \href@noop {} {\emph {\bibinfo {booktitle} {Journal of
  Physics: Conference Series}}},\ Vol.\ \bibinfo {volume} {598}\ (\bibinfo
  {organization} {IOP Publishing},\ \bibinfo {year} {2015})\ p.\ \bibinfo
  {pages} {012004}\BibitemShut {NoStop}%
\bibitem [{\citenamefont {Kim}(2015)}]{kim2015review}%
  \BibitemOpen
  \bibfield  {author} {\bibinfo {author} {\bibfnamefont {S.-B.}\ \bibnamefont
  {Kim}},\ }in\ \href@noop {} {\emph {\bibinfo {booktitle} {Proceedings of the
  2nd International Symposium on Science at J-PARC—Unlocking the Mysteries of
  Life, Matter and the Universe—}}}\ (\bibinfo {year} {2015})\ p.\ \bibinfo
  {pages} {023005}\BibitemShut {NoStop}%
\bibitem [{\citenamefont {Fogli}\ \emph {et~al.}(2006)\citenamefont {Fogli},
  \citenamefont {Lisi}, \citenamefont {Marrone},\ and\ \citenamefont
  {Palazzo}}]{fogli2006global}%
  \BibitemOpen
  \bibfield  {author} {\bibinfo {author} {\bibfnamefont {G.}~\bibnamefont
  {Fogli}}, \bibinfo {author} {\bibfnamefont {E.}~\bibnamefont {Lisi}},
  \bibinfo {author} {\bibfnamefont {A.}~\bibnamefont {Marrone}}, \ and\
  \bibinfo {author} {\bibfnamefont {A.}~\bibnamefont {Palazzo}},\ }\href@noop
  {} {\bibfield  {journal} {\bibinfo  {journal} {Progress in Particle and
  Nuclear Physics}\ }\textbf {\bibinfo {volume} {57}},\ \bibinfo {pages} {742}
  (\bibinfo {year} {2006})}\BibitemShut {NoStop}%
\bibitem [{\citenamefont {Fogli}\ \emph {et~al.}(2002)\citenamefont {Fogli},
  \citenamefont {Lisi}, \citenamefont {Montanino},\ and\ \citenamefont
  {Palazzo}}]{fogli2002supernova}%
  \BibitemOpen
  \bibfield  {author} {\bibinfo {author} {\bibfnamefont {G.~L.}\ \bibnamefont
  {Fogli}}, \bibinfo {author} {\bibfnamefont {E.}~\bibnamefont {Lisi}},
  \bibinfo {author} {\bibfnamefont {D.}~\bibnamefont {Montanino}}, \ and\
  \bibinfo {author} {\bibfnamefont {A.}~\bibnamefont {Palazzo}},\ }\href@noop
  {} {\bibfield  {journal} {\bibinfo  {journal} {Physical Review D}\ }\textbf
  {\bibinfo {volume} {65}},\ \bibinfo {pages} {073008} (\bibinfo {year}
  {2002})}\BibitemShut {NoStop}%
\bibitem [{\citenamefont {Maki}\ \emph {et~al.}(1962)\citenamefont {Maki},
  \citenamefont {Nakagawa},\ and\ \citenamefont {Sakata}}]{maki1962remarks}%
  \BibitemOpen
  \bibfield  {author} {\bibinfo {author} {\bibfnamefont {Z.}~\bibnamefont
  {Maki}}, \bibinfo {author} {\bibfnamefont {M.}~\bibnamefont {Nakagawa}}, \
  and\ \bibinfo {author} {\bibfnamefont {S.}~\bibnamefont {Sakata}},\
  }\href@noop {} {\bibfield  {journal} {\bibinfo  {journal} {Progress of
  Theoretical Physics}\ }\textbf {\bibinfo {volume} {28}},\ \bibinfo {pages}
  {870} (\bibinfo {year} {1962})}\BibitemShut {NoStop}%
\bibitem [{\citenamefont {Nakamura}\ and\ \citenamefont
  {Petcov}(2010)}]{nakamura2010neutrino}%
  \BibitemOpen
  \bibfield  {author} {\bibinfo {author} {\bibfnamefont {K.}~\bibnamefont
  {Nakamura}}\ and\ \bibinfo {author} {\bibfnamefont {S.}~\bibnamefont
  {Petcov}},\ }\href@noop {} {\bibfield  {journal} {\bibinfo  {journal} {K.
  Nakamura et al.(Particle Data Group), J. Phys. G}\ }\textbf {\bibinfo
  {volume} {37}},\ \bibinfo {pages} {075021} (\bibinfo {year}
  {2010})}\BibitemShut {NoStop}%
\bibitem [{\citenamefont {Kodama}\ \emph {et~al.}(2001)\citenamefont {Kodama},
  \citenamefont {Ushida}, \citenamefont {Andreopoulos}, \citenamefont
  {Saoulidou}, \citenamefont {Tzanakos}, \citenamefont {Yager}, \citenamefont
  {Baller}, \citenamefont {Boehnlein}, \citenamefont {Freeman}, \citenamefont
  {Lundberg} \emph {et~al.}}]{kodama2001observation}%
  \BibitemOpen
  \bibfield  {author} {\bibinfo {author} {\bibfnamefont {K.}~\bibnamefont
  {Kodama}}, \bibinfo {author} {\bibfnamefont {N.}~\bibnamefont {Ushida}},
  \bibinfo {author} {\bibfnamefont {C.}~\bibnamefont {Andreopoulos}}, \bibinfo
  {author} {\bibfnamefont {N.}~\bibnamefont {Saoulidou}}, \bibinfo {author}
  {\bibfnamefont {G.}~\bibnamefont {Tzanakos}}, \bibinfo {author}
  {\bibfnamefont {P.}~\bibnamefont {Yager}}, \bibinfo {author} {\bibfnamefont
  {B.}~\bibnamefont {Baller}}, \bibinfo {author} {\bibfnamefont
  {D.}~\bibnamefont {Boehnlein}}, \bibinfo {author} {\bibfnamefont
  {W.}~\bibnamefont {Freeman}}, \bibinfo {author} {\bibfnamefont
  {B.}~\bibnamefont {Lundberg}},  \emph {et~al.},\ }\href@noop {} {\bibfield
  {journal} {\bibinfo  {journal} {Physics Letters B}\ }\textbf {\bibinfo
  {volume} {504}},\ \bibinfo {pages} {218} (\bibinfo {year}
  {2001})}\BibitemShut {NoStop}%
\bibitem [{\citenamefont {Danby}\ \emph {et~al.}(1962)\citenamefont {Danby},
  \citenamefont {Gaillard}, \citenamefont {Goulianos}, \citenamefont
  {Lederman}, \citenamefont {Mistry}, \citenamefont {Schwartz},\ and\
  \citenamefont {Steinberger}}]{danby1962observation}%
  \BibitemOpen
  \bibfield  {author} {\bibinfo {author} {\bibfnamefont {G.}~\bibnamefont
  {Danby}}, \bibinfo {author} {\bibfnamefont {J.}~\bibnamefont {Gaillard}},
  \bibinfo {author} {\bibfnamefont {K.}~\bibnamefont {Goulianos}}, \bibinfo
  {author} {\bibfnamefont {L.}~\bibnamefont {Lederman}}, \bibinfo {author}
  {\bibfnamefont {N.}~\bibnamefont {Mistry}}, \bibinfo {author} {\bibfnamefont
  {M.}~\bibnamefont {Schwartz}}, \ and\ \bibinfo {author} {\bibfnamefont
  {J.}~\bibnamefont {Steinberger}},\ }\href@noop {} {\bibfield  {journal}
  {\bibinfo  {journal} {Physical Review Letters}\ }\textbf {\bibinfo {volume}
  {9}},\ \bibinfo {pages} {36} (\bibinfo {year} {1962})}\BibitemShut {NoStop}%
\bibitem [{\citenamefont {Reines}(1996)}]{reines1996neutrino}%
  \BibitemOpen
  \bibfield  {author} {\bibinfo {author} {\bibfnamefont {F.}~\bibnamefont
  {Reines}},\ }\href@noop {} {\bibfield  {journal} {\bibinfo  {journal}
  {Reviews of Modern Physics}\ }\textbf {\bibinfo {volume} {68}},\ \bibinfo
  {pages} {317} (\bibinfo {year} {1996})}\BibitemShut {NoStop}%
\bibitem [{\citenamefont {Aguilar-Arevaloe}\ \emph {et~al.}()\citenamefont
  {Aguilar-Arevaloe}, \citenamefont {Andersonp}, \citenamefont {Bartoszekg},
  \citenamefont {Bazarkom}, \citenamefont {Briceg}, \citenamefont {Browng},
  \citenamefont {Bugele}, \citenamefont {Caod}, \citenamefont {Coneye},
  \citenamefont {Conrade} \emph {et~al.}}]{aguilarpreprint}%
  \BibitemOpen
  \bibfield  {author} {\bibinfo {author} {\bibfnamefont {A.}~\bibnamefont
  {Aguilar-Arevaloe}}, \bibinfo {author} {\bibfnamefont {C.}~\bibnamefont
  {Andersonp}}, \bibinfo {author} {\bibfnamefont {L.}~\bibnamefont
  {Bartoszekg}}, \bibinfo {author} {\bibfnamefont {A.}~\bibnamefont
  {Bazarkom}}, \bibinfo {author} {\bibfnamefont {S.}~\bibnamefont {Briceg}},
  \bibinfo {author} {\bibfnamefont {B.}~\bibnamefont {Browng}}, \bibinfo
  {author} {\bibfnamefont {L.}~\bibnamefont {Bugele}}, \bibinfo {author}
  {\bibfnamefont {J.}~\bibnamefont {Caod}}, \bibinfo {author} {\bibfnamefont
  {L.}~\bibnamefont {Coneye}}, \bibinfo {author} {\bibfnamefont
  {J.}~\bibnamefont {Conrade}},  \emph {et~al.},\ }\href@noop {} {\
  }\BibitemShut {NoStop}%
\bibitem [{\citenamefont {Adri{\'a}n-Mart{\'\i}nez}\ \emph
  {et~al.}(2016)\citenamefont {Adri{\'a}n-Mart{\'\i}nez}, \citenamefont
  {Ageron}, \citenamefont {Aharonian}, \citenamefont {Aiello}, \citenamefont
  {Albert}, \citenamefont {Ameli}, \citenamefont {Anassontzis}, \citenamefont
  {Androulakis}, \citenamefont {Anghinolfi}, \citenamefont {Anton} \emph
  {et~al.}}]{adrian2016prototype}%
  \BibitemOpen
  \bibfield  {author} {\bibinfo {author} {\bibfnamefont {S.}~\bibnamefont
  {Adri{\'a}n-Mart{\'\i}nez}}, \bibinfo {author} {\bibfnamefont
  {M.}~\bibnamefont {Ageron}}, \bibinfo {author} {\bibfnamefont
  {F.}~\bibnamefont {Aharonian}}, \bibinfo {author} {\bibfnamefont
  {S.}~\bibnamefont {Aiello}}, \bibinfo {author} {\bibfnamefont
  {A.}~\bibnamefont {Albert}}, \bibinfo {author} {\bibfnamefont
  {F.}~\bibnamefont {Ameli}}, \bibinfo {author} {\bibfnamefont
  {E.}~\bibnamefont {Anassontzis}}, \bibinfo {author} {\bibfnamefont
  {G.}~\bibnamefont {Androulakis}}, \bibinfo {author} {\bibfnamefont
  {M.}~\bibnamefont {Anghinolfi}}, \bibinfo {author} {\bibfnamefont
  {G.}~\bibnamefont {Anton}},  \emph {et~al.},\ }\href@noop {} {\bibfield
  {journal} {\bibinfo  {journal} {The European Physical Journal C}\ }\textbf
  {\bibinfo {volume} {76}},\ \bibinfo {pages} {54} (\bibinfo {year}
  {2016})}\BibitemShut {NoStop}%
\bibitem [{\citenamefont {collaboration}\ \emph {et~al.}(2012)\citenamefont
  {collaboration} \emph {et~al.}}]{lbne2012long}%
  \BibitemOpen
  \bibfield  {author} {\bibinfo {author} {\bibfnamefont {L.}~\bibnamefont
  {collaboration}} \emph {et~al.},\ }\href@noop {} {\bibfield  {journal}
  {\bibinfo  {journal} {arXiv preprint arXiv:1204.2295}\ } (\bibinfo {year}
  {2012})}\BibitemShut {NoStop}%
\bibitem [{\citenamefont {Auger}\ \emph {et~al.}(2012)\citenamefont {Auger},
  \citenamefont {Auty}, \citenamefont {Barbeau}, \citenamefont {Bartoszek},
  \citenamefont {Baussan}, \citenamefont {Beauchamp}, \citenamefont
  {Benitez-Medina}, \citenamefont {Breidenbach}, \citenamefont {Chauhan},
  \citenamefont {Cleveland} \emph {et~al.}}]{auger2012exo}%
  \BibitemOpen
  \bibfield  {author} {\bibinfo {author} {\bibfnamefont {M.}~\bibnamefont
  {Auger}}, \bibinfo {author} {\bibfnamefont {D.}~\bibnamefont {Auty}},
  \bibinfo {author} {\bibfnamefont {P.}~\bibnamefont {Barbeau}}, \bibinfo
  {author} {\bibfnamefont {L.}~\bibnamefont {Bartoszek}}, \bibinfo {author}
  {\bibfnamefont {E.}~\bibnamefont {Baussan}}, \bibinfo {author} {\bibfnamefont
  {E.}~\bibnamefont {Beauchamp}}, \bibinfo {author} {\bibfnamefont
  {C.}~\bibnamefont {Benitez-Medina}}, \bibinfo {author} {\bibfnamefont
  {M.}~\bibnamefont {Breidenbach}}, \bibinfo {author} {\bibfnamefont
  {D.}~\bibnamefont {Chauhan}}, \bibinfo {author} {\bibfnamefont
  {B.}~\bibnamefont {Cleveland}},  \emph {et~al.},\ }\href@noop {} {\bibfield
  {journal} {\bibinfo  {journal} {Journal of Instrumentation}\ }\textbf
  {\bibinfo {volume} {7}},\ \bibinfo {pages} {P05010} (\bibinfo {year}
  {2012})}\BibitemShut {NoStop}%
\bibitem [{\citenamefont {Avrorin}\ \emph {et~al.}(2011)\citenamefont
  {Avrorin}, \citenamefont {Aynutdinov}, \citenamefont {Belolaptikov},
  \citenamefont {Berezhnev}, \citenamefont {Bogorodsky}, \citenamefont
  {Budnev}, \citenamefont {Danilchenko}, \citenamefont {Domogatsky},
  \citenamefont {Doroshenko}, \citenamefont {Dyachok} \emph
  {et~al.}}]{avrorin2011gigaton}%
  \BibitemOpen
  \bibfield  {author} {\bibinfo {author} {\bibfnamefont {A.}~\bibnamefont
  {Avrorin}}, \bibinfo {author} {\bibfnamefont {V.}~\bibnamefont {Aynutdinov}},
  \bibinfo {author} {\bibfnamefont {I.}~\bibnamefont {Belolaptikov}}, \bibinfo
  {author} {\bibfnamefont {S.}~\bibnamefont {Berezhnev}}, \bibinfo {author}
  {\bibfnamefont {D.}~\bibnamefont {Bogorodsky}}, \bibinfo {author}
  {\bibfnamefont {N.}~\bibnamefont {Budnev}}, \bibinfo {author} {\bibfnamefont
  {I.}~\bibnamefont {Danilchenko}}, \bibinfo {author} {\bibfnamefont
  {G.}~\bibnamefont {Domogatsky}}, \bibinfo {author} {\bibfnamefont
  {A.}~\bibnamefont {Doroshenko}}, \bibinfo {author} {\bibfnamefont
  {A.}~\bibnamefont {Dyachok}},  \emph {et~al.},\ }\href@noop {} {\bibfield
  {journal} {\bibinfo  {journal} {Nuclear Instruments and Methods in Physics
  Research Section A: Accelerators, Spectrometers, Detectors and Associated
  Equipment}\ }\textbf {\bibinfo {volume} {639}},\ \bibinfo {pages} {30}
  (\bibinfo {year} {2011})}\BibitemShut {NoStop}%
\bibitem [{\citenamefont {Barabanov}\ \emph {et~al.}(1999)\citenamefont
  {Barabanov}, \citenamefont {Beresnev}, \citenamefont {Kornoukhov},
  \citenamefont {Yanovich}, \citenamefont {Zatsepin}, \citenamefont {Danilov},
  \citenamefont {Korpusov}, \citenamefont {Kostikova}, \citenamefont {Krylov},\
  and\ \citenamefont {Yakshin}}]{barabanov1999rare}%
  \BibitemOpen
  \bibfield  {author} {\bibinfo {author} {\bibfnamefont {I.}~\bibnamefont
  {Barabanov}}, \bibinfo {author} {\bibfnamefont {V.}~\bibnamefont {Beresnev}},
  \bibinfo {author} {\bibfnamefont {V.}~\bibnamefont {Kornoukhov}}, \bibinfo
  {author} {\bibfnamefont {E.}~\bibnamefont {Yanovich}}, \bibinfo {author}
  {\bibfnamefont {G.}~\bibnamefont {Zatsepin}}, \bibinfo {author}
  {\bibfnamefont {N.}~\bibnamefont {Danilov}}, \bibinfo {author} {\bibfnamefont
  {G.}~\bibnamefont {Korpusov}}, \bibinfo {author} {\bibfnamefont
  {G.}~\bibnamefont {Kostikova}}, \bibinfo {author} {\bibfnamefont
  {Y.}~\bibnamefont {Krylov}}, \ and\ \bibinfo {author} {\bibfnamefont
  {V.}~\bibnamefont {Yakshin}},\ }\href@noop {} {\bibfield  {journal} {\bibinfo
   {journal} {arXiv preprint physics/9908005}\ } (\bibinfo {year}
  {1999})}\BibitemShut {NoStop}%
\bibitem [{\citenamefont {Bolton}(2005)}]{bolton2005braidwood}%
  \BibitemOpen
  \bibfield  {author} {\bibinfo {author} {\bibfnamefont {T.}~\bibnamefont
  {Bolton}},\ }\href@noop {} {\bibfield  {journal} {\bibinfo  {journal}
  {Nuclear Physics B-Proceedings Supplements}\ }\textbf {\bibinfo {volume}
  {149}},\ \bibinfo {pages} {166} (\bibinfo {year} {2005})}\BibitemShut
  {NoStop}%
\bibitem [{\citenamefont {Buck}\ \emph {et~al.}(2017)\citenamefont {Buck},
  \citenamefont {Lindner},\ and\ \citenamefont {Roca}}]{buck2017scintillation}%
  \BibitemOpen
  \bibfield  {author} {\bibinfo {author} {\bibfnamefont {C.}~\bibnamefont
  {Buck}}, \bibinfo {author} {\bibfnamefont {M.}~\bibnamefont {Lindner}}, \
  and\ \bibinfo {author} {\bibfnamefont {C.}~\bibnamefont {Roca}},\ }in\
  \href@noop {} {\emph {\bibinfo {booktitle} {Journal of Physics: Conference
  Series}}},\ Vol.\ \bibinfo {volume} {888}\ (\bibinfo {organization} {IOP
  Publishing},\ \bibinfo {year} {2017})\ p.\ \bibinfo {pages}
  {012101}\BibitemShut {NoStop}%
\bibitem [{\citenamefont {Dornelas}\ \emph {et~al.}(2016)\citenamefont
  {Dornelas}, \citenamefont {Ara{\'u}jo}, \citenamefont {Cerqueira},
  \citenamefont {Costa},\ and\ \citenamefont
  {N{\'o}brega}}]{dornelas2016front}%
  \BibitemOpen
  \bibfield  {author} {\bibinfo {author} {\bibfnamefont {T.}~\bibnamefont
  {Dornelas}}, \bibinfo {author} {\bibfnamefont {F.}~\bibnamefont
  {Ara{\'u}jo}}, \bibinfo {author} {\bibfnamefont {A.}~\bibnamefont
  {Cerqueira}}, \bibinfo {author} {\bibfnamefont {J.}~\bibnamefont {Costa}}, \
  and\ \bibinfo {author} {\bibfnamefont {R.}~\bibnamefont {N{\'o}brega}},\
  }\href@noop {} {\bibfield  {journal} {\bibinfo  {journal} {Journal of
  Instrumentation}\ }\textbf {\bibinfo {volume} {11}},\ \bibinfo {pages}
  {P07018} (\bibinfo {year} {2016})}\BibitemShut {NoStop}%
\bibitem [{\citenamefont {Kindin}\ \emph {et~al.}(2015)\citenamefont {Kindin},
  \citenamefont {Amelchakov}, \citenamefont {Barbashina}, \citenamefont
  {Bogdanov}, \citenamefont {Burtsev}, \citenamefont {Chernov}, \citenamefont
  {Khokhlov}, \citenamefont {Khomyakov}, \citenamefont {Kokoulin},
  \citenamefont {Kompaniets} \emph {et~al.}}]{kindin2015measuring}%
  \BibitemOpen
  \bibfield  {author} {\bibinfo {author} {\bibfnamefont {V.}~\bibnamefont
  {Kindin}}, \bibinfo {author} {\bibfnamefont {M.}~\bibnamefont {Amelchakov}},
  \bibinfo {author} {\bibfnamefont {N.}~\bibnamefont {Barbashina}}, \bibinfo
  {author} {\bibfnamefont {A.}~\bibnamefont {Bogdanov}}, \bibinfo {author}
  {\bibfnamefont {V.}~\bibnamefont {Burtsev}}, \bibinfo {author} {\bibfnamefont
  {D.}~\bibnamefont {Chernov}}, \bibinfo {author} {\bibfnamefont
  {S.}~\bibnamefont {Khokhlov}}, \bibinfo {author} {\bibfnamefont
  {V.}~\bibnamefont {Khomyakov}}, \bibinfo {author} {\bibfnamefont
  {R.}~\bibnamefont {Kokoulin}}, \bibinfo {author} {\bibfnamefont
  {K.}~\bibnamefont {Kompaniets}},  \emph {et~al.},\ }in\ \href@noop {} {\emph
  {\bibinfo {booktitle} {Journal of Physics: Conference Series}}},\ Vol.\
  \bibinfo {volume} {632}\ (\bibinfo {organization} {IOP Publishing},\ \bibinfo
  {year} {2015})\ p.\ \bibinfo {pages} {012015}\BibitemShut {NoStop}%
\bibitem [{\citenamefont {Kuze}\ \emph {et~al.}(2005)\citenamefont {Kuze},
  \citenamefont {Collaboration} \emph {et~al.}}]{kuze2005kaska}%
  \BibitemOpen
  \bibfield  {author} {\bibinfo {author} {\bibfnamefont {M.}~\bibnamefont
  {Kuze}}, \bibinfo {author} {\bibfnamefont {K.}~\bibnamefont {Collaboration}},
   \emph {et~al.},\ }\href@noop {} {\bibfield  {journal} {\bibinfo  {journal}
  {Nuclear Physics B-Proceedings Supplements}\ }\textbf {\bibinfo {volume}
  {149}},\ \bibinfo {pages} {160} (\bibinfo {year} {2005})}\BibitemShut
  {NoStop}%
\bibitem [{\citenamefont {Labare}(2017)}]{labare2017solid}%
  \BibitemOpen
  \bibfield  {author} {\bibinfo {author} {\bibfnamefont {M.}~\bibnamefont
  {Labare}},\ }in\ \href@noop {} {\emph {\bibinfo {booktitle} {Journal of
  Physics: Conference Series}}},\ Vol.\ \bibinfo {volume} {888}\ (\bibinfo
  {organization} {IOP Publishing},\ \bibinfo {year} {2017})\ p.\ \bibinfo
  {pages} {012180}\BibitemShut {NoStop}%
\bibitem [{\citenamefont {McKinsey}\ and\ \citenamefont
  {Coakley}(2005)}]{mckinsey2005neutrino}%
  \BibitemOpen
  \bibfield  {author} {\bibinfo {author} {\bibfnamefont {D.~N.}\ \bibnamefont
  {McKinsey}}\ and\ \bibinfo {author} {\bibfnamefont {K.}~\bibnamefont
  {Coakley}},\ }\href@noop {} {\bibfield  {journal} {\bibinfo  {journal}
  {Astroparticle Physics}\ }\textbf {\bibinfo {volume} {22}},\ \bibinfo {pages}
  {355} (\bibinfo {year} {2005})}\BibitemShut {NoStop}%
\bibitem [{\citenamefont {Osmanov}(2011)}]{osmanov2011minerva}%
  \BibitemOpen
  \bibfield  {author} {\bibinfo {author} {\bibfnamefont {B.}~\bibnamefont
  {Osmanov}},\ }\href@noop {} {\bibfield  {journal} {\bibinfo  {journal} {arXiv
  preprint arXiv:1109.2855}\ } (\bibinfo {year} {2011})}\BibitemShut {NoStop}%
\bibitem [{\citenamefont {Takei}(2009)}]{takei2009scibar}%
  \BibitemOpen
  \bibfield  {author} {\bibinfo {author} {\bibfnamefont {H.}~\bibnamefont
  {Takei}},\ }in\ \href@noop {} {\emph {\bibinfo {booktitle} {Journal of
  Physics: Conference Series}}},\ Vol.\ \bibinfo {volume} {160}\ (\bibinfo
  {organization} {IOP Publishing},\ \bibinfo {year} {2009})\ p.\ \bibinfo
  {pages} {012034}\BibitemShut {NoStop}%
\bibitem [{\citenamefont {Walding}(2007)}]{walding2007muon}%
  \BibitemOpen
  \bibfield  {author} {\bibinfo {author} {\bibfnamefont {J.}~\bibnamefont
  {Walding}},\ }in\ \href@noop {} {\emph {\bibinfo {booktitle} {AIP Conference
  Proceedings}}},\ Vol.\ \bibinfo {volume} {967}\ (\bibinfo {organization}
  {AIP},\ \bibinfo {year} {2007})\ pp.\ \bibinfo {pages} {289--291}\BibitemShut
  {NoStop}%
\bibitem [{\citenamefont {Wilkes}(2005)}]{wilkes2005uno}%
  \BibitemOpen
  \bibfield  {author} {\bibinfo {author} {\bibfnamefont {R.~J.}\ \bibnamefont
  {Wilkes}},\ }\href@noop {} {\bibfield  {journal} {\bibinfo  {journal} {arXiv
  preprint hep-ex/0507097}\ } (\bibinfo {year} {2005})}\BibitemShut {NoStop}%
\bibitem [{\citenamefont {Hara{\'n}czyk}(2017)}]{haranczyk2017icarus}%
  \BibitemOpen
  \bibfield  {author} {\bibinfo {author} {\bibfnamefont {M.}~\bibnamefont
  {Hara{\'n}czyk}},\ }in\ \href@noop {} {\emph {\bibinfo {booktitle} {Journal
  of Physics: Conference Series}}},\ Vol.\ \bibinfo {volume} {798}\ (\bibinfo
  {organization} {IOP Publishing},\ \bibinfo {year} {2017})\ p.\ \bibinfo
  {pages} {012162}\BibitemShut {NoStop}%
\bibitem [{\citenamefont {Alimonti}\ \emph {et~al.}(2009)\citenamefont
  {Alimonti}, \citenamefont {Arpesella}, \citenamefont {Back}, \citenamefont
  {Balata}, \citenamefont {Bartolomei}, \citenamefont {De~Bellefon},
  \citenamefont {Bellini}, \citenamefont {Benziger}, \citenamefont
  {Bevilacqua}, \citenamefont {Bondi} \emph {et~al.}}]{alimonti2009borexino}%
  \BibitemOpen
  \bibfield  {author} {\bibinfo {author} {\bibfnamefont {G.}~\bibnamefont
  {Alimonti}}, \bibinfo {author} {\bibfnamefont {C.}~\bibnamefont {Arpesella}},
  \bibinfo {author} {\bibfnamefont {H.}~\bibnamefont {Back}}, \bibinfo {author}
  {\bibfnamefont {M.}~\bibnamefont {Balata}}, \bibinfo {author} {\bibfnamefont
  {D.}~\bibnamefont {Bartolomei}}, \bibinfo {author} {\bibfnamefont
  {A.}~\bibnamefont {De~Bellefon}}, \bibinfo {author} {\bibfnamefont
  {G.}~\bibnamefont {Bellini}}, \bibinfo {author} {\bibfnamefont
  {J.}~\bibnamefont {Benziger}}, \bibinfo {author} {\bibfnamefont
  {A.}~\bibnamefont {Bevilacqua}}, \bibinfo {author} {\bibfnamefont
  {D.}~\bibnamefont {Bondi}},  \emph {et~al.},\ }\href@noop {} {\bibfield
  {journal} {\bibinfo  {journal} {Nuclear Instruments and Methods in Physics
  Research Section A: Accelerators, Spectrometers, Detectors and Associated
  Equipment}\ }\textbf {\bibinfo {volume} {600}},\ \bibinfo {pages} {568}
  (\bibinfo {year} {2009})}\BibitemShut {NoStop}%
\bibitem [{\citenamefont {Shirai}(2013)}]{shirai2013kamland}%
  \BibitemOpen
  \bibfield  {author} {\bibinfo {author} {\bibfnamefont {J.}~\bibnamefont
  {Shirai}},\ }\href@noop {} {\bibfield  {journal} {\bibinfo  {journal}
  {Nuclear Physics B-Proceedings Supplements}\ }\textbf {\bibinfo {volume}
  {237}},\ \bibinfo {pages} {28} (\bibinfo {year} {2013})}\BibitemShut
  {NoStop}%
\bibitem [{\citenamefont {Robertson}(2013)}]{robertson2013katrin}%
  \BibitemOpen
  \bibfield  {author} {\bibinfo {author} {\bibfnamefont {R.}~\bibnamefont
  {Robertson}},\ }\href@noop {} {\bibfield  {journal} {\bibinfo  {journal}
  {arXiv preprint arXiv:1307.5486}\ } (\bibinfo {year} {2013})}\BibitemShut
  {NoStop}%
\bibitem [{\citenamefont {Khatun}\ \emph {et~al.}(2018)\citenamefont {Khatun},
  \citenamefont {Thakore},\ and\ \citenamefont {Agarwalla}}]{khatun2018can}%
  \BibitemOpen
  \bibfield  {author} {\bibinfo {author} {\bibfnamefont {A.}~\bibnamefont
  {Khatun}}, \bibinfo {author} {\bibfnamefont {T.}~\bibnamefont {Thakore}}, \
  and\ \bibinfo {author} {\bibfnamefont {S.~K.}\ \bibnamefont {Agarwalla}},\
  }\href@noop {} {\bibfield  {journal} {\bibinfo  {journal} {arXiv preprint
  arXiv:1801.00949}\ } (\bibinfo {year} {2018})}\BibitemShut {NoStop}%
\bibitem [{\citenamefont {Bhattacharya}\ \emph {et~al.}(2014)\citenamefont
  {Bhattacharya}, \citenamefont {Pal}, \citenamefont {Majumder},\ and\
  \citenamefont {Mondal}}]{bhattacharya2014error}%
  \BibitemOpen
  \bibfield  {author} {\bibinfo {author} {\bibfnamefont {K.}~\bibnamefont
  {Bhattacharya}}, \bibinfo {author} {\bibfnamefont {A.~K.}\ \bibnamefont
  {Pal}}, \bibinfo {author} {\bibfnamefont {G.}~\bibnamefont {Majumder}}, \
  and\ \bibinfo {author} {\bibfnamefont {N.~K.}\ \bibnamefont {Mondal}},\
  }\href@noop {} {\bibfield  {journal} {\bibinfo  {journal} {Computer Physics
  Communications}\ }\textbf {\bibinfo {volume} {185}},\ \bibinfo {pages} {3259}
  (\bibinfo {year} {2014})}\BibitemShut {NoStop}%
\bibitem [{\citenamefont {Jediny}\ \emph {et~al.}(2017)\citenamefont {Jediny}
  \emph {et~al.}}]{jediny2017nova}%
  \BibitemOpen
  \bibfield  {author} {\bibinfo {author} {\bibfnamefont {F.}~\bibnamefont
  {Jediny}} \emph {et~al.},\ }\href@noop {} {\bibfield  {journal} {\bibinfo
  {journal} {PoS}\ ,\ \bibinfo {pages} {025}} (\bibinfo {year}
  {2017})}\BibitemShut {NoStop}%
\bibitem [{\citenamefont {Zaborov}\ \emph {et~al.}(2018)\citenamefont {Zaborov}
  \emph {et~al.}}]{zaborov2018km3net}%
  \BibitemOpen
  \bibfield  {author} {\bibinfo {author} {\bibfnamefont {D.}~\bibnamefont
  {Zaborov}} \emph {et~al.},\ }\href@noop {} {\bibfield  {journal} {\bibinfo
  {journal} {arXiv preprint arXiv:1803.08017}\ } (\bibinfo {year}
  {2018})}\BibitemShut {NoStop}%
\bibitem [{\citenamefont {Collaboration}(2016)}]{collaboration2016km3net}%
  \BibitemOpen
  \bibfield  {author} {\bibinfo {author} {\bibfnamefont {K.}~\bibnamefont
  {Collaboration}},\ }\href@noop {} {\bibfield  {journal} {\bibinfo  {journal}
  {Phys. G: Nucl. Part. Phys}\ }\textbf {\bibinfo {volume} {43}},\ \bibinfo
  {pages} {084001} (\bibinfo {year} {2016})}\BibitemShut {NoStop}%
\bibitem [{\citenamefont {Sakharov}(1967)}]{sakharov1967violation}%
  \BibitemOpen
  \bibfield  {author} {\bibinfo {author} {\bibfnamefont {A.~D.}\ \bibnamefont
  {Sakharov}},\ }\href@noop {} {\bibfield  {journal} {\bibinfo  {journal} {JETP
  lett.}\ }\textbf {\bibinfo {volume} {5}},\ \bibinfo {pages} {24} (\bibinfo
  {year} {1967})}\BibitemShut {NoStop}%
\bibitem [{\citenamefont {Brunner}(2013)}]{brunner2013measurement}%
  \BibitemOpen
  \bibfield  {author} {\bibinfo {author} {\bibfnamefont {J.}~\bibnamefont
  {Brunner}},\ }\href@noop {} {\bibfield  {journal} {\bibinfo  {journal}
  {Advances in High Energy Physics}\ }\textbf {\bibinfo {volume} {2013}}
  (\bibinfo {year} {2013})}\BibitemShut {NoStop}%
\bibitem [{\citenamefont
  {V.I.~Garkusha}(2015)}]{гаркушаисследование}%
  \BibitemOpen
  \bibfield  {author} {\bibinfo {author} {\bibfnamefont {F.~N. . A.~S.}\
  \bibnamefont {V.I.~Garkusha}},\ }\href@noop {} {\bibfield  {journal}
  {\bibinfo  {journal} {IHEP Preprint 2015-5}\ } (\bibinfo {year}
  {2015})}\BibitemShut {NoStop}%
\bibitem [{\citenamefont {et~al.}(2010)}]{U-70}%
  \BibitemOpen
  \bibfield  {author} {\bibinfo {author} {\bibfnamefont {N.~E.~T.}\
  \bibnamefont {et~al.}},\ }\href@noop {} {\emph {\bibinfo {title} {Facility
  for intense hadron beams}}},\ edited by\ \bibinfo {editor} {\bibfnamefont
  {e.~a.}\ \bibnamefont {A.~M.~Zaytsev}, \bibfnamefont {V.~A.~Petrov}}\
  (\bibinfo  {publisher} {News and Problems of Fundamental Physics 2 (9)},\
  \bibinfo {year} {2010})\BibitemShut {NoStop}%
\bibitem [{\citenamefont {Wolfenstein}(1978)}]{wolfenstein1978neutrino}%
  \BibitemOpen
  \bibfield  {author} {\bibinfo {author} {\bibfnamefont {L.}~\bibnamefont
  {Wolfenstein}},\ }\href@noop {} {\bibfield  {journal} {\bibinfo  {journal}
  {Physical Review D}\ }\textbf {\bibinfo {volume} {17}},\ \bibinfo {pages}
  {2369} (\bibinfo {year} {1978})}\BibitemShut {NoStop}%
\bibitem [{\citenamefont {Mikheyev}(1985)}]{mikheyev1985sp}%
  \BibitemOpen
  \bibfield  {author} {\bibinfo {author} {\bibfnamefont {S.}~\bibnamefont
  {Mikheyev}},\ }\href@noop {} {\bibfield  {journal} {\bibinfo  {journal} {Sov.
  J. Nucl. Phys.}\ }\textbf {\bibinfo {volume} {42}},\ \bibinfo {pages} {913}
  (\bibinfo {year} {1985})}\BibitemShut {NoStop}%
\bibitem [{\citenamefont {Katz}(2014)}]{katz2014orca}%
  \BibitemOpen
  \bibfield  {author} {\bibinfo {author} {\bibfnamefont {U.~F.}\ \bibnamefont
  {Katz}},\ }\href@noop {} {\bibfield  {journal} {\bibinfo  {journal} {arXiv
  preprint arXiv:1402.1022}\ } (\bibinfo {year} {2014})}\BibitemShut {NoStop}%
\bibitem [{\citenamefont {Migliozzi}\ \emph {et~al.}(2016)\citenamefont
  {Migliozzi}, \citenamefont {Collaboration} \emph
  {et~al.}}]{migliozzi2016high}%
  \BibitemOpen
  \bibfield  {author} {\bibinfo {author} {\bibfnamefont {P.}~\bibnamefont
  {Migliozzi}}, \bibinfo {author} {\bibfnamefont {K.}~\bibnamefont
  {Collaboration}},  \emph {et~al.},\ }in\ \href@noop {} {\emph {\bibinfo
  {booktitle} {Journal of Physics: Conference Series}}},\ Vol.\ \bibinfo
  {volume} {718}\ (\bibinfo {organization} {IOP Publishing},\ \bibinfo {year}
  {2016})\ p.\ \bibinfo {pages} {052024}\BibitemShut {NoStop}%
\bibitem [{\citenamefont {Kouchner}(2016)}]{kouchner2016km3net}%
  \BibitemOpen
  \bibfield  {author} {\bibinfo {author} {\bibfnamefont {A.}~\bibnamefont
  {Kouchner}},\ }in\ \href@noop {} {\emph {\bibinfo {booktitle} {Journal of
  Physics: Conference Series}}},\ Vol.\ \bibinfo {volume} {718}\ (\bibinfo
  {organization} {IOP Publishing},\ \bibinfo {year} {2016})\ p.\ \bibinfo
  {pages} {062030}\BibitemShut {NoStop}%
\bibitem [{\citenamefont {Bruijn}\ and\ \citenamefont {van
  Eijk}(2016)}]{bruijn2016km3net}%
  \BibitemOpen
  \bibfield  {author} {\bibinfo {author} {\bibfnamefont {R.}~\bibnamefont
  {Bruijn}}\ and\ \bibinfo {author} {\bibfnamefont {D.}~\bibnamefont {van
  Eijk}},\ }in\ \href@noop {} {\emph {\bibinfo {booktitle} {The 34th
  International Cosmic Ray Conference}}},\ Vol.\ \bibinfo {volume} {236}\
  (\bibinfo {organization} {SISSA Medialab},\ \bibinfo {year} {2016})\ p.\
  \bibinfo {pages} {1157}\BibitemShut {NoStop}%
\bibitem [{\citenamefont {Aartsen}\ \emph {et~al.}(2017)\citenamefont
  {Aartsen}, \citenamefont {Abraham}, \citenamefont {Ackermann}, \citenamefont
  {Adams}, \citenamefont {Aguilar}, \citenamefont {Ahlers}, \citenamefont
  {Ahrens}, \citenamefont {Altmann}, \citenamefont {Andeen}, \citenamefont
  {Anderson} \emph {et~al.}}]{aartsen2017pingu}%
  \BibitemOpen
  \bibfield  {author} {\bibinfo {author} {\bibfnamefont {M.}~\bibnamefont
  {Aartsen}}, \bibinfo {author} {\bibfnamefont {K.}~\bibnamefont {Abraham}},
  \bibinfo {author} {\bibfnamefont {M.}~\bibnamefont {Ackermann}}, \bibinfo
  {author} {\bibfnamefont {J.}~\bibnamefont {Adams}}, \bibinfo {author}
  {\bibfnamefont {J.}~\bibnamefont {Aguilar}}, \bibinfo {author} {\bibfnamefont
  {M.}~\bibnamefont {Ahlers}}, \bibinfo {author} {\bibfnamefont
  {M.}~\bibnamefont {Ahrens}}, \bibinfo {author} {\bibfnamefont
  {D.}~\bibnamefont {Altmann}}, \bibinfo {author} {\bibfnamefont
  {K.}~\bibnamefont {Andeen}}, \bibinfo {author} {\bibfnamefont
  {T.}~\bibnamefont {Anderson}},  \emph {et~al.},\ }\href@noop {} {\bibfield
  {journal} {\bibinfo  {journal} {Journal of Physics G: Nuclear and Particle
  Physics}\ }\textbf {\bibinfo {volume} {44}},\ \bibinfo {pages} {054006}
  (\bibinfo {year} {2017})}\BibitemShut {NoStop}%
\bibitem [{\citenamefont {Clark}(2016)}]{clark2016pingu}%
  \BibitemOpen
  \bibfield  {author} {\bibinfo {author} {\bibfnamefont {K.}~\bibnamefont
  {Clark}},\ }\href@noop {} {\bibfield  {journal} {\bibinfo  {journal} {Nucl.
  Part. Phys. Proc.}\ }\textbf {\bibinfo {volume} {273}},\ \bibinfo {pages}
  {1870} (\bibinfo {year} {2016})}\BibitemShut {NoStop}%
\bibitem [{\citenamefont {Aartsen}\ \emph {et~al.}(2015)\citenamefont
  {Aartsen}, \citenamefont {Ackermann}, \citenamefont {Adams}, \citenamefont
  {Aguilar}, \citenamefont {Ahlers}, \citenamefont {Ahrens}, \citenamefont
  {Altmann}, \citenamefont {Anderson}, \citenamefont {Arguelles}, \citenamefont
  {Arlen} \emph {et~al.}}]{aartsen2015determining}%
  \BibitemOpen
  \bibfield  {author} {\bibinfo {author} {\bibfnamefont {M.}~\bibnamefont
  {Aartsen}}, \bibinfo {author} {\bibfnamefont {M.}~\bibnamefont {Ackermann}},
  \bibinfo {author} {\bibfnamefont {J.}~\bibnamefont {Adams}}, \bibinfo
  {author} {\bibfnamefont {J.}~\bibnamefont {Aguilar}}, \bibinfo {author}
  {\bibfnamefont {M.}~\bibnamefont {Ahlers}}, \bibinfo {author} {\bibfnamefont
  {M.}~\bibnamefont {Ahrens}}, \bibinfo {author} {\bibfnamefont
  {D.}~\bibnamefont {Altmann}}, \bibinfo {author} {\bibfnamefont
  {T.}~\bibnamefont {Anderson}}, \bibinfo {author} {\bibfnamefont
  {C.}~\bibnamefont {Arguelles}}, \bibinfo {author} {\bibfnamefont
  {T.}~\bibnamefont {Arlen}},  \emph {et~al.},\ }\href@noop {} {\bibfield
  {journal} {\bibinfo  {journal} {Physical Review D}\ }\textbf {\bibinfo
  {volume} {91}},\ \bibinfo {pages} {072004} (\bibinfo {year}
  {2015})}\BibitemShut {NoStop}%
\bibitem [{\citenamefont {Gonzalez-Garcia}(2004)}]{gonzalez2004mc}%
  \BibitemOpen
  \bibfield  {author} {\bibinfo {author} {\bibfnamefont {M.}~\bibnamefont
  {Gonzalez-Garcia}},\ }\href@noop {} {\bibfield  {journal} {\bibinfo
  {journal} {Phys. Rev. D}\ }\textbf {\bibinfo {volume} {70}},\ \bibinfo
  {pages} {093005} (\bibinfo {year} {2004})}\BibitemShut {NoStop}%
\bibitem [{\citenamefont {Barger}\ \emph {et~al.}(2012)\citenamefont {Barger},
  \citenamefont {Gandhi}, \citenamefont {Ghoshal}, \citenamefont {Goswami},
  \citenamefont {Marfatia}, \citenamefont {Prakash}, \citenamefont {Raut},\
  and\ \citenamefont {Sankar}}]{barger2012neutrino}%
  \BibitemOpen
  \bibfield  {author} {\bibinfo {author} {\bibfnamefont {V.}~\bibnamefont
  {Barger}}, \bibinfo {author} {\bibfnamefont {R.}~\bibnamefont {Gandhi}},
  \bibinfo {author} {\bibfnamefont {P.}~\bibnamefont {Ghoshal}}, \bibinfo
  {author} {\bibfnamefont {S.}~\bibnamefont {Goswami}}, \bibinfo {author}
  {\bibfnamefont {D.}~\bibnamefont {Marfatia}}, \bibinfo {author}
  {\bibfnamefont {S.}~\bibnamefont {Prakash}}, \bibinfo {author} {\bibfnamefont
  {S.~K.}\ \bibnamefont {Raut}}, \ and\ \bibinfo {author} {\bibfnamefont
  {S.~U.}\ \bibnamefont {Sankar}},\ }\href@noop {} {\bibfield  {journal}
  {\bibinfo  {journal} {Physical review letters}\ }\textbf {\bibinfo {volume}
  {109}},\ \bibinfo {pages} {091801} (\bibinfo {year} {2012})}\BibitemShut
  {NoStop}%
\bibitem [{\citenamefont {Chatterjee}\ \emph {et~al.}(2013)\citenamefont
  {Chatterjee}, \citenamefont {Ghoshal}, \citenamefont {Goswami},\ and\
  \citenamefont {Raut}}]{chatterjee2013octant}%
  \BibitemOpen
  \bibfield  {author} {\bibinfo {author} {\bibfnamefont {A.}~\bibnamefont
  {Chatterjee}}, \bibinfo {author} {\bibfnamefont {P.}~\bibnamefont {Ghoshal}},
  \bibinfo {author} {\bibfnamefont {S.}~\bibnamefont {Goswami}}, \ and\
  \bibinfo {author} {\bibfnamefont {S.~K.}\ \bibnamefont {Raut}},\ }\href@noop
  {} {\bibfield  {journal} {\bibinfo  {journal} {Journal of High Energy
  Physics}\ }\textbf {\bibinfo {volume} {2013}},\ \bibinfo {pages} {10}
  (\bibinfo {year} {2013})}\BibitemShut {NoStop}%
\bibitem [{\citenamefont {Agarwalla}\ \emph {et~al.}(2013)\citenamefont
  {Agarwalla}, \citenamefont {Prakash},\ and\ \citenamefont
  {Sankar}}]{agarwalla2013resolving}%
  \BibitemOpen
  \bibfield  {author} {\bibinfo {author} {\bibfnamefont {S.~K.}\ \bibnamefont
  {Agarwalla}}, \bibinfo {author} {\bibfnamefont {S.}~\bibnamefont {Prakash}},
  \ and\ \bibinfo {author} {\bibfnamefont {S.~U.}\ \bibnamefont {Sankar}},\
  }\href@noop {} {\bibfield  {journal} {\bibinfo  {journal} {Journal of High
  Energy Physics}\ }\textbf {\bibinfo {volume} {2013}},\ \bibinfo {pages} {131}
  (\bibinfo {year} {2013})}\BibitemShut {NoStop}%
\bibitem [{\citenamefont {Winter}(2016)}]{winter2016atmospheric}%
  \BibitemOpen
  \bibfield  {author} {\bibinfo {author} {\bibfnamefont {W.}~\bibnamefont
  {Winter}},\ }\href@noop {} {\bibfield  {journal} {\bibinfo  {journal}
  {Nuclear Physics B}\ }\textbf {\bibinfo {volume} {908}},\ \bibinfo {pages}
  {250} (\bibinfo {year} {2016})}\BibitemShut {NoStop}%
\bibitem [{\citenamefont {Rubbia}(1977)}]{rubbia1977liquid}%
  \BibitemOpen
  \bibfield  {author} {\bibinfo {author} {\bibfnamefont {C.}~\bibnamefont
  {Rubbia}},\ }\href@noop {} {\emph {\bibinfo {title} {The liquid argon time
  projection chamber: a new concept for neutrino detectors}}},\ \bibinfo {type}
  {Tech. Rep.}\ (\bibinfo {year} {1977})\BibitemShut {NoStop}%
\bibitem [{\citenamefont {Nygren}(1974)}]{nygren1974proposal}%
  \BibitemOpen
  \bibfield  {author} {\bibinfo {author} {\bibfnamefont {D.~R.}\ \bibnamefont
  {Nygren}},\ }\href@noop {} {\  (\bibinfo {year} {1974})}\BibitemShut
  {NoStop}%
\bibitem [{\citenamefont {Miyajima}\ \emph {et~al.}(1974)\citenamefont
  {Miyajima}, \citenamefont {Takahashi}, \citenamefont {Konno}, \citenamefont
  {Hamada}, \citenamefont {Kubota}, \citenamefont {Shibamura},\ and\
  \citenamefont {Doke}}]{miyajima1974average}%
  \BibitemOpen
  \bibfield  {author} {\bibinfo {author} {\bibfnamefont {M.}~\bibnamefont
  {Miyajima}}, \bibinfo {author} {\bibfnamefont {T.}~\bibnamefont {Takahashi}},
  \bibinfo {author} {\bibfnamefont {S.}~\bibnamefont {Konno}}, \bibinfo
  {author} {\bibfnamefont {T.}~\bibnamefont {Hamada}}, \bibinfo {author}
  {\bibfnamefont {S.}~\bibnamefont {Kubota}}, \bibinfo {author} {\bibfnamefont
  {H.}~\bibnamefont {Shibamura}}, \ and\ \bibinfo {author} {\bibfnamefont
  {T.}~\bibnamefont {Doke}},\ }\href@noop {} {\bibfield  {journal} {\bibinfo
  {journal} {Physical Review A}\ }\textbf {\bibinfo {volume} {9}},\ \bibinfo
  {pages} {1438} (\bibinfo {year} {1974})}\BibitemShut {NoStop}%
\bibitem [{\citenamefont {Hofmann}\ \emph {et~al.}(1976)\citenamefont
  {Hofmann}, \citenamefont {Klein}, \citenamefont {Schulz}, \citenamefont
  {Spengler},\ and\ \citenamefont {Wegener}}]{hofmann1976production}%
  \BibitemOpen
  \bibfield  {author} {\bibinfo {author} {\bibfnamefont {W.}~\bibnamefont
  {Hofmann}}, \bibinfo {author} {\bibfnamefont {U.}~\bibnamefont {Klein}},
  \bibinfo {author} {\bibfnamefont {M.}~\bibnamefont {Schulz}}, \bibinfo
  {author} {\bibfnamefont {J.}~\bibnamefont {Spengler}}, \ and\ \bibinfo
  {author} {\bibfnamefont {D.}~\bibnamefont {Wegener}},\ }\href@noop {}
  {\bibfield  {journal} {\bibinfo  {journal} {Nuclear Instruments and Methods}\
  }\textbf {\bibinfo {volume} {135}},\ \bibinfo {pages} {151} (\bibinfo {year}
  {1976})}\BibitemShut {NoStop}%
\end{thebibliography}%

\end{document}